\documentclass[twoside,a4paper,10pt]{book}

\usepackage{amsmath}
\usepackage{amssymb}
\usepackage{graphicx}
\usepackage{color}
\usepackage{fancyhdr}              	
\usepackage{multirow}

\let\a=\alpha \let\b=\beta  \let\g=\gamma  \let\d=\delta \let\e=\varepsilon
  \let\h=\eta   \let\th=\theta  
\let\m=\mu    \let\n=\nu             \let\r=\rho
\let\s=\sigma \let\t=\tau   \let\f=\varphi \let\ph=\varphi
   
\let\G=\Gamma \let\D=\Delta   
    \let\Si=\Sigma     
\let\O=\Omega 
\let\ee=\epsilon

\let\==\equiv
\let\io=\infty 
\def\ie{{i.e. }}
\def\eg{{e.g. }}
\let\dpr=\partial

 \def\HH{{\cal H}}

\def\GG{{\cal G}} \def\SS{{\cal S}}
  
\def\ZZ{{\cal Z}}

\def\erf{\text{erf}}

\def\ol#1{{\overline #1}}

\def\qed{\raise1pt\hbox{\vrule height5pt width5pt depth0pt}}

\def\to{\rightarrow}
\def\la{\left\langle}
\def\ra{\right\rangle}

\def\wt{\widetilde}

\mathchardef\aa   = "050B
\mathchardef\bb   = "050C
\mathchardef\ggg  = "050D
\mathchardef\xxx  = "0518
\mathchardef\zzzzz= "0510
\mathchardef\oo   = "0521
\mathchardef\lll  = "0515
\mathchardef\mm   = "0516
\mathchardef\Dp   = "0540
\mathchardef\H    = "0548
\mathchardef\FFF  = "0546
\mathchardef\ppp  = "0570
\mathchardef\nn   = "0517
\mathchardef\ff   = "0527
\mathchardef\pps  = "0520
\mathchardef\FFF  = "0508
\mathchardef\nnnnn= "056E

\newcommand{\beq}{\begin{equation}}
\newcommand{\eeq}{\end{equation}}
\newcommand{\bea}{\begin{eqnarray}}
\newcommand{\eea}{\end{eqnarray}}

\begin{document}

\frontmatter

\begin{titlepage}

\thispagestyle{empty}
\begin{center}
{\large \textsc{ECOLE NORMALE SUPERIEURE DE PARIS}} \\

\vspace{1cm}

\includegraphics[width=3cm]{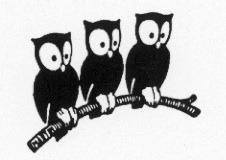}

\vspace{1cm}

{\large \textbf{D\'epartement de Physique}} \\[.5cm]
{\large \textbf{Th\`ese d'habilitation \`a diriger des recherches}}\\[.25cm]

\vspace{2.5cm}

{\huge

\textbf{Theory of Simple Glasses} 

} 
\end{center}

\vspace{2.8cm}
{\large

\noindent Committee: \hfill Candidate:

\vspace{.2cm}

\noindent Prof.~\textsl{Werner Krauth} (president) 
\hfill Dr.~\textsl{Francesco Zamponi}\\[1mm]
\noindent Prof.~\textsl{Christoph Dellago} \\[1mm]
\noindent Prof.~\textsl{Benoit Doucot} \\[1mm]
\noindent Prof.~\textsl{Juan P. Garrahan} (referee) \\[1mm]
\noindent Prof.~\textsl{Enzo Marinari} (referee) \\[1mm]
\noindent Prof.~\textsl{Gilles Tarjus} (referee) \\[1mm]
}

\vspace{1.2cm}

\begin{center}
{\large {Discussed on November 30, 2012}}
\end{center}

\end{titlepage}

\pagestyle{empty}
\pagenumbering{roman}
\mbox{}
\newpage

\setcounter{page}{1}
\tableofcontents
\newpage
\pagestyle{fancy}

\addcontentsline{toc}{chapter}{Introduction}

\chapter*{Introduction}

In the French academic system, the {\it Habilitation \`a diriger des recherches}, or HDR, is a title that allows one to supervise the research of PhD students.
Therefore, to obtain it, the candidate is required to prove that he has a wide enough view of his
research subjects, and the capability to build a comprehensive research project, such that
he can guide his research, and that of others, in the short and long runs.
At {\it Ecole Normale Sup\'erieure}, the length of the thesis is supposed to be around 40 pages.

Given these constraints, I decided to focus on one of my research subjects only, on which I have been working continuously
over the last eight years, and where I strongly contributed to the design of the research. 
This is the theory of the glass and jamming transitions of hard spheres.
On this topic, I recently wrote a very detailed review paper~\cite{PZ10}
as well as a long research paper~\cite{BJZ11}. I refer to these two papers for most of the technical details.
Here I only tried to explain this subject in a concise and self-contained way,
highlighting what I consider the main motivations for its study and trying to put my work in a broader perspective.
Basically I tried to tell this story in the same way as I would tell it to a student who contacted me to start a PhD thesis on these topics.

The glass problem has a reputation of being controversial. 
It has even been said that ``There are more theories of the glass transition than there are theorists who propose them''~\cite{NYT}. 
Part of this controversy is fully justified. The problem is indeed extremely challenging, and it has long been difficult for theory 
to provide precise predictions that could be experimentally tested, in order to obtain conclusive answers~\cite{Ta11}.
A lot of progress has nonetheless been achieved thanks to the impressive numerical and experimental developments of the last couple of decades, and
theory is rapidly evolving in the aim to match this effort.

Yet part of the controversy may also be philosophical\footnote{
I also fear that another part of the problem may be systemic. The institutional organization of scientific research seems sometimes designed to promote the radicalization of scientific discussions. Most of us experience a strong pressure to highlight the difference between one's theory and those of others, in order to publish our papers, raise more funds, etc. For example, a journalist was recently interested in writing about the publication of one of my papers for a prestigious journal, but the journal's editor refused to publish the news piece when he was told that our results did not generate sufficient controversy. Our results partially agreed with an existing theory, so they were considered not interesting enough!
}. 
There seems to be some disagreement about what one expects from a ÒtheoryÓ of the glass transition. Before proceeding, I thus find it useful to briefly review this issue, in order to clarify my scientific intent.
I will use a convenient definition of ``scientific theory'' that has been discussed by Lucio Russo in~\cite[Sec~1.3]{Russo} (together with a nice discussion of its limits that is not reproduced
here). According to this definition, a ``scientific theory'' must have the three following properties:
\begin{enumerate}
\item Its statements do not concern concrete objects pertaining to the real world, but specific abstract objects. 
\item It has a deductive structure: it is made by a few postulates concerning its objects, and by a method to derive from
them a potentially infinite number of consequences.
\item Its application to the real world is based on a series of ``correspondence rules'' between the objects of the theory and those of the real world.
\end{enumerate}
According to Russo, a useful criterion to determine whether a theory has these properties is to check if one can compile a collection of exercises,
that can be solved within the theory. 
Indeed, solving a problem in the context of the theory is nothing else but an (arbitrarily difficult) ``exercise''.
This definition is of course very restrictive, as it rules out all phenomenological and empirical knowledge in science. The aim here is not to claim
any superiority of ``scientific theories'' with respect to other more empirical or descriptive forms of science, but to provide
{\it a} precise definition of a theory, in order to avoid useless controversies based on disagreement on what kind of ``theory'' one is looking for.

``Scientific theories'' are powerful for two reasons: 
\begin{itemize}
\item[{\it (i)}] Working on two parallel
levels (the ``model'' and the real world) allows for a very flexible way of reasoning, and in particular one can guarantee the ``truth''
of scientific statements by limiting them to the domain of the theory. 
\item[{\it (ii)}] The theory can be extended, by using the deductive method and introducing new correspondence rules,
to treat situations that were not a priori included in the initial objectives for which the theory was developed.
\end{itemize}
At the same time, it is important to keep in mind that {\it any ``scientific theory'' has a limited utility, as in general it can only be used
to model phenomena that are not too ``far'' from those that motivated its elaboration.  Theories that become inadequate to describe a new phenomenology
must be, for this reason, substituted; they remain however, according to our definition, ``scientific theories'' and one can continue to use them in their domain
of validity}\footnote{For instance, Newtonian mechanics did not become useless once quantum mechanics was developed, and the fact that it gives incorrect
predictions --e.g. the instability of the hydrogen atom-- does not mean that it is plain wrong.}~\cite{Russo}.
This last statement is particularly important because it reminds us that
{\it the} theory of a given class of phenomena --e.g. the glass transition-- will never 
exist\footnote{
Contemporary popular literature on science sometimes forgets this point, e.g. when talking about particle physics.
}: 
scientific theories are never unique nor everlasting. 
They are models of reality, and there is no problem in using different models of the same phenomenon, 
and in replacing current models with more powerful
ones when they are found.

I will argue in this thesis that the so-called Random First Order Transition, or RFOT, theory --on which my work is based-- is indeed {\it a} scientific theory 
of the glass transition according to the definition above.
In particular, I will present a striking example of the usefulness of property {\it (ii)} above: RFOT theory, 
originally born as a theory of the thermal glass transition, can be surprisingly
extended to describe the physics of the athermal jamming transition that happens inside the glass phase.
One of the main criticism to RFOT theory, which has been a source of endless debates, 
is that it predicts the existence of a phase transition (the Kauzmann transition) that, in the real world, is at best unobservable and
at worst non existent. However, according to the definition above, this is not a big issue: it just means that this ``abstract object'' is not related by a correspondence
rule to a real world phenomenon, and is just a tool that is used within the theory to model other, observable, 
phenomena\footnote{A similar situation was present in the standard model of particle physics until a few months ago, when the discovery of the Higgs boson
was announced.
But even if the Higgs boson was not discovered, the standard model would have retained its validity as a theory of low-energy particle physics.
The Higgs boson would have been just an object of the theory, without any real world correspondent.
}.
Other theories of the glass transition, for instance those based on the idea of dynamic facilitation, 
provide in my opinion nice complementary views
on the problem. 
Overall, 
exciting progresses have been made, and even if we don't have --and will never have-- {\it the} theory of the glass transition,
we learned how to build consistent models of many phenomena related to it and there is reason to hope that even bigger progresses will be made in the near future.

In the following,
Chap.~\ref{chap1} provides an introductory account of RFOT theory, while Chap.~\ref{chap2} reviews my original work on this problem.
A final comment on the title of this thesis is needed here. The approach to the glass transition that I will present is really similar in spirit to the one of the celebrated
book {\it Theory of simple liquids} by Hansen and McDonald~\cite{Hansen}. This approach is in fact based on a natural extension of the techniques used in that book to describe
``simple'' liquids --monoatomic ones, modeled as hard spheres or point particles interacting through simple potentials such as the Lennard-Jones-- to the glassy
states that are obtained by cooling or compressing such liquids.
I hope that soon enough the theory will be so much developed that someone will write a book with this title, which would be the natural ``volume 2'' of a series on the amorphous
states of matter,  ``volume 1'' being~\cite{Hansen}.

\mainmatter

\pagenumbering{arabic}
\setcounter{page}{1}


\chapter{A theory of the glass transition}
\label{chap1}


\section{Basic facts}
\label{sec:basic_facts}

\subsection{Viscosity and structural relaxation time}
\label{sec:viscosity}

\begin{figure}[t]
\centering
\includegraphics[height=230pt]{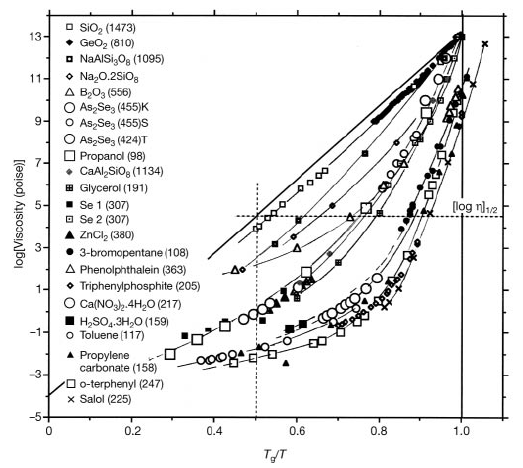}
\includegraphics[height=230pt]{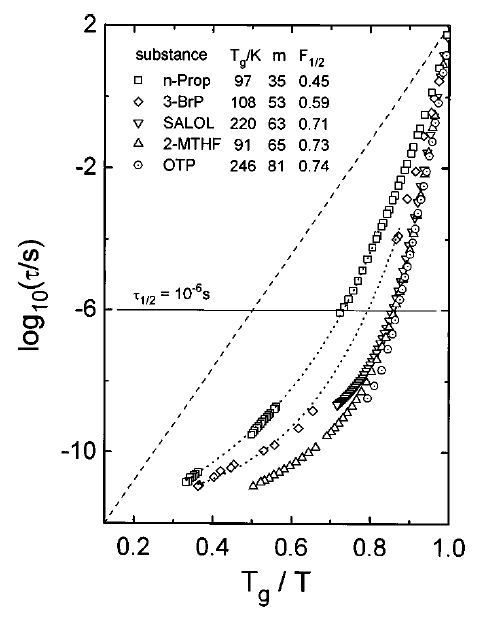}
\caption[Viscosity and relaxation time of many glass formers]
{(Left, from~\cite{MA01}) Viscosity data for many glass forming liquids.
The logarithm of the viscosity measured in Poise is reported as a function
of $T_g/T$. The (calorimetric) $T_g$ is defined as the temperature at which 
the enthalpy relaxation time is $\sim 200$s, and its value is 
reported in parenthesis in the key of the figure. 
Note that for some systems the value of the calorimetric
$T_g$ does not satisfy exactly the condition $\h(T_g)=10^{13}$ Poise. 
Fragility is the slope of the curves in $T_g/T=1$.
(Right, from~\cite{RA98}) Structural relaxation time obtained from 
dielectric relaxation measurements.
The dashed line indicates Arrhenius behavior. 
The values of $T_g$, obtained from $\t_\a(T_g)=100$s,
are reported in the key.}
\label{fig1:rel}
\end{figure}

Although liquids normally crystallize on cooling, there
are members of all liquids types (including molecular, polymeric, ionic and
metallic) that can be supercooled below the melting temperature
$T_m$ and then solidify at some temperature $T_g$, the 
{\it glass transition temperature}. 
The viscosity $\h(T)$ of the liquid increases continuously
but very fast below $T_m$ and at some point reaches values so high that
the liquid does not flow anymore and can be considered a solid for
all practical purposes: at low temperatures, an amorphous solid
phase is observed.
The temperature $T_g$ marking the transition between the liquid and
the glass is often conventionally defined by the condition $\h(T_g)=10^{13}$ Poise, 
but many other definitions are possible.
As an example of this phenomenon, in Fig.~\ref{fig1:rel} the 
viscosity of many glass forming liquids is reported as a function 
of the temperature. Following Angell~\cite{MA01}, the quantity
$\log_{10} \left[\frac{\h(T)}{\rm Poise}\right]$ is reported as a 
function of $T_g/T$.
The viscosity increases of about $17$ orders of magnitude on decreasing the
temperature by a factor $2$. Note that because the increase of viscosity
is so fast, the dependence of $T_g$ on the particular value of viscosity
($10^{13}$ Poise) which is chosen to define it is very weak.
The viscosity around $T_g$ is often described by the 
Vogel--Fulcher--Tamman (VFT) law,
\beq
\label{VFT}
\h(T) = \h_\io e^{\frac{\D}{T-T_0}} \ ,
\eeq
where $\h_\io$, $\D$ and $T_0$ are system--dependent parameters.
If $T_0 = 0$ this relation reduces to the Arrhenius law; otherwise,
the extrapolation of the viscosity below $T_g$ leads to a divergence
at $T=T_0$. 

The viscosity is related to the {\it structural relaxation time} $\t_\a$
by the Maxwell relation, $\h = G_\io \t_a$, where $G_\io$ is the 
infinite--frequency shear modulus of the liquid.
The structural relaxation time is related to the decorrelation of
density fluctuations. In glass forming liquids, for $T_m \gg T \gtrsim T_g$,
the decorrelation of density fluctuations happens on two well separated
time scales: a ``fast'' time scale ($\t_0 \sim 10^{-12} s$), which is related to 
vibrations of the particles around the disordered instantaneous positions,
and a ``slow'' time scale $\t_\a$, which is related to cooperative 
rearrangements of the disordered structure around which the fast vibrations
take place. Through the Maxwell relation, the fast increase of viscosity
around $T_g$ is then related to a marked slowing down of the structural
dynamics; usually, at $T_g$ one has $\t_\a \sim 100s \sim 10^{14} \t_0$, while in the liquid
phase $\t_\a \sim \m s$.
The structural relaxation time, obtained from dielectric relaxation data,
of some fragile glass forming liquids is reported in the right panel
of Fig.~\ref{fig1:rel}. The behavior of $\t_\a(T)$ is also described
by a VFT law with an apparent divergence at $T=T_0$. This leads to the
interpretation of $T_0$ as a temperature at which a structural arrest
takes place.

Whether such a divergence really exist in glasses is hotly debated.
The reason is that because of the strong divergence of Eq.~\eqref{VFT}, the $T_g$
below which the system cannot be equilibrated is quite far from $T_0$: typically,
$T_g \approx \frac43 T_0$. Hence it is very hard to approach the putative divergence point
while being at equilibrium. For comparison, in standard critical phenomena the relaxation
time goes as $\t \sim \t_0 |T/T_c-1|^{-z \nu}$ with an exponent $z \nu \sim 1$~\cite{HH77}. 
This means that if one is
able to equilibrate on a time scale $\t/\t_0 = 10^{14}$, then one can approach the critical point at a distance
$|T-T_c| \sim 10^{-14} T_c$. Hence, in this case one can equilibrate the system arbitrarily close to the critical point,
the only limitation being the precision on temperature control.

\subsection{Correlation functions}

Consider a simple glass former made by $N$ particles with positions $ x_j$.
An important observation is that, if
$\hat \r_{ k} = N^{-1/2} \sum_j e^{i  k \,  x_j}$ is the Fourier transform of the density
profile, static correlation functions such as the structure factor
\beq
S(k) = \lim_{N\to\io} \frac{1}{N}\sum_{jl}
\langle e^{i  k \, ( x_j - x_l  )} \rangle = 
 \lim_{N\to\io} \langle \hat\r_{ k} \hat\r_{- k} \rangle
\eeq
do not exhibit any particular change of behavior at the glass transition: they are very smooth functions of temperature
in all the range of temperatures above and below $T_g$.

The most spectacular signature of the incoming transition can be found in time-dependent correlations, such
as the {\it mean-square displacement}
\beq\label{eq:MSDdef}
\la \D r^2(t) \ra = \lim_{N\to\io} \frac1N \sum_j \la |  x_j(t) - x_j(0) |^2 \ra
\eeq
and the so-called {\it coherent} and {\it incoherent scattering functions}, respectively defined by:
\beq\label{eq1:Fk}
\begin{split}
F( k, t) &= \lim_{N\to\io} \frac{1}{N}\sum_{jl}
\langle e^{i  k \, [ \,  x_j(t)- x_l(0) \, ]} \rangle = 
 \lim_{N\to\io} \langle \hat\r_{ k}(t) \hat\r_{- k}(0) \rangle \ , \\
F_s( k, t) &= \lim_{N\to\io} \frac{1}{N}\sum_{j}
\langle e^{i  k \, [ \,  x_j(t)- x_j(0) \, ]} \rangle  \ . \\
\end{split}
\eeq
Here the average is over the dynamical process, {\it at equilibrium in the liquid phase}.

\begin{figure}[t]
\centering
\includegraphics[width=.49\textwidth]{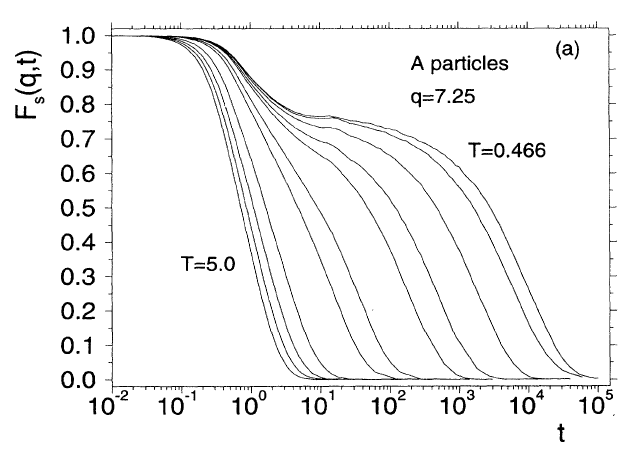}
\includegraphics[width=.49\textwidth]{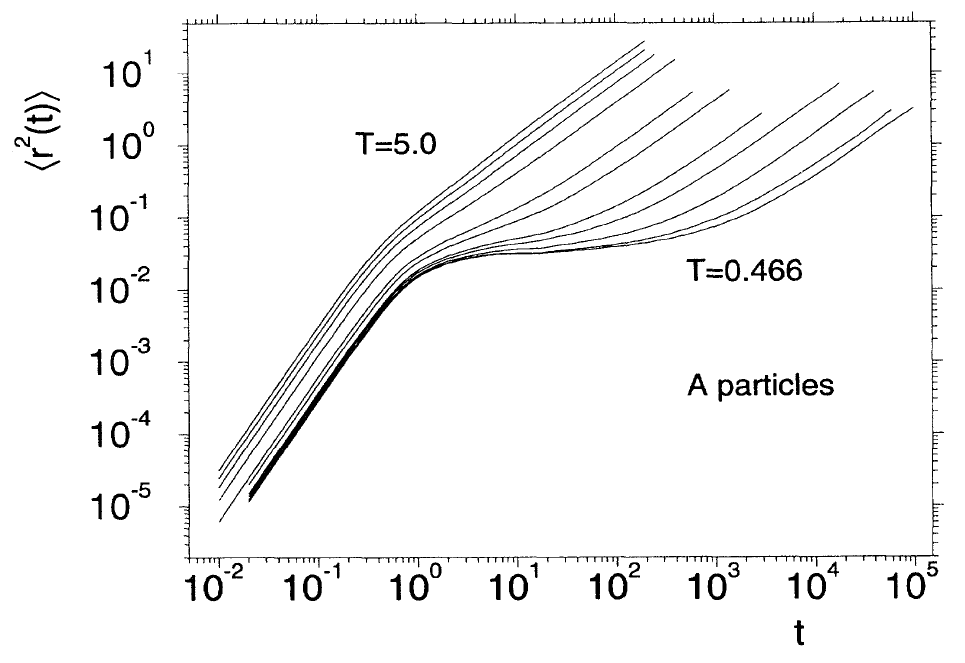}
\caption[Dynamical correlations]
{
Dynamical correlations from a numerical simulation of a binary mixture of particles interacting through Lennard-Jones potentials, with parameters
chosen in such a way to avoid crystallization and obtain a good glass former.
(Left, from~\cite{KA95b}) Incoherent scattering function at a fixed wavevector (here denoted by $q$) as a function of time and several temperatures above the glass transition.
(Right, from~\cite{KA95a}) The mean square displacement as a function of time for the same temperatures.
The {\it plateau} at intermediate times is seen to appear at the lower temperatures.
}
\label{fig1:Fs}
\end{figure}

An example of such correlations is given in Fig.~\ref{fig1:Fs}, for a numerical simulation of a
Lennard-Jones system~\cite{KA95a,KA95b}.
Close to the glass transition, time-dependent correlations display two distinct relaxations,
separated by a region where they depend weakly on time (a {\it plateau}).
Based on these results,
a common pictorial interpretation of the dynamics of glass forming liquids
above $T_g$ is the following: for short times the particles are ``caged'' by
their neighbors and vibrate around a local structure on a nanometric scale;
the structural relaxation is then interpreted as a slow cooperative rearrangement
of the cages. Note that on the time scale of the structural relaxation time $\t_\a$,
the root mean square displacement of the particles is of the order of $10\%$ of the particle radius,
so one cannot think to the structural relaxation as a process of single--particle 
``jumps'' between adjacent cages.

At $T_g$ and below, the structural relaxation time is much larger than the experimentally or numerically
accessible window and the {\it plateau} extends to the largest accessible times. Hence one can define
a {\it caging order parameter} as
\beq
\label{nonergdyn}
\begin{split}
&A = \frac{1}{2d} \lim_{t \to \io} \la \D r^2(t) \ra \ , \\
&f(k) = \lim_{t\to\io} F(k,t)/S(k) \ ,
\end{split}
\eeq
where the limit $t\to\io$ has to be intended as taking the largest accessible times.
$A$ is often called {\it cage radius} while $f(k)$ is often called {\it non-ergodicity parameter}.
If the relaxation time were really to diverge at some $T_0$,
then this {\it ideal glass transition} would be characterized by the (discontinuous) appearance of truly finite long-time limits of
dynamical correlations, signaling the complete structural arrest.

It is also interesting to consider the out of equilibrium dynamics following a fast quench into the glass phase.
Suppose that an equilibrium liquid prepared at some temperature $T_i> T_g$ is rapidly cooled (quenched) at
a temperature $T_f<T_g$. Define $t=0$ as the time at which the cooling procedure ends and $T=T_f$. One can
wait a time $t_w$ after the quench and then measure the dynamical correlations 
$F( k; t, t_w) = \lim_{N\to\io} \langle \hat\r_{ k}(t+t_w) \hat\r_{- k}(t_w) \rangle$.
Then, it is observed that increasing $t_w$ (the age of the system) has roughly the same effect as decreasing temperature at equilibrium.
One observes a two-step relaxation with a structural relaxation time that increases with $t_w$ until the structure
becomes arrested. This process is known as {\it aging} and it is another important characterization of the glass phase.

\subsection{Configurational entropy}

The idea that the dynamics in the supercooled phase is separated in a fast
intra--cage motion and in a slow cooperative rearrangement of the structure
suggests to split the total entropy of the liquid in a ``vibrational''
contribution, related to the volume of the cages,
and a ``configurational'' contribution, that counts the
number of different disordered structures that the liquid can assume~\cite{Ka48}:
\beq
S_{liq}(T) \sim S_{vib}(T) + S_c(T) \ .
\eeq
To estimate the vibrational contribution to the entropy of the liquid, it is often
assumed that it is roughly of the order of the entropy of the corresponding
crystal. Despite the fact that this idea is plain wrong for hard spheres and similar system
where excluded volume effects play an important role, this approximation is not so bad
in systems --such as the Lennard-Jones potential-- where interactions are smoother 
and have longer range.
It is then possible to estimate the configurational entropy $S_c(T)$ as
\beq
\label{ScDEF}
S_c(T) = S_{liq}(T)-S_{cryst}(T)=\D S_m - \int_T^{T_m} \frac{d T'}{T'} \,
\big[ C_{liq}(T') - C_{cryst}(T') \big] \ ,
\eeq
where $\D S_m \equiv S_{liq}(T_m)-S_{cryst}(T_m)$ 
is the entropy difference between the liquid and the crystal at
the melting temperature $T_m$, and 
$C(T) = T \frac{\dpr S}{\dpr T}$ is the specific heat.
Note that in experiments one usually works at constant pressure, $C=C_p$, while
in numerical simulations and in theoretical computations one usually works 
at constant volume, $C=C_v$. 

\begin{figure}[t]
\centering
\includegraphics[width=450pt]{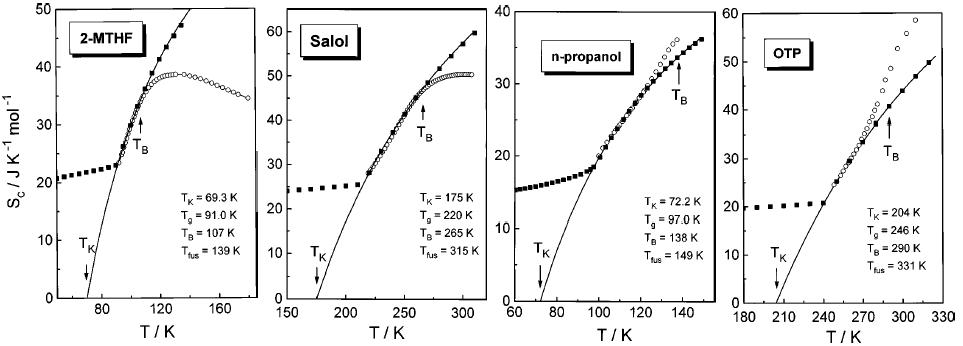}
\caption[Configurational entropy of four different fragile glass formers]
{(From~\cite{RA98} and references therein) Configurational entropy $S_c(T)$ of four 
fragile glass formers. The black squares are obtained from calorimetric measurements
of the specific heat of the liquid and of the crystal, see Eq.~(\ref{ScDEF}). Below
$T_g$ (reported in the key) the liquid falls out of equilibrium. The black line
is the extrapolation according to Eq.~(\ref{ScFIT}) of the equilibrium data for 
$T \geq T_g$ below $T_g$, that goes to zero at $T=T_K$. 
The open white circles are another estimate derived from the dielectric relaxation data of 
Fig.~\ref{fig1:rel} using the Adam--Gibbs relation~\cite{AG65}.
}
\label{fig1:Sconf}
\end{figure}

In Fig.~\ref{fig1:Sconf} the estimate of $S_c$, obtained from calorimetric measurements
of the specific heat and using Eq.~(\ref{ScDEF}), is reported
for four different fragile glass formers. Below $T_g$ the liquid falls out of equilibrium
as the structural relaxation time becomes of the order of the 
experimental time scale ($\sim 100$s).
This means that the structural rearrangements are ``frozen'' on the 
experimental time scale and the only contribution to the specific heat comes
from the intra--cage vibrational motion; in this situation the specific heat of
the liquid becomes of the order of the crystal's one and $S_c(T)$ approaches
a constant value. However, one can ask what would happen if the time scale of the
experiment were much bigger, say $10^6$s. In this case, the glass transition temperature
$T_g$ would be lower and the {\it plateau} would be reached at smaller values
of $S_c$. If one assumes to be able to perform an {\it infinitely slow} experiment,
one can imagine to follow the extrapolation of the data collected above $T_g$ to
lower temperatures. For fragile liquids, it is found that a good extrapolation
is
\beq
\label{ScFIT}
S_c(T) = S_\io \left( 1 - \frac{T_K}{T} \right) \ ,
\eeq
where the parameters $S_\io$ and $T_K$ are fitted from the data above $T_g$.
This extrapolation is reported as a full line in Fig.~\ref{fig1:Sconf}.

The outcome of this procedure is that the configurational entropy seems to vanish
at a finite temperature $T_K$. Because $S_c$ counts the number of different structures
that the liquid can access, it is not expected to become negative.
A possible explanation of this paradoxical behavior was proposed by 
Kauzmann~\cite{Ka48}, who argued that at some temperature between $T_g$ and $T_K$
the free energy barrier for crystal nucleation becomes of the order of the free
energy barrier between different structures of the liquid. This means that
the time scale for crystal nucleation becomes of the order of the structural
relaxation time $\t_\a$ of the liquid, and one cannot think anymore of
an ``equilibrium'' liquid because crystallization will occur on the same time scale
needed to equilibrate the liquid. The extrapolation of $S_c(T)$ down to $T_K$ is
then meaningless, and the paradox is solved. This argument has been recently 
reconsidered, see \eg~\cite{Ca09}, and its implications are still under 
investigation.

Alternatively, one can assume that the existence of the crystal is irrelevant,
because crystallization can be in some way strongly inhibited: for instance, by
considering binary mixtures~\cite{KA95a,KA95b}, or --in numerical simulations-- by adding
a potential term to the Hamiltonian that inhibits nucleation~\cite{DAPR00}.
If crystallization is neglected, the extrapolation of $S_c$
suggests that at $T_K$ a phase transition happens, at which the number
of structures available to the liquid is no more exponential, as $S_c=0$,
and the system is frozen in one amorphous structure which can be called
an {\it ideal glass}. Below $T_K$, the only contribution to the entropy
of the ideal glass is the vibrational one, so the specific heat has a
{\it downward} jump at $T_K$, corresponding to the freezing of some degrees of freedom
at the transition. 
The transition is expected to be of second order
from a thermodynamical point of view.

An evidence supporting this picture is the fact that in almost all the fragile
glass formers it is found that $T_K \sim T_0$. For instance, in~\cite{An97} some
30 cases where $T_0 = T_K$ with an error of order $3\%$ are reported 
(but the reliability of these extrapolation has been questioned in~\cite{ECG09}).
The equality of (non-zero) $T_0$ and $T_K$ implies that 
both the structural relaxation time and the viscosity diverge at
$T_K$, so that the structures that are reached at $T_K$ are thermodynamically
stable, being associated to an infinite structural relaxation time.
Of course, this ideal glass transition, that would occur {\it in equilibrium}, would not be
observable: at some temperature $T_g > T_K$ where $\t_\a(T_g) = \t_{exp}$
a {\it real} glass transition, freezing the system in a nonequilibrium 
amorphous state (a {\it real} glass), happens. The value of $T_g$, as well as
the properties of the nonequilibrium glass (density, structure, etc.) depend
on the value of $\t_{exp}$, which is usually $\sim 100$s as already discussed.

\subsection{Dynamical heterogeneities}

\begin{figure}[t]
\centering
\includegraphics[width=.5\textwidth]{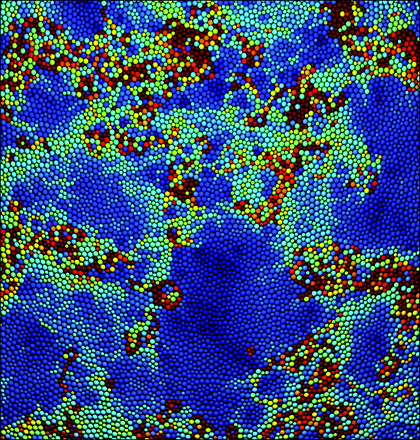}
\caption[Dynamical heterogeneities]{
(From the cover of PNAS, September 8, 2009)
Dynamical heterogeneity in a structural glass-forming liquid. Shown is a spaceÐtime rendering of the equilibrium dynamics of a two-dimensional 
super-cooled fluid mixture of 10000 particles after a fraction of a structural relaxation time. 
Particles are colored according to their overlap with their initial positions: a particle that is displaced by more than one particle 
diameter is dark red; a particle that has no displacement is dark blue; intermediate colors in the visible spectrum 
coincide with intermediate displacements. The color variation illustrates significant dynamic heterogeneity. 
Yet the spatial arrangement of particles at a given time seems perfectly homogeneous when coarse grained over only one of two particle diameters. 
The juxtaposition shows that the dynamics of this system are highly correlated, but the structure is seemingly not. 
}
\label{fig1:hete}
\end{figure}

An important question is whether the slow relaxation observed on approaching $T_g$ is produced by 
independent relaxation events that are due to locally high potential energy barrier, or by cooperative effects
that make relaxation difficult.
To quantify this, it is convenient to introduce a correlation function that measures how much
structural relaxation is correlated
in space.

Consider the real-space density profiles $\hat\r(x,t) = \sum_{i=1}^N \d(x-x_i(t))$
at time zero and at time $t$, respectively given by 
$\hat\rho(x,0)$ and $\hat\rho(x,t)$. One can define a local similarity measure of these 
configurations as 
\beq\label{ove}
\hat C(r,t) = \int dx f(x) \hat\r\left(r+\frac{x}2,t\right) \hat\r\left(r-\frac{x}2,0\right)- \r^2 \ ,
\eeq
where $f(x)$ is an arbitrary ``smoothing"
function of the density field with some short range $r_0$.
In experiments, $f(x)$ could describe the resolution of the detection system
and can be for instance a Gaussian of width $r_0$. 
Call $C(t) =V^{-1} \int dr \langle \hat C(r,t) \rangle$ the spatially and thermally averaged correlation function.
On approaching the glass transition, this function behaves exactly as the coherent scattering function defined in
Eq.~(\ref{eq1:Fk}).

The correlation function $\hat C(r,t)$ probes the dynamics in the vicinity of point $r$.
Spatial correlations in the dynamics
can be quantified by introducing the correlation function of $\hat C(r,t)$, \ie a four-point dynamical correlation
\beq\label{eq1:G4}
\begin{split}
&G_4(r,t)=
 \langle 
\hat C(r,t) \hat C(0,t) 
\rangle -\langle 
\hat C(r,t)\rangle\langle \hat C(0,t) 
\rangle \ .
\end{split}\eeq
The latter decays as $G_4(r,t)\sim \exp(-r / \xi(t))$ thus defining a ``dynamical correlation length'' $\xi(t)$.
In a variety of glass formers,
it is found that $\xi(t)$ grows when $t$ approaches values corresponding to the {\it plateau}, and reaches
its maximum $\overline{\xi}$ on the scale of the structural relaxation. The value of $\overline{\xi}$ is found
to increase upon decreasing temperature towards $T_g$~\cite{BBBCS11}.

A large value of $\overline{\xi}$ indicates that if a given region in space is ``mobile'' --\ie it has a structural relaxation
faster than the average-- then neighboring regions will also be mobile, and similarly for slow regions -- those with a structural
relaxation slower than the average.
This fact indicates that dynamics is cooperative and characterized by a ``facilitation'' mechanism~\cite{Ga02,RS03,WG04,KHGGC11}: a mobile region can speed up
the dynamics of neighboring regions thanks to dynamical correlations. Often such cooperatively rearranging regions, whose presence is now well established, 
are called ``dynamical heterogeneities"~\cite{BBBCS11}.
More pictorial measures of dynamical heterogeneities are possible, one of them is reported in Fig.~\ref{fig1:hete}.

\begin{figure}[t]
\centering
\includegraphics[width=.6\textwidth]{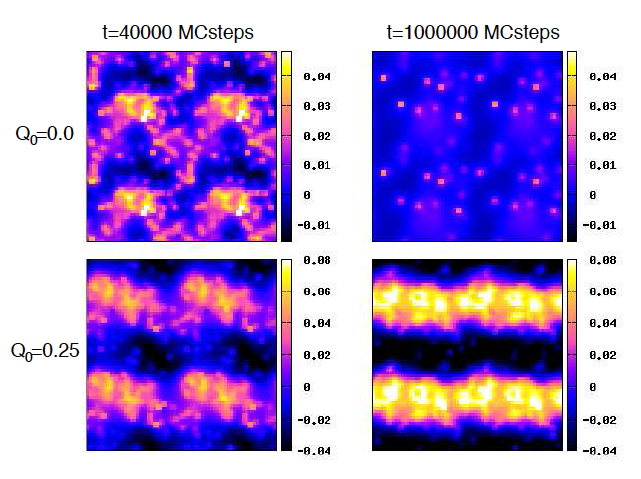}
\caption[Static correlation length]{
(From~\cite{CCGGGPV10}) 
Fluctuations of the overlap field, $\d \hat C(r, t) = \hat C(r, t) - C(t)$,
for a two dimensional slice of the three-dimensional simulated system. 
Upper panels: unconstrained system. 
Lower panels: constrained system ($\overline{C} =0.25$). 
Left panels: $t = \t_\a$. Right panels: large times. Phase separation is observed
in the constrained system.
}
\label{fig1:staticQ}
\end{figure}

An important question is whether dynamical heterogeneities are purely dynamical in nature or if they are induced, or accompanied, 
by the growth of some static correlation length. As previously discussed, simple measures of structural correlations do not show any
particular anomaly at the glass transition, while the dynamics displays a strong slowing down and increasing heterogeneity.
There have been several attempts to define and measure such a static correlation length, leading in particular to the introduction
of the so-called {\it point-to-set} correlation length~\cite{BB04,MS06,FM07} which has been measured \eg in~\cite{CGV07,BBCGV08}.
Perhaps the most spectacular result is the one obtained in~\cite{CCGGGPV10}. 
Recall that $C(t)$ measures the average ``similarity" (overlap) between the density profile at time $t=0$ and $t$. 
In this numerical simulation of a Lennard-Jones--like glass former, a constraint has been added to
the system, imposing that $C(t) \geq \overline{C}$, a fixed constant. 
In other word, the overlap with the initial configuration (which is just an equilibrium reference configuration) 
must be bigger than some prefixed constant. Hence, this constraint is {\it static} in nature: the system can visit all
configurations with overlap larger than $\overline{C}$ according to its equilibrium distribution.
The result of this investigation, reproduced in Fig.~\ref{fig1:staticQ}, show that the system will phase-separate
into regions where the overlap is very low and regions where overlap is very high. This indicates that some {\it static} 
``surface tension" between regions of high and low mobility exists, and is responsible for dynamical heterogeneities.
Similar results have been obtained by imposing the overlap constraint in different ways, see \eg~\cite{BK12,CB12}.

\subsection{Glass and jamming transitions}
\label{sec:defjamming}

Besides the glass transition, another ``rigidity'' transition has attracted a lot of attention: the so-called ``jamming'' transition
of granular matter. Observing a jamming transition is an everyday experience, similarly to the glass transition. An athermal amorphous assembly
of hard objects --such as nuts, oranges, tennis balls-- will be mechanically stable, meaning that it can support finite stresses, 
if its density is large enough. In absence of friction, for three-dimensional spheres, typically around $\sim 64\%$ of
space is filled, while in presence of friction the density can be lower.
For comparison, in three dimensions the closest packing of an assembly of hard spheres is such that
$\sim 74\%$ of space is occupied, corresponding to a bcc/fcc crystalline structure.

Because granular matter is typically made by hard objects, we need to introduce a 
model system that can interpolate between finite temperature 
glasses and hard spheres. A very convenient one is the 
model of frictionless harmonic spheres: 
$N$ spherical particles of diameter $\sigma$ are enclosed 
in a volume $V$ in $d$ spatial dimensions, and interact with a soft 
harmonic repulsion of finite range:
\beq
v(r) = \epsilon (1-r/\sigma)^2 \, \theta(\sigma-r) \ ,
\label{eq1:pot}
\eeq 
where $r \geq 0$ is the interparticle distance, $\th(r)$ is the Heaviside step function and $\epsilon$ controls the strength of the repulsion.
In this model, there is no force between particles if they do not overlap, while there is a harmonic repulsion if the particles overlap. This is supposed
to mimic rigid particles with a finite elasticity.
The model (\ref{eq1:pot}), originally proposed to describe wet foams, has become a paradigm in numerical studies of the $T=0$ jamming transition~\cite{OLLN02,He10,LNSW10}. 
It has also been studied at finite temperatures~\cite{BW08,BW09,ZXCYAAHLNY09}, and finds experimental realizations in emulsions, soft 
colloids and grains~\cite{bookmicrogel}. The model 
has the two needed control parameters to explore the jamming phase diagram:
the temperature $T$ (expressed in units of $\epsilon$), and the fraction of volume occupied by the particles in the absence of overlap, the {\it packing
fraction} $\f = \r V_d(\sigma/2)$, where $V_d(r)$ is the volume of a $d$-dimensional sphere of radius $r$ and $\r = N/V$ is the density.

\begin{figure}[t]
\centering
\includegraphics[width=.5\textwidth]{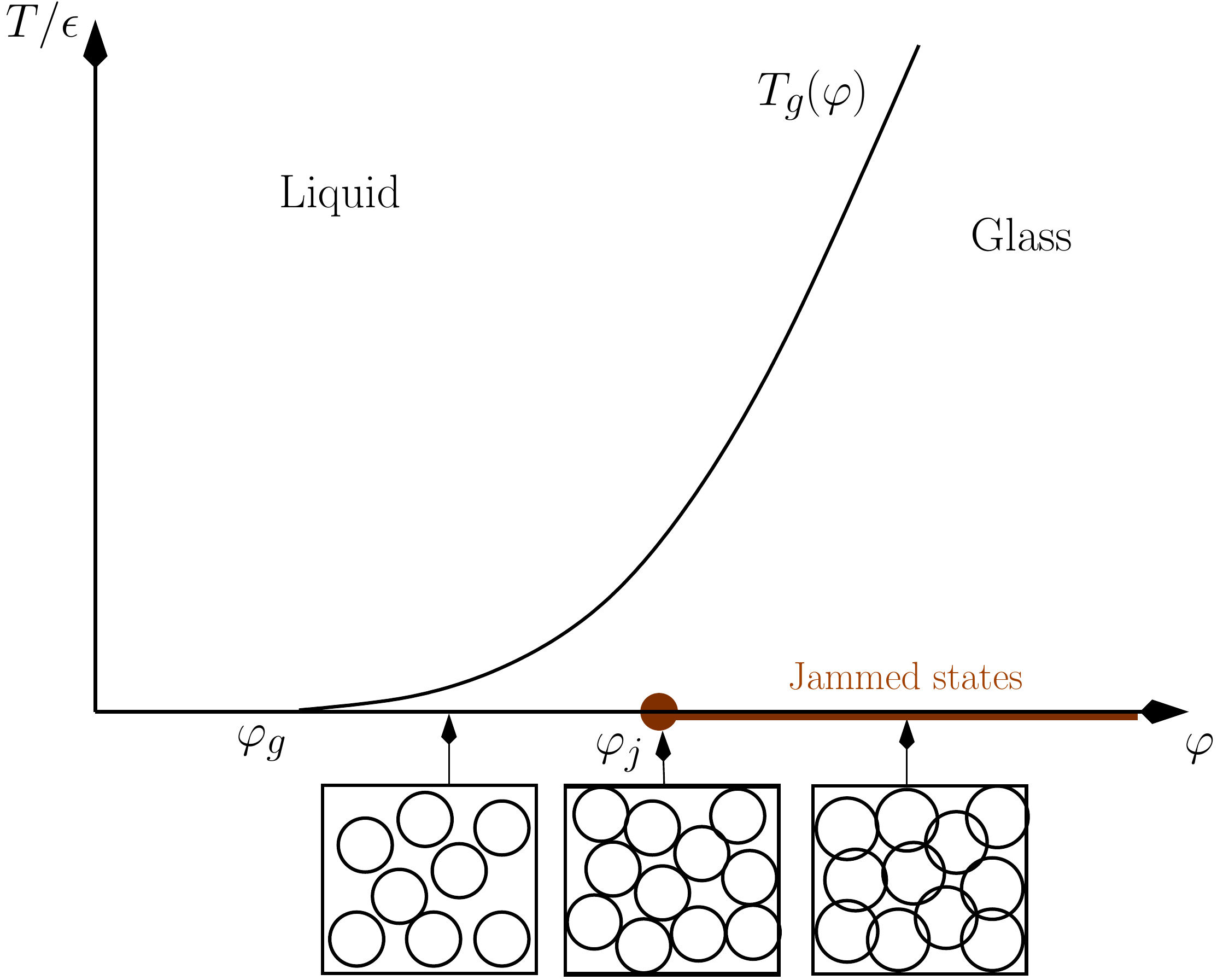}
\caption[Jamming phase diagram]
{
A schematic phase diagram for the glassy states of the 
model defined by Eq.~(\ref{eq1:pot}) showing both the 
glass transition line at $T_g(\ph)$ and the jamming point 
at $T=0$ and $\ph = \ph_{j}$.
Below $\ph_j$, particles do not 
overlap, as in hard sphere configurations. 
Above $\ph_j$, there is a finite density of overlaps,
and the energy and pressure are finite. In this region, 
hard sphere configurations cannot be found.
}
\label{fig:PDsch}
\end{figure}

Over the last decade, a large number of numerical observations have been reported for this model~\cite{He10,LNSW10}. A jamming transition is observed at $T=0$ at some critical volume fraction $\ph_j$, the density above which the packings carry a finite density of particle overlaps. This transition is pictorially represented in Fig.~\ref{fig:PDsch}.
Numerically, the zero-temperature energy density, $e_{\rm GS}$, and pressure, $P$, are found to increase continuously from zero above $\ph_j$ as power laws~\cite{OLLN02}. The pair correlation function of the density fluctuations~\cite{Hansen}, $g(r)$, develops singularities near $\ph_j$~\cite{DTS05,SLN06}, which are smoothed by thermal excitations~\cite{ZXCYAAHLNY09}. In particular, $g( 1 ) = \infty$ at $\ph=\ph_j$ and $T=0$. This behavior implies that the density of contacts between particles, $z$, jumps discontinuously from $0$ to a finite value, $z_j$, at $\ph_j$, and increases further algebraically above $\ph_j$~\cite{OLLN02,He10}. Thus, the jamming transition appears as a phase transition taking place in the absence of thermal motion, with a very peculiar critical behavior and physical consequences observable experimentally in a large number of materials.

To make a connection with the glass phenomenology discussed above, note that
the liquid undergoes a glass transition at some temperature $T_{g}(\ph)$. The temperature $T_g(\ph)$ vanishes at a volume fraction $\ph_g \sim 58\%$ which corresponds to the glass
transition of hard spheres. This is because when $\ee \gg T$, the harmonic spheres model reduces to hard spheres, provided the density is low enough that overlaps can be avoided.
The point $\ph_g$ {\it is not the jamming transition}: indeed, the ground state energy and pressure remain zero across $\ph_g$.
Above $\ph_g$, the system enters at zero temperature 
a {\it hard sphere glassy state}. 
In this state, particles vibrate near well-defined (but random) positions, 
and the system is not yet jammed. Jamming happens when
the glass reaches its close packing density, which is then identified with 
the jamming transition at $\f_j$.
In Fig.~\ref{fig:PDsch} the density interval $[\ph_g, \ph_{j}] \sim [0.58, 0.64]$
is simply the amorphous 
analog of the interval $\ph \in [\ph_m,\ph_{cp}]\sim [0.54,0.74]$ for ordered 
states of hard spheres where a compressible crystalline structure
exists at thermal equilibrium.

The crucial difference between the hard sphere glass phase for $\f \in [\f_g,\f_j]$ and the jammed phase for $\f > \f_j$ is the following.
At any finite temperature, both phases are rigid, in the sense that they have finite elastic moduli, pressure, etc.
However, the rigidity of the glass phase is {\it entropic} in nature: it is due to thermal vibrations of the particles, that induce collisions
between neighboring particles and allow them to exchange momentum. Indeed, pressure and all elastic moduli are proportional to
temperature in this regime, and they vanish at zero temperature~\cite{Hansen,IBS12}.
On the contrary, the rigidity of the jammed phase is due to overlaps between particles, that produce interaction forces even at strictly
zero temperature. Above $\f_j$, pressure and elastic moduli are finite even at $T=0$~\cite{OLLN02,He10,LNSW10}.
In summary, $\f_j$ marks a transition {\it inside the glass phase} from entropic to mechanical rigidity. Below $\f_j$, some noise (\eg due
to thermal agitation) is needed to stabilize the system, that would otherwise be mechanically unstable. Above $\f_j$, the system is rigid
in absence of any noise.

The main theoretical challenge in studying the jamming 
transition at $\ph_{j}$ is that it happens deep inside the glass phase, hence out of equilibrium. 
Therefore, an accurate theoretical
description of the glass phase is needed. 
However, once this is achieved, {\it the jamming transition provides a unique opportunity to perform very detailed tests of 
theories of the glass structure}.
Indeed, while the structural signatures of the glass transition are quite elusive (no signature at all in the pair correlation $g(r)$, 
weak singularities in the thermodynamics),
the structural signatures of the jamming transition are extremely easy to detect numerically: at $\f_j$ marked changes happen
in $g(r)$, and pressure and energy have strong singularities.

\subsection{Summary}
\label{sec:basic_facts_summary}

In summary, the liquid-glass transition is a very strange transition, much different from standard critical phenomena.
It is characterized by:
\begin{itemize}
\item A strong increase of relaxation time upon lowering temperature gives
the impression of a divergence at a finite temperature $T_0$. However the divergence is so strong that the system falls out of equilibrium
at a $T_g$ which is quite far from $T_0$, in such a way that approaching the putative critical point while being at equilibrium is impossible.
\item A separation of time scales between fast vibrational dynamics ($\t_0 \sim 10^{-12} s$) and slow structural relaxation $\t_\a \gg \t_0$.
Dynamical correlations exhibit a characteristic {\it plateau} for $\t_0 \ll t \ll \t_\a$. At $T_g$ the length of the {\it plateau} exceeds the accessible
time scales and the system is effectively frozen (in this sense, $T_g$ effectively acts as $T_0$, except for the fact that the system falls out of equilibrium).
The value of mean square displacement at this plateau defines the {\it cage radius} $A$ at the glass transition, while the long time limit of the coherent
scattering function defines the non-ergodic parameter $f( k)$, Eq.~\eqref{nonergdyn}.
\item A strongly heterogeneous dynamics, with well-separated regions containing mobile or slow particles. Although everybody agrees on the
existence of these regions as defined from the dynamics, some have argued that they have a static origin, and have given correspondingly numerical evidence
for a growing static correlation length.
\item 
The equilibrium configurational entropy of the liquid, $S_c(T)$, is found to decrease quickly around $T_g$, and to be correlated with the increasing
of relaxation time (as expressed by the Adam-Gibbs relation~\cite{AG65}). If one believes in extrapolations, the 
extrapolation of $S_c(T)$ to zero defines a
temperature $T_K$ which is found, for several systems, to be quite close to the extrapolated divergence of relaxation time at $T_0$.
Moreover, the freezing out of structural degrees of freedom at $T_g$ induces a sudden drop in specific heat (or in compressibility for hard spheres).
\end{itemize}
In addition, for systems with a finite radius of interaction, a {\it jamming} transition happens inside the glassy region of the temperature-density phase
diagram, at which the rigidity of the glass changes in origin. A correct description of the glass phase
should allow one to describe this transition and its geometrical and structural signatures.


\section{The random first order transition theory of the glass transition}

The random first order transition (RFOT) theory is a very ambitious program to obtain a rather complete theory
of the liquid-glass transition and of the glass state. It basically consists in following the same approach that led 
to the classical theory of phase transitions, 
adapting it to take into account the presence of disorder.
Here a very schematic presentation of RFOT theory will be given, based on the analogies of Tab.~\ref{tab1:RFOT}. 
More detailed presentations, and a comparison with other theories of
glasses, can be found in~\cite{LW07,Ca09,BBBCS11,BB11,WL12}.

\begin{table}
\begin{tabular}{|c|c|c|}
\hline
& Theory of second order PT (gas-liquid)  & RFOT theory of the liquid-glass transition \\
\hline
\multirow{4}{*}{Qualitative MFT}  & Landau theory (1937)  &  Replica theory (Parisi, 1979; KTW, 1987) \\
						& {\it Spontaneous $Z_2$ symmetry breaking}	&   	{\it Spontaneous replica symmetry breaking} \\
                                   			& {\it Scalar order parameter} 				&	{\it Order parameter: overlap matrix $q_{ab}$} \\
		             			& {\it Critical slowing down}				&	{\it Dynamical transition ``\`a la MCT''} \\
\hline
						&	{\it Liquid-gas:} $ \b p/\r = 1/(1-\r b) - \b a \r $ 	&						 					\\
	Quantitative MFT		&	(Van der Waals 1873) 					&  	Kirkpatrick and Wolynes 1987					\\
	(exact for $d\to\io$) 		&	{\it Magnetic:} $m=\tanh(\b J m)$			&  	{\bf Sec.~\ref{sec:highd}} 	          \\
					 	&	(Curie-Weiss 1907)						& 											\\	
						&   	Bethe approximation (1935)				&	Random graphs, cavity method				\\				
\hline
						&	Liquid theory (1950s)					&	{\it DFT} (Stoessel-Wolynes, 1984)				\\
	Quantitative theory		&	{\it Hypernetted Chain (HNC)} 				&  	{\it MCT} (Bengtzelius-G\"otze-Sjolander 1984)		\\
	in finite $d$ 	 		&	{\it Percus-Yevick (PY)} 					&  	{\it Replicas} (M\'ezard-Parisi 1996) 		 	          \\
					 	&										& 	Jamming -- {\bf Sec.~\ref{sec:finited}} 		\\					
\hline
	Corrections 				  	& {\it Ginzburg criterion, $d_u=4$} (1960) &   	{\it Ginzburg criterion, $d_u=8$} (2011)   		\\
	around MFT					& {\it Renormalization group} (1970s)	&	{\it Renormalization group} (2010--)   		\\
								&  {\it Nucleation theory} (Langer, 1960)	&	 {\it Nucleation theory} (KTW 1987) 	\\
\hline
\end{tabular}
\caption{A very schematic view of the analogy between RFOT theory and the theory of standard phase transitions. 
References in bold point to later sections where original material will be presented.}
\label{tab1:RFOT}
\end{table}

\subsection{An overview of the classical theory of phase transitions}

\subsubsection{Mean field theory}

The classical theory of phase transitions started from a very schematic theory, the Landau theory~\cite{La37}.
Landau theory allows one to identify the correct order parameter, which for the liquid-gas transition is
a scalar order parameter that signals the onset of spontaneous $Z_2$ symmetry breaking.
It is a schematic description that generalizes previous theories such as the Van der Waals
equation of state of liquids~\cite{VdW73} and the Curie-Weiss description of magnetism~\cite{We07,VV45}. These two descriptions have also the advantage
of being exact in some well defined limit of given microscopic models~\cite{LP66,PL71}. In particular, they are both exact when
the spatial dimension goes to infinity. Take for instance the Ising model in $d$ dimensions, 
with a ferromagnetic coupling $J/(2d)$. For $d\to \io$, the solution of this model is given exactly by the Curie-Weiss
equation $m = \tanh(\b J m)$~\cite{ParisiBook}. The reason is that when $d\to\io$, the number of neighbors of a given spin grows quickly,
and correlations among them are correspondingly reduced, because each of them is connected to a large number of other
neighbors. At the leading order in $1/d$, the model on a hypercubic lattice is well approximated by the model defined on a
tree of the same connectivity, which is known as the Bethe approximation~\cite{Be35} and
provides an improved version of the Curie-Weiss theory.
Similarly, the Van der Waals equation provides an exact theory of the liquid-gas transition when $d\to\io$ for liquids with
a hard core and a properly scaled attractive interaction.
Because all these theories are based on the idea that neighbors of a given particle are uncorrelated and provide a ``mean field''
acting on the particle, to be determined self-consistently, they are known as {\it mean field theories} (MFT).

\subsubsection{Approximate theories in finite dimensions}

The Van der Waals theory can be seen as a simple resummation of the low-density virial expansion.
Improved theories of the liquid phase (and sometimes also of the liquid-gas transition) can be obtained
by more careful resummations of the virial expansion. This leads to self-consistent integral equations for
the correlation function $g(r)$ of the liquid. The most popular of such approximations are the Hypernetted Chain (HNC) or Percus-Yevick (PY)
closures~\cite{Hansen}. Although these theories provide a much better description than the mean field theory,
they retain a mean field flavor as they somehow neglect three-body correlations and provide a self-consistent
equation for the pair correlations. Improved HNC and PY equations, that contain three-body correlations,
can be obtained. These theories provide reasonable quantitative characterizations of the liquid phase for a
given choice of the microscopic interaction potential, provided one is not too close to a critical point.

\subsubsection{Corrections around mean field theory}

A somehow different approach focuses on the description of the vicinity of the critical point, where the system
is characterized by a large correlation length and some of its properties become universal, in the sense that
they do not depend on the microscopic details but only on the type of symmetry that is broken at the transition.
It is observed that mean field theory gives good predictions for the universal properties for large enough $d$, but
the predictions become poor below some {\it upper critical dimension} $d_u$.
Hence in this case a good approach is to start from the qualitative Landau-Ginzburg theory and try to perform
a systematic inclusion of non--mean field effects.
The simplest calculation that one can do is a one loop perturbative calculation. Asking that the one loop corrections
remain smaller than the leading mean field results leads to the {\it Ginzburg criterion} (see~\cite{Am74} and references therein). 
The latter states that above dimension
$d_u=4$, mean field theory gives the correct critical exponents. Still loop corrections will be relevant for non-universal quantities if one is sufficiently close
to the critical point. How close to the critical point one should be to see deviations from mean field predictions
depends on the details of the model, but in any case when $d\to\io$ the interval where corrections are important shrinks to zero.
On the contrary, below $d_u=4$ loop corrections affect also the universal
predictions of mean field theory, \ie the critical exponents. In this case the Ginzburg criterion is however universal: it states that deviations from mean field
become visible when the correlation length of the system is larger than a universal number.
In this case the mean field predictions for the critical exponents, and in some cases also the predictions for non-universal quantities,
can be improved through the renormalization group approach~\cite{Wi75,ParisiBook,De07}.

Another important direction to improve mean field theory is to add the effect of nucleation below the critical point, where a first order transition
is observed. Here, starting from the mean field theory, one can take into account the effect of nucleation by non-perturbative instantonic 
calculations or via phenomenological theory~\cite{La69}.

\subsubsection{Summary}

In summary, the classical approach to phase transitions starts from the qualitative Landau theory, that provides a framework to understand
the universal features of the transition in terms of spontaneous symmetry breaking and the associated order parameter.
Then, the theory can be improved in two independent directions. On the one hand, one can try to be more quantitative by keeping the mean
field nature of the theory but taking into account more microscopic details of the system: this is the basis of the liquid theory developed in the 1950s,
and it works well far enough from the transition. On the other hand, one can remain qualitative (i.e. forget about the microscopic details)
but try to include non-mean field effects: mainly the effect of critical fluctuations at a second order phase transitions, that are included perturbatively
through the renormalization group, and the effect of nucleation at a first order transition point, that are non-perturbative in nature. This approach
is presented in Tab.~\ref{tab1:RFOT}.

Of course one would like in the end to develop, for a given system, a complete theory capable of describing correctly both its qualitative (universal) properties
and quantitative (non-universal) properties. However, already for simple liquids this seems to be a formidably difficult task.

\subsection{Qualitative mean field theory of the glass transition}
\label{sec:MFTqual}

The RFOT theory aims at repeating the same steps discussed above for standard phase transitions in the case of the liquid-glass transition.
The first step is then to identify a qualitative, Landau-like theory that reproduces the basic facts discussed in Sec.~\ref{sec:basic_facts}.
Next, one would like to construct quantitative approximate theories in finite dimensions, and finally understand non--mean-field corrections
related to both critical fluctuations and nucleation -- as it will turn out that both corrections simultaneously contribute in glasses.
Before proceeding, it is worth to stress that of course nothing guarantees that the same program that is successful for standard phase transitions
should work for the glass transition. Maybe the latter is a completely different phenomenon that requires an entirely new theoretical approach. Therefore,
the predictions of the RFOT theory must be, as usual in physics, tested a posteriori against experiment and numerical simulations. The result of such tests is encouraging 
as discussed \eg in~\cite{LW07,WL12,PZ10,Go09} and in Sec.~\ref{sec:finited}.
Of course this does not mean that alternative --and often complementary-- approaches to the glass transition are less interesting, see \eg~\cite{BB11,Ta11} for a review.

\subsubsection{The $p$-spin universality class}

The qualitative mean field theory of the glass transition was constructed by 
Kirkpatrick, Thirumalai and Wolynes (KTW) in a series of pioneering papers appeared from 1987 to 1989~\cite{KT87,KT87b,KT88,KT89,KTW89,KW87,KW87b}.
Building on the concept of replica symmetry breaking developed by Parisi in the solution of the mean field 
Sherrington-Kirkpatrick (SK) model of a spin glass~\cite{Pa79}, 
KTW realized that a simple generalization of the SK model has the desired phenomenology.
This is the so-called $p$-spin glass, whose Hamiltonian is given by
\begin{equation}
\label{Hpspin}
H_p(\s) = -\sum_{i_1 < \cdots < i_p} J_{i_1,\cdots,i_p} \sigma_{i_1} \cdots \sigma_{i_p} \ ,
\end{equation}
where $\sigma_i$ are either real variables subject to a spherical constraint 
$\sum_i \sigma^2_i =N$, or Ising variables, $\s_i=\pm 1$,
and $ J_{i_1,\cdots,i_p}$ are independent quenched random Gaussian variables with zero 
mean and variance $p! J^2/(2 N^{p-1})$. The sum is over all the 
{\it ordered} $p$-uples of indices $i_1 < \cdots < i_p$.
It is a mean field model because each degree of freedom interact with all the 
others with a strength that vanishes in the thermodynamic limit, in order to have
an extensive average energy.

\begin{figure}[t]
\centering
\includegraphics[width=.5\textwidth]{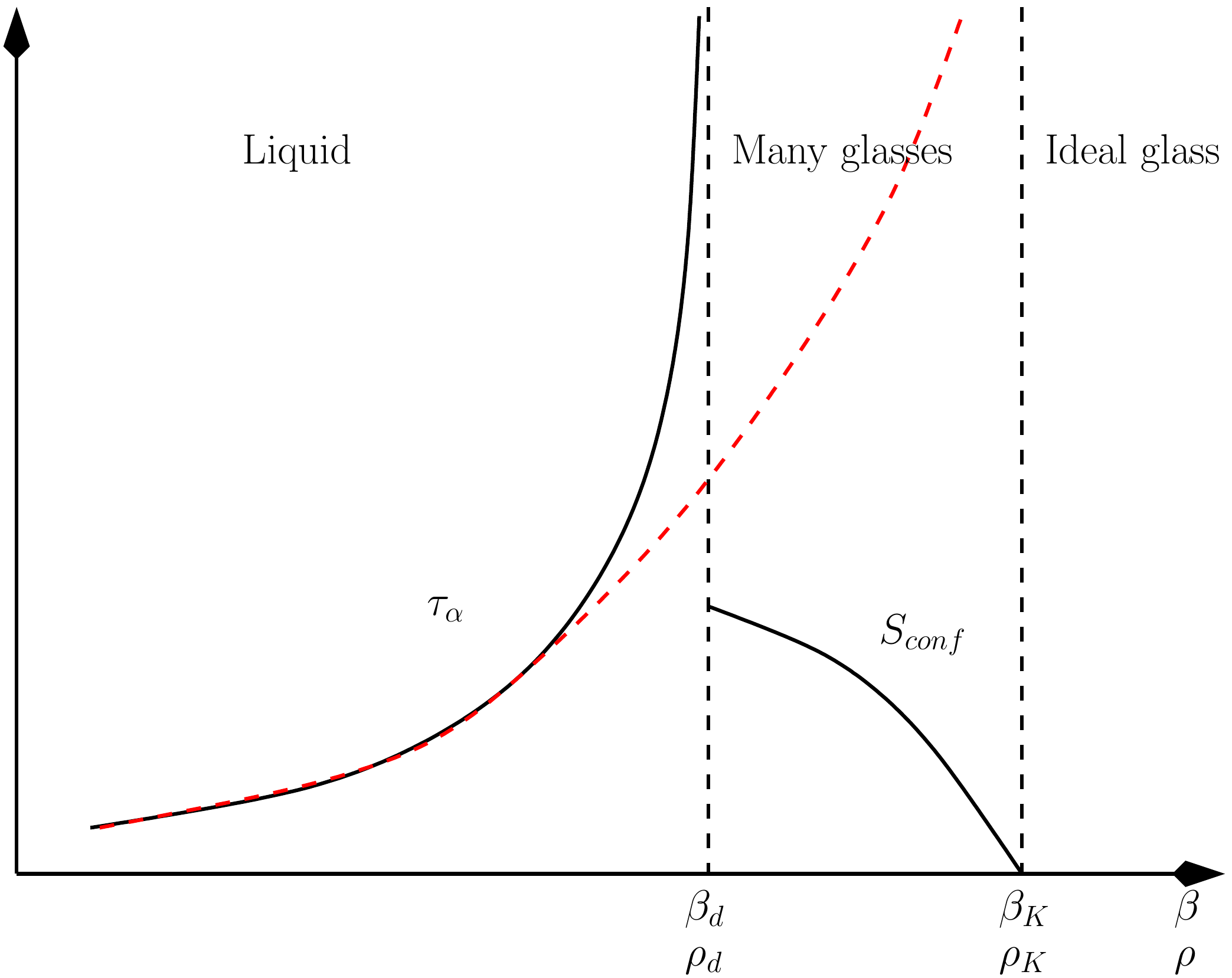}
\caption[Mean field phase diagram of the $p$-spin model]
{
Qualitative mean field phase diagram of RFOT theory, as described in Sec.~\ref{sec:MFTqual}.
A dynamical transition is observed at $T_d$, where $\t_\a\sim (T-T_d)^{-\g}$. Below $T_d$, 
the configurational entropy is finite and many glasses dominate the thermodynamics, while the
dynamics is not ergodic. At $T_K$, the configurational entropy vanishes and the number of glasses
becomes finite. It is reasonable to conjecture that in finite dimensional systems activated processes
restore ergodicity below $T_d$ so that $\t_\a$ only diverges at $T_K$ according to a VFT--like law (red
dashed line). The same phase diagram can be obtained as a function of increasing density instead of
decreasing temperature in Bethe lattice glass models~\cite{BM01,MKK08}.
}
\label{fig:MFTsch}
\end{figure}

This system shows (Fig.~\ref{fig:MFTsch}) an {\it equilibrium} Kauzmann
transition at a finite temperature $T_K$, where the configurational
entropy vanishes, the specific heat jumps downward and the order parameter 
discontinuously jumps to a finite value. Its dynamics is very similar
to the one of glass forming liquids in the region of temperature 
$T_m > T \gg T_g$, but the VFT behavior of the relaxation time is
not reproduced by these models: instead, a power law divergence of the
relaxation time is found at a temperature $T_d>T_K$. Although
this scaling is due to the mean field nature of this model, it
is not completely unrelated to what is observed in glass forming liquids, where 
the behavior of $\t_\a$ at temperature $T$ not too close to $T_g$ can also be described by a power law.
Indeed, the equations that describe the dynamics of the $p$-spin glass model
are formally very similar to the so-called {\it Mode--Coupling} equations~\cite{Go09}
that describe well the dynamics of supercooled liquids in a range of temperature
below $T_m$ but not too close to $T_g$~\cite{Cu02}. The dynamics displays {\it aging}
if the system is prepared at equilibrium above $T_d$ and quenched below this temperature~\cite{CK93}.
This model also displays dynamical heterogeneities: one can define a dynamical susceptibility
of the form of Eq.~\eqref{eq1:G4}, which diverges on approaching $T_d$~\cite{KT88,FP00}. 

The study of the
$p$-spin model provided a crucial connections between statics and dynamics, because the dynamical 
transition is found to be related to the emergence of {\it metastable states} of the system below $T_d$, separated
by infinite barriers~\cite{KT88,KT89,KW87}.
These metastable states have infinite life time in the mean field framework, and they are responsible for the divergence
of the relaxation time at $T_d$. It is important at this point to anticipate that in finite dimension, it is reasonable to expect
that the life time of the metastable states will be finite. Activated processes of barrier-crossing will restore the ergodicity below
$T_d$, and the relaxation time will only diverge at $T_K$. A relatively simple nucleation theory shows that the relaxation time
should diverge in a VFT-like way at $T_K$~\cite{KTW89}, and predicts the
existence of a static divergent correlation~\cite{BB04,FM07}.

Although $p$-spin models seem quite abstract and very different from structural glasses, one can construct mean field models
that belong to the same universality class and are formulated in terms of interacting particles~\cite{BM01,MKK08}.
This is interesting because these particle models also display a {\it jamming} transition with a phase diagram identical to the
one in Fig.~\ref{fig:PDsch}.
Moreover, these models do not display any quenched disorder, like structural glasses,
hence suggesting that the presence of quenched disorder in the $p$-spin Hamiltonian is not an essential ingredient to produce
a glassy phenomenology.

The Landau theory that describes the glass transition of the $p$-spin universality class
is formulated, by using the replica trick, 
in terms of an overlap matrix $q_{ab}$, which is the 
order parameter of the transition. It is the self-overlap between two different identical copies
(replicas) of the system, labeled $a$ and $b$:
\beq
\label{overlapeq}
q_{ab} \equiv \frac{1}{N} \sum_{i=1}^N \s_i^a \s_i^b \ ,
\eeq
and plays the role, in the context of spin glass theory, of the nonergodicity
factor (\ref{nonergdyn}). If $q_{ab}(x)$ is a local version of the overlap in point $x$,
the Landau free energy has the form~\cite{Fr05,DSW05,CBTT11,FPRR11}
\beq\label{eq:LandauTh}
\begin{split}
F[q] &= \int dx \left\{ \frac12 \sum_{ab} (\nabla q_{ab}(x))^2 + V[q(x)] \right\} \ , \\
V(q) &= \sum_{ab} \left( \frac{t}2 q_{ab}^2 - \frac{u+w}3 q_{ab}^3 + \frac{y}4 q_{ab}^4 \right)
-\frac{u}3 \sum_{abc} q_{ab}q_{bc}q_{ca} \ .
\end{split} \eeq
For the moment the number of replicas is left unspecified, this point will be discussed below.

In summary, the basic facts listed in Sec.~\ref{sec:basic_facts_summary} are reproduced by this mean field theory.
Moreover, the theory is predictive: the existence of dynamical
heterogeneities~\cite{KT88,FP00} and of a static correlation length~\cite{BB04,FM07}
were first predicted based on the analysis of $p$-spin--like models, and later observed numerically.
Excellent reviews on the properties of the $p$-spin--like models have been recently
published~\cite{CC05,BCKM97,Cu02}; in the following only the main results will be reviewed,
referring to these works for all the details.

\subsubsection{The TAP free energy}

To better understand what is going on in these models one has to investigate
the structure of their phase space. In particular, one wishes to characterize
the {\it equilibrium states} in order to understand the nature of the 
thermodynamical transition at $T_K$, as well as the structure of the
{\it metastable states} that trap the system at $T_d$ and are
responsible for the existence of a dynamical transition.

It turns out that at the mean field level a {\it pure state} $\a$ (equilibrium or metastable) is completely determined by the set of local
magnetizations $m^\a_i$, $i=1,\cdots,N$. 
The local magnetizations are minima of the
{\it Thouless--Anderson--Palmer (TAP) free energy}~\cite{MPV87,TAP77} which is defined by
\beq
F(m_i) = -T \log \left\{ \sum_{\s} e^{-\b H[\s] + \b \sum_i h_i (\s_i - m_i)}   \right\}
\eeq
where the auxiliary fields $h_i$ are introduced to fix the local magnetizations and
are determined by the condition $d F/d h_i = 0$ at fixed $m_i$. Its explicit calculation
relies on high temperature or large dimension expansions~\cite{GY91,Ri92}.
The weight $w_\a$ of state $\a$ is proportional to $\exp [-\b N f_\a]$, where
$f_\a = F(m_i^\a)/N$. Local minima of $F$ having a free energy
density $f > f_{min}$ for $N\to \io$ are {\it metastable} states.
The TAP free energy $F(m_i)$ depends, in general, explicitly
on the temperature, so the whole structure of the states may depend strongly
on temperature.

In mean field $p$-spin models, 
at low enough temperature, the number of states of given free energy
density $f$ is $\O(f)=\exp [ N \Si(f)]$. The function $\Si(f)$ is a decreasing function of $f$,
that vanishes continuously at $f=f_{min}$ and jumps discontinuously to zero above $f=f_{max}$.
A similar behavior is found in all models belonging to the $p$-spin universality class. 
The main peculiarity of $p$-spin models is that an 
{\it exponential number} of metastable states is present at low 
enough temperature.

One can write the partition function $Z$, at low enough temperature and for $N \to \io$, 
in the following way:
\beq
\label{Zm1}
\begin{split}
Z = e^{-\b N F(T)} \sim \sum_\a e^{-\b N f_\a}
= \int_{f_{min}}^{f_{max}}df \, e^{N [\Si(f)-\b f]}
\sim e^{N [\Si(f^*)-\b f^*]} \ ,
\end{split}
\eeq
where $f^* \in [f_{min},f_{max}]$ is such that $f - T \Si(f)$ 
is minimum, \ie it is the solution of
\beq
\frac{d\Si}{df} = \frac{1}{T} \ ,
\eeq
provided that it belongs to the interval $[f_{min},f_{max}]$.
Starting from high temperature, one encounters three temperature regions (Fig.~\ref{fig:MFTsch}):
\begin{itemize}
\item For $T > T_d$, the free energy density of the paramagnetic state is
smaller than $f - T\Si(f)$ for any $f\in [f_{min},f_{max}]$, so the paramagnetic
state dominates and coincides with the Gibbs state (in this region the decomposition
(\ref{Zm1}) is meaningless).
\item For $T_d\geq T \geq T_K$, a value $f^* \in [f_{min},f_{max}]$ is found, such that
$f^* - T \Si(f^*)$ is equal to $f_{para}$. This means that the paramagnetic state
is obtained from the superposition of an
{\it exponential number} of pure states of {\it higher} individual free energy
density $f^*$. The Gibbs measure is split on this exponential number of
contributions: however, no phase transition happens at $T_d$ because of the
equality $f^* - T \Si(f^*)=f_{para}$ which guarantees that the free energy is
analytic on crossing~$T_d$. 
\item For $T < T_K$, the partition function is dominated by the lowest free 
energy states, $f^* = f_{min}$, with $\Si(f_{min})=0$ and 
$F(T)=f_{min} - T \Si(f_{min}) = f_{min}$. At $T_K$ a phase transition occurs,
corresponding to the 1-step replica symmetry breaking transition found in the
replica computation. The order parameter $q_{ab}$ jumps discontinuously at the
transition.
\end{itemize}
In the range of temperatures $T_d > T > T_K$, the phase space of the model
is disconnected in an exponentially large number of states, giving a contribution
$\Si(T) \equiv \Si(f^*(T))$ to the total entropy of the system.
This means that the entropy $S(T)$ for $T_d > T > T_K$ can be written as
\beq
S(T) = \Si(T) + S_{vib}(T) \ ,
\eeq
$S_{vib}(T)$ being the individual entropy of a state of free energy $f^*$.
From the latter relation it turns out that the complexity $\Si(T)$ is the $p$-spin
analogue of the configurational entropy $S_c(T)$ of supercooled 
liquids\footnote{In the interpretation of experimental data one should remember that
in experiments $S_{vib}$ can be estimated only by the entropy of the crystal. However,
the vibrational properties of the crystal can be different from the vibrational 
properties of an amorphous glass, see \cite{RS01} for a review. 
Corrections due to this fact must be taken into account: in some cases it has been claimed that 
the difference is reduced to a proportionality factor between 
$S_c$ and $\Si$~\cite{SW83,corezzi}.
On the contrary, in numerical simulations $\Si(T)$ and $S_{vib}(T)$ can be measured directly~\cite{SKT99,CMPV99,AF07,XFL11}.
}.

The TAP approach provides also a pictorial explanation of the presence of
a dynamical transition at $T_d$. If the system is equilibrated at high 
temperature in the paramagnetic phase, and suddenly quenched below $T_d$,
the energy density start to decrease toward its equilibrium value.
This relaxation process can be represented as a descent in the free 
energy landscape at fixed temperature starting from high values of $f$.
When the system reaches the value $f_{max}$ it becomes
trapped in the highest metastable state and is unable to relax to the
equilibrium states of free energy $f^*$, as the free energy barriers between
different states cannot be crossed in mean field~\cite{BCKM97,Cu02}. 
For this reason below $T_d$ the system is unable to equilibrate.

\subsubsection{Real replica method}

For systems that belong to the $p$-spin universality class,
two very powerful and general methods to compute the complexity as a
function of the free energy of the states without directly solving the TAP
equations exist; they have been developed in 
\cite{KT89,MP99,FP95,Mo95,FP97,BFP97,Me99,MP00}.
They go under the name of the {\it real replica method}~\cite{Mo95} 
and the {\it potential method}~\cite{FP95}.
Both methods consider a number of copies of the system coupled
by a small field conjugated to the order parameter~$q_{ab}$.
Here, only a very short account of the real replica method will be presented.

The idea of \cite{KT89,Mo95,MP99} is to consider $m$ copies of the original system, 
coupled by a small attractive term added to the Hamiltonian.
The coupling is then switched off after the thermodynamic limit has been taken.
For $T<T_d$, the small attractive coupling is enough to constrain
the $m$ copies to be in the same TAP state.
At low temperatures, the partition function of the replicated system is 
then
\beq
\label{Zm}
Z_m = e^{-\b N \Phi(m,T)} \sim \sum_\a e^{-\b N m f_\a}
= \int_{f_{min}}^{f_{max}}df \, 
e^{N [\Si(f)-\b m f]}
\sim  e^{N [\Si(f^*)-\b m f^*]} \ ,
\eeq
where now $f^*(m,T)$ is such that $m f - T \Si(f)$ is minimum and
satisfies the equation
$\frac{d\Si}{df} = \frac{m}{T}$.
If $m$ is allowed to assume real values by an analytical continuation, 
the complexity can be computed from the knowledge
of the function $\Phi(m,T)=m f^*(m,T) - T \Si(f^*(m,T))$. 
Indeed, it is easy to show that
\beq
\label{mcomplexity}
\begin{split}
&f^*(m,T) = \frac{\partial \, \Phi(m,T)}{\partial m} \ , \\
&\Si(m,T) = \Si(f^*(m,T)) = m^2 \frac{\partial \,[ m^{-1} \b \Phi(m,T)]}{\partial m} = 
m \b f^*(m,T) - \b \Phi(m,T) \ .
\end{split}
\eeq
The function $\Si(f)$ can be reconstructed from the parametric plot of $f^*(m,T)$ 
and $\Si(m,T)$ by varying $m$ at fixed temperature. In other words, $\Phi(m,T)$ is the Legendre
transform of the function $\Si(f)$. Its knowledge allows one to obtain the function $\Si(f)$ and from it
reconstruct all the thermodynamic properties of the system, including those of metastable states
and the dynamical transition $T_d$.
In~\cite{Me99} this method was applied to the spherical $p$-spin 
system and it was shown that the method reproduces all the results obtained from the
explicit TAP computation.
Note that replicas are introduced here to take into account the existence of multiple equilibrium
states, without any reference to quenched disorder.

\subsubsection{Dynamics}

The Langevin dynamics of the model can also be solved exactly~\cite{KT88,CK93,Cu02}.
The Martin-Siggia-Rose formalism~\cite{MSR73,DD78} allows one to write it in terms of a closed equation for the
time-dependent spin-spin correlation function
\beq
C(t)=\lim_{N\to\io} \frac1N \sum_i \langle \s_i(t) \s_i(0) \rangle \ .
\eeq
Note that the latter measures the overlap between the configuration at time 0 and that at time $t$.
The self-consistent equation for $C(t)$ reads~\cite{Cu02,CC05}
\beq\label{eq:DYNpspin}
\frac{dC(t)}{dt} = -T C(t) - \frac{p}{2T} \int_0^t du C(t-u)^{p-1} \frac{dC(u)}{du} \ .
\eeq

From this equation, one can show that
a dynamical transition
is found at the same temperature $T_d$ that comes from the TAP computation.
Upon approaching $T_d$, $C(t)$ has the same form as in Fig.~\ref{fig1:Fs}.
It shows a two-step relaxation, with well separated time scales and a structural
relaxation time that
shows a power-law divergence for $T \to T_d$, $\t_\a \sim |T-T_d|^{-\g}$. 
A dynamical order parameter
can be defined as
\beq
\label{overlapdyn}
q_d =  \lim_{t \to \io} C(t) \ ;
\eeq
it is the analogue of the dynamical nonergodicity factor defined in (\ref{nonergdyn})
and jumps to a finite value at $T_d$. Below $T_d$ the system is no more able to
equilibrate with the thermal bath and enters a {\it nonequilibrium} regime.
This result confirms that {\it metastable} states are responsible for the slowing down of
the dynamics and for the dynamical transition at $T_d$.

As already mentioned, a four-point dynamical susceptibility associated to $C(t)$ can be
defined, and one can show that it diverges at $T_d$~\cite{KT88,FP00}.


\subsection{Quantitative mean field theory (theories) of the glass transition}

Even before the coherent RFOT theory was formulated by KTW, several independent attempts to construct quantitative approximate theories,
in the same spirit of the HNC or PY theory of liquids, had been formulated.
Here a short review of these theories will be given, focusing in particular on their connection with the broader RFOT picture.

\subsubsection{Density functional theory}

A density functional theory (DFT) of glassy states was first proposed in~\cite{SW84,SSW85} and later improved by 
several authors~\cite{KT88,KT89,DV99,CKDKS05,KM03,YYO07}. DFT is the analog of the TAP approach for particle systems.
One writes the free energy of the system as a functional of the density field. A commonly used approximate form is
\beq\label{eq:DFT}
-\b F[\r(x)] =  \int \r(x) [ 1 - \log \r(x) ] +\frac{1}2 \int dx dx' \r(x) \r(x') c(x-x') \ , 
\eeq
where $c(x)$ is the direct correlation function of the liquid~\cite{Hansen}. 
The density profile is determined by minimization of the free energy.
Because in general this is a too difficult task, one looks for a parametrization of the density field as a sum of Gaussian terms:
\beq\label{eq:DFTrhoG}
\r(x) = \sum_{j=1}^N \frac{e^{-\frac{(x-R_j)^2}{2 A}}}{\sqrt{2 \pi A}^d} \ .
\eeq
Now the parameters $A$ (the ``cage radius'') and $R_j$ (the ``equilibrium positions'' in the glass) have to be determined by inserting
this parametrization of the density in Eq.~(\ref{eq:DFT}) and minimizing the result, which has the form
\beq
-\b F[\r(x)] / N = \frac{d}2 \log(2\pi A) + \frac{1}2 \int \frac{dq}{(2\pi)^d} c(q) S_0(q) e^{-A q^2} \ ,
\eeq
where $S_0(q) = N^{-1} \sum_{jk} e^{i q (R_j - R_k)}$ is the structure factor of the equilibrium positions.

At this point one can either choose to approximate $S_0(q)$ with the liquid structure factor, $S_0(q) \sim S(q)$, and determine only $A$ by minimization~\cite{SW84,SSW85}
or try to perform a full minimization including the $R_j$ for a finite system size $N$~\cite{DV99,CKDKS05,KM03,YYO07}.
In both cases, amorphous solutions with a finite $A$ are found at high enough density or low enough temperature.
However, DFT has several problems:
\begin{itemize}
\item The approximations that are made, starting from the definition of the density functional in Eq.~\eqref{eq:DFT}, are quite crude and for this reason the quantitative
accuracy of DFT is very poor. Of course one could improve these approximations, but a satisfactory quantitative DFT has not been obtained for the moment.
\item Even if some amorphous solutions can be found numerically at finite $N$, counting their number and its scaling with $N$ is a formidable task, hence a proper
thermodynamic characterization of the glassy states and their configurational entropy $\Si(f)$ is impossible.
\end{itemize}

For these reasons, DFT remains for the moment only useful to illustrate the basic mechanism of the glass transition in the RFOT framework: the discontinuous 
appearance of amorphous free energy minima~\cite{SW84,SSW85,KT88,KT89}. A satisfactory quantitative description of this phenomenon in $d=3$ requires
considerable improvement of DFT.

\subsubsection{Mode-coupling theory}

Mode-coupling theory (MCT) has been initially developed to describe the critical slowing down at a second order phase transition~\cite{OK76},
and later generalized to describe the glass transition~\cite{BGS84,Go09}. MCT equations are closed equations for the coherent scattering
function defined in Eq.~\eqref{eq1:Fk}. If $\phi(k,t) = F(k,t)/S(k)$ is the normalized correlator such that $\phi(k,t=0)=1$, the MCT equations read~\cite{SS10,Go09}
\beq\label{eq:MCT}
\begin{split}
&\g_k \frac{d \phi(k,t)}{dt} = -  \phi(k,t) - \int_0^t du \, m(k,t-u) \frac{d \phi(k,u)}{du}  \ , \\
&m(k,t) = \frac{\Omega_{d-1}}{(4 \pi)^d} \int_0^\io dp \int_{|k-p|}^{k+p}dq V(k,p,q) \phi(p,t) \phi(q,t) \ ,
\end{split}\eeq
where $\Omega_d$ is the $d$-dimensional solid angle and $\g_k$ is a $k$-dependent microscopic relaxation rate related to $S(k)$.
The kernel $V(k,p,q)$ is a function of the static liquid correlation functions $S(q)$ and $c(q)$~\cite{SS10}.

The analogy between MCT equations~\eqref{eq:MCT} and the exact equations that describe the Langevin dynamics of $p$-spin models,
Eq.~\eqref{eq:DYNpspin}, is striking. In particular choosing $p=3$ one sees that the only difference is the fact that the MCT equations are complicated
by the $q$-dependence of all correlation functions, that is clearly due to the finite dimensional nature of MCT. This similarity has led many to conjecture that MCT is the
finite-dimensional formulation of the dynamic theory of RFOT~\cite{KW87,KT88,KT89,BCKM97,Cu02}.
However, for the moment, the MCT equations have always been derived by performing some ``uncontrolled'' approximations, in the sense that no
systematic improvement of the derivation can be obtained. A derivation of MCT in terms of a systematic expansion (possibly a low-density or 
high-dimensional expansion)~\cite{ABL06,KiKa07,JvW11} 
would help clarify its relation with the general RFOT scenario. 
It has been shown that the basic phenomenology of MCT is insensitive to the detailed form of the kernel, suggesting that indeed MCT can be seen as
a Landau theory of the dynamics at a glass transition~\cite{ABB09}.

Despite the fact that the direct connection between MCT and the RFOT picture remains partially unclear, 
the MCT equations can be solved both numerically and analytically in some asymptotic regimes and their predictions are identical to those
obtained from the mean field Eq.~\eqref{eq:DYNpspin}: they predict a sharp discontinuous transition at a temperature $T_d$, where the relaxation
time of $\phi(k,t)$ diverges as a power-law. Below $T_d$, the non-ergodic parameter $\lim_{t \to \io} \phi(k,t) = f(k)$ and the cage radius $A$
are both non-zero as in Eq.~\eqref{nonergdyn}. MCT gives quantitative predictions for $T_d$ and $f(k)$. Moreover
on approaching $T_d$ from above, it predicts a number of scaling laws.
Detailed tests of these predictions have been made both numerically~\cite{KA95a,KA95b} and experimentally~\cite{MUP91,Go99,LKKL12}.
There is quite general consensus that MCT predictions work very well in a range of temperatures slightly above the glass transition $T_g$, while
on approaching $T_g$ deviations from MCT behavior become more pronounced.

An important point is that MCT did not limit itself to provide quantitative results for already known quantities. It also provides new predictions for 
previously unknown phenomena, that were later confirmed by numerical simulations and experiments. The most impressive one is probably
the prediction of a re-entrant glass transition line and the existence of a glass-liquid-glass and a direct glass-glass transition in attractive 
colloids~\cite{Sc02}.

\subsubsection{Replica theory}

Despite early attempts~\cite{KT88}, a quantitative replica theory (RT) of finite dimensional glasses has been developed much later than DFT and MCT, mainly
thanks to the work of M\'ezard and Parisi~\cite{MP96,MP09} who found a suitable formulation of the real replica method of Monasson~\cite{Mo95} for particle
systems.

The dynamical transition can be described in a static framework
by introducing a replicated version of the system~\cite{Mo95,MP96}: 
for every particle one introduces $m-1$ additional particles identical to the first one. 
In this way one obtains $m$ copies of the original system, labeled by $a=1,\ldots,m$.
The interaction potential between two particles belonging to replicas $a,b$ is $v_{ab}(r)$, where
$v_{aa}(r) = v(r)$,
the original potential, and $v_{ab}(r)$ for $a\neq b$ is an attractive potential that constrains
the replicas to be in the same metastable state.
The basic fields are the one and two point density functions
\beq\label{physical_fields}
\begin{split}
&\hat\rho_a(x)=\sum_{i=1}^N\delta(x-x_i^{a}) \ , \\
&\hat\rho_{ab}^{(2)}(x,y)=\hat\rho_a(x)\hat\rho_b(y)-\hat \rho_a(x)\delta_{ab}\delta(x-y) \ .
\end{split}
\eeq

To detect the dynamical transition one has to study the two point correlation functions when
$v_{ab}(r)\to 0$ for $a\neq b$, and in the limit $m\to 1$ 
which reproduces the original model~\cite{Mo95,MP96}.
In this limit, the two-replica correlation function is, for $a\neq b$:
\beq
\langle  \hat C_{ab}(r) \rangle = \int d x f(x)  \la \hat\r_a\left(r+\frac{x}2 \right) \hat\r_b\left(r-\frac{x}2\right) \ra - \r^2 \ .
\eeq
Because of the limit $v_{ab}(r)\to 0$, the two replicas fall in the same state but are otherwise uncorrelated inside the state,
therefore $\langle  \hat C_{ab}(r) \rangle = \langle  \hat C(r,t \to\io) \rangle $ which provides the crucial identification
between replicas and dynamics~\cite{FPRR11,PZ10}. 
Similar mappings can be obtained for four-point correlations.

One can introduce for convenience an external field $\nu_a(x)$ (that is derived from a space-dependent chemical potential),
in such a way that the density correlation functions
can be obtained by taking the derivative of the free-energy with respect to it~\cite{Hansen}. 
The free energy is defined as the logarithm of the partition function, and
its double Legendre transform defines the Gibbs free energy $\Gamma[\{\rho_a(x)\},\{\rho^{(2)}_{ab}(x,y)\}]$~\cite{Hansen,MH61}:
\beq\label{eq:HNCrep}
\begin{split}
&\G =  \frac 12 \sum_{a,b} \int d x d y \left[ \r^{(2)}_{ab}(x,y) \ln \left( \frac{\r^{(2)}_{ab}(x,y)}{\r_a(x) \r_b(y)}
 \right) 
-  \r^{(2)}_{ab}(x,y) + \r_a(x) \r_b(y) \right]  + \sum_a \int dx \r_a(x) \left[ \ln \r_a(x) - 1 \right] \\
&+ \hskip-10pt  \sum_{n \ge 3, a_1,\ldots,a_n} \hskip-10pt \frac{(-1)^n}{2n}
\int d x_1 \cdots d x_n \, \r_{a_1}(x_1) h_{a_1 a_2}(x_1,x_2) 
 \cdots \r_{a_n}(x_n) h_{a_n a_1}(x_n,x_1)  + \G_{\rm 2PI} \ ,
\end{split}
\eeq
where $h_{ab}(x,y) = \r^{(2)}_{ab}(x,y)/\r_a(x) \r_b(y) - 1$ and 
$\G_{\rm 2PI}$ is the sum of 2-line irreducible diagrams~\cite{MH61}.
The average values of the fields in Eq.~\eqref{physical_fields}, namely $\overline\rho_a(x)$ and $\overline\rho_{ab}(x,y)$, can be obtained by solving the saddle point equations
\begin{equation}\label{SP}
\left.\frac{\delta \Gamma[\{\rho_a\},\{\rho^{(2)}_{ab}\}]}{\delta \rho_{ab}^{(2)}(x,y)}\right|_{\overline \rho_{ab}(x,y)}=\frac{1}{2}v_{ab}(x,y) \ ,
\end{equation}
and similarly for $\rho_a(x)$.

The crucial advantage of replica theory with respect to DFT is that the whole system of $m$ replicas is in its liquid phase, hence it is {\it homogeneous}.
In fact, introducing replicas allows one to sum over all amorphous metastable glasses, thus recovering translational invariance.
Therefore, one can restrict the considered solutions to those where $\rho_a(x)=\rho$, a choice that allows one to greatly simplify the computation with respect to DFT,
where instead one has to work with the full amorphous density profile $\rho(x)$.

A concrete implementation of this procedure amounts to neglecting the 2PI diagrams, therefore obtaining a replicated version of the HNC equations~\cite{MP96}.
In this way one obtains predictions for $T_d$, $T_K$, the complexity function, and the non-ergodic parameter $f(k)$~\cite{MP96}.
The results for $T_d$ and $T_k$ are reasonably good, however the prediction for $f(k)$ is much worse than the MCT one. Moreover, the complexity is found to
be quite a bit smaller than the numerical estimate, and this scheme gives inconsistent predictions in the glass phase below $T_K$~\cite{MP96}.

A better approximation scheme in replica theory is obtained by considering a slightly different formulation, in which the basic degrees of freedom are 
{\it molecules} built of an atom from each replica. The physical reason for this is that replicas are assumed to be in the same state, hence the density profile
of the different replicas is assumed to be similar, and this can be described by assuming that atom 1 in replica 1 is very close to atom 1 in replica 2, and
so on. Furthermore, the average distance between atoms in the molecule, which is related to the cage radius $A$,
is assumed to be small and one performs a systematic expansion in powers of $A$ starting from the standard formulation of molecular liquid theory
(\eg in the HNC approximation)~\cite{MP99,MP09}.
This formulation is much more effective for treating the glass phase at $T_K$ and below, where the cages are well formed.
Indeed, this approach gives good quantitative predictions for the configurational entropy above $T_K$, and for the specific heat and the
energy of the glass phase below $T_K$~\cite{MP99,CMPV99}.
Moreover, this approach allows one to obtain a very detailed theory of the jamming transition that happens deep in the glass phase. 
This will be the subject of Sec.~\ref{sec:finited}.
The main drawback of this approach is that because molecules are assumed to be stable and have small radius, the process of ``molecule dissociation''
that in replica language leads to the disappearance of metastable states at $T_d$ is not very well described by this approach. Depending on the details,
either $T_d$ is not found, or it is largely overestimated.

The construction of a ``mixed'' replica approach, capable of describing both the region around and below $T_K$ where cages are well formed, and the
region around $T_d$ where cages are unstable, would be an important achievement, that could be obtained by mixing the diagrams of the replicated
HNC with those of the molecular liquid formulation.

\subsubsection{Towards a consistent finite dimensional implementation of RFOT theory}

The different aspects of the RFOT scenario for mean-field $p$-spin glasses all find a counterpart in the study of the liquid-glass transition in finite dimension.
The equivalent of the TAP approach is DFT: in both cases one tries to find the metastable states by minimizing a suitable free energy functional.
The dynamics of $p$-spin models is described by an equation that corresponds to the so-called ``schematic'' limit of MCT, \ie the limit where the wavevector
dependence is neglected. Finally, replica theory can be formulated for $p$-spin models and liquids along the same lines.

Yet, the important difference is that in mean field $p$-spin models, it can be shown exactly that the TAP, dynamic and replica approaches give completely
equivalent descriptions of the same phenomenon, which is the emergence of glassy metastable states~\cite{CC05}. While in finite dimensions, DFT, MCT and RT
are necessarily approximate theories that use different approximations and lead to quite different quantitative predictions, with marked discrepancies
\eg on the value of $T_d$, on the shape of the non-ergodic parameter, and so on.
This is not particularly worrying as far as $d=3$ is concerned, because already in liquid theory different approximations schemes (\eg HNC and PY)
obviously lead to different predictions~\cite{Hansen}. 

However, for the RFOT scenario to be consistent, one would like to show that in the limit $d\to\io$, the static and dynamic descriptions of the transition
become equivalent, exactly in the same way as the HNC and PY equations both converge to the Van der Waals equation of state.
This consistency check was attempted in one of the earliest works on RFOT theory~\cite{KW87}. There it was shown that under a Gaussian approximation
for the non-ergodic parameter, DFT and MCT lead to equivalent predictions for the dynamical transition point of hard spheres.
Another interesting attempt in this direction is the work of Szamel who derived the MCT equations from a static replica approach~\cite{Sz10}.
However, despite these apparent successes, the limit $d\to\io$ turns out to be much more complicated: the solution of full MCT (without the Gaussian approximation)~\cite{SS10,IM10}
leads to a strong discrepancy with RT~\cite{PZ10} and DFT~\cite{KW87}.
Therefore despite these results, the problem of obtaining a consistent RFOT theory of the glass transition in $d\to\io$ remains open.
This will be discussed in Sec.~\ref{sec:highd}.

\subsection{Corrections around mean field theory}

After a consistent mean field theory of the glass transition has been obtained,
in finite dimensions one would like to compute the corrections due to fluctuations around the mean field approximation.
When this program is carried out, one finds that (according to the mean field theory itself) 
there are two important sources 
of corrections to the mean field scenario. The first corrections originate from critical
fluctuations (the dynamical heterogeneities) that become important around the dynamical transition $T_d$ below the upper critical
dimension, as in any standard critical phenomenon~\cite{BB07,FPRR11}.
The second corrections are non-perturbative phenomena related to activated processes, that are important below $T_d$, close
to the experimental glass transition $T_g$ and down to $T_K$.
They can be taken into account by a phenomenological approach,
leading to a number of predictions that are in good agreement with experiment~\cite{LW07}.
This approach can be partially justified via a renormalization group analysis~\cite{CDFMP10,CBTT11}:
however, this part of the theory remains the most controversial~\cite{BB09}.

One big source of controversy is the very existence of the Kauzmann temperature $T_K$, which relies on a dubious extrapolation of
the relaxation time below $T_g$. In fact, many other functional forms that do not display a singularity at finite temperature are compatible
with the data, see \eg~\cite{ECG09}.
Many theoretical arguments have indeed been proposed
against its existence in finite dimensions~\cite{St88,DTS06}. None of them is however completely convincing, leaving the question open to debate. 
The disappearance of the Kauzmann transition in any finite dimensions (or possibly below some lower critical dimension larger than 3)
could be due to a non-perturbative process that is not taken into account in the mean field theory. For this reason, the present approach
(start from MFT and then consider corrections around it) seems unable to capture such hypothetical processes.
In any case, although the existence or not of $T_K$ is certainly
an important question from the theoretical point of view, it is not so important for the theory of everyday glasses. In these systems, one is
always quite far from $T_K$, and the static and dynamic length scales never reach too large values (actually, they typically are of the order
of a few interparticle distances at $T_g$). Therefore, the really important question for the theory of glasses is {\it not} whether MFT retains its validity
at all length scales, down to the transition point. The important question is whether MFT (plus corrections) retains its validity down to $T_g$, and whether
it can explain the phenomenology of the glasses we observe in nature. It might well be that MFT provides a good description in a region where the relevant
length scales are smaller than, say, 100 interparticle distances, and for larger length scales some unknown process completely washes away the mean field
phenomenology. If this is the case, MFT would be an excellent theory of real glasses, because one never observes such large length scales and 
deviations from mean field would remain unobserved. Such a situation would be somehow similar to the BCS theory of superconductivity, which is a mean field theory but yet
provides excellent qualitative and quantitative predictions. This is due to the fact that the pairing length $a$ is so large that each Cooper pair effectively 
interact with a large number of other pairs. Hence the Cooper pair correlation length $\xi$ is much smaller than $a$, $\xi/a\ll 1$, 
making MFT a very good description. Indeed, the associated Ginzburg criterion tells that one has to be
extremely close to the transition to observe deviations from BCS theory.
The aim of the rest of this discussion will be therefore to give criteria of validity of MFT self-consistently within MFT itself.

\subsubsection{Critical fluctuations around MCT}

Within the static formulation of RFOT theory, 
the dynamical transition at $T_d$ is a spinodal point at which the metastable states become unstable and disappear~\cite{FPRR11,KZ11a}.
Associated with the instability one finds a diverging correlation length (akin to the diverging length that is found at a spinodal point) 
that gives rise to dynamical heterogeneity~\cite{FP00}.
As for a standard spinodal point, in its vicinity one can reformulate the field theory in Eq.~\eqref{eq:LandauTh} as a cubic theory.
It has been shown in~\cite{FPRR11} that the relevant theory is a $\f^3$ theory in a random field. 
Performing a systematic loop expansion around the mean field saddle point in such a theory, one can obtain a Ginzburg criterion
and compute the upper critical dimension $d_u$.
This allows one to establish that $d_u=8$ (for the $\f^3$ theory $d_u=6$, but the random field provides dimensional reduction)~\cite{FPRR11}.
Using the HNC replicated free energy given in Eq.~\eqref{eq:HNCrep}, a quantitative estimate of the Ginzburg criterion has been
given in~\cite{FJPUZ12} for $d<d_u$. The result of this analysis is that in $d=3$, one needs a dynamical correlation length of order
1 (in units of interparticle distance) to observe deviation from the mean field behavior. A comparison with numerical data obtained in~\cite{SA08}
shows that the dynamical correlation becomes of order 1 only slightly below $T_d$. Hence in the full range where the MCT
phenomenology is observed, the Ginzburg criterion indicates that non--mean-field corrections remain small. Below $T_d$, these corrections might
become important, however at that point activated processes start to dominate over the MCT phenomenology. 
One concludes that
in $d=3$ critical fluctuations around the dynamical transition do not play any observable role. 

Such critical fluctuations have been also described within MCT~\cite{BBMR06,BB07,BBBKMR07a,BBBKMR07b}, leading to 
a very detailed characterization of the dynamical heterogeneity and to similar results for what concerns the range of validity of mean field theory.
In particular, in~\cite{BB09} a phenomenological Ginzburg criterion for MCT has been proposed, that leads to the same conclusion as the analysis
of~\cite{FJPUZ12}. 

These results seem to indicate that developing a systematic RG theory to discuss critical fluctuations around the dynamic transition might not
be worth the effort, because deviations from mean field should be very difficult to observe at least in $d=3$. The situation might be different
in higher dimensions where nucleation is suppressed and the MCT regime is more clearly observed~\cite{CIPZ12}.

\subsubsection{Activated processes and nucleation theory at low temperatures}

The last, and actually most important, ingredient of the RFOT scenario is a phenomenological treatment of the
activated processes that restore ergodicity between $T_d$ and $T_K$. These processes are at the origin of the
slowing down at $T_g$ and in particular they are responsible for the exponential divergence of the relaxation time,
Eq.~\eqref{VFT}.
In the MFT, the transition at $T_K$ has a mixed nature. It is a first order transition from the point of view of the order
parameter: in fact, at $T_K$ the free energy of two phases --the one where replicas are uncorrelated and the one
where replicas are in the same state-- cross. However, it is a second order transition from the thermodynamic point of
view. This mixed character makes the Kauzmann transition a very peculiar one.
Indeed, its description requires setting up a nucleation theory~\cite{KTW89,LW07,WL12,BB04,BB09} --due to the first order nature of the mean field transition--
but at the same time it predicts the existence of a diverging correlation length as in second order transitions, with associated
critical exponents. In the following a short account of this theory will be presented.

As usual, in mean field theory nucleation processes --where a nucleus of a phase forms inside another phase, paying a surface
tension cost but gaining in bulk free energy-- 
are not taken into account, because in infinite dimensions
the surface and volume terms have the same scaling ($L^d$ is the same as $L^{d-1}$ when $d\to\io$).
Hence, in mean field theory
metastable states
with an {\it intensive} free energy higher than the free energy of the ground states,
$f > f_{min}$, still have an infinite life time.
These states are responsible for the existence of a finite
complexity. Their lifetime is infinite, so they are able to trap the system 
below $T_d$. This is the reason why the dynamical transition, \ie the divergence
of the structural relaxation time, happens at a temperature $T_d > T_K$.
On the contrary,
in a model with short range interactions, metastable states have a finite lifetime
due to the nucleation of bubbles of the stable phase inside the metastable one,
so they are not thermodynamically stable.
One should expect the existence of well defined states with $f>f_{min}$
to be impossible; but the analogy between mean field models and real glasses
is based on the analogy between the complexity $\Si(T)$ and the
configurational entropy $S_c(T)$. How can one explain the existence of a finite
configurational entropy, related to well defined metastable states, in a short
range system?

Moreover, the observed crossover of the relaxation time from a power--law behavior
to a Vogel--Fulcher--Tamman law (\ref{VFT}), that happens around $T_d$,
is not explained by the mean field theory, which predicts a strict
power--law divergence of $\t_\a$ for $T \to T_d^+$. 
The observation of a finite relaxation time below $T_d$ is again related to the finite
lifetime of metastable states. The system, instead of being trapped forever into a
state, is able to escape, due to nucleation processes; it is then trapped by
another state, and so on. These processes of jump between metastable states are
{\it activated} processes: the system has to cross some free energy barrier in order
to jump from one state to another. The relaxation time is then expected to scale
as
$\t_\a(T) \sim \t_\io \exp \big[ \b \D F(T) \big]$,
$\D F(T)$ being the typical free energy barrier that the system has to cross at
temperature $T$.
The VFT law and the observation that $T_0 \sim T_K$ suggest that the barrier should
diverge at $T_K$, $\D F(T) \sim (T-T_K)^{-1}$; more generally, the Adam--Gibbs formula~\cite{AG65} relates
this divergence to the vanishing configurational entropy, $\D F(T) \sim S_c(T)^{-1}$.
It is then essential to understand what is really the meaning of $S_c(T)$ in finite dimension
and why it is related to the free energy barrier for nucleation.

A crucial observation (actually, a theorem~\cite{MS06})
is that the divergence of the relaxation time at $T_K$, in short range
systems, is possible {\it only} if the cooperative processes of structural rearrangement involve
atoms that are correlated on a typical length scale $\xi$, which diverges at $T_K$.
If no divergent length scale is present in the system, it is always possible to divide it
in {\it finite} subsystems, each one relaxing {\it independently} of the others: and the relaxation
of a finite system is bound to happen in finite time, if the interactions are finite and have short range.
A simple idea that follows from the above observation and can explain how the mean field picture 
is modified in short range systems is the following~\cite{KTW89,XW00,LW07,BB04,Fr05,DSW05}.
There exists a typical length scale $\xi(T)$ over which structural relaxation processes
happen. If one looks at smaller length scales, the system behaves as if it were mean field:
metastable states are stable for $l < \xi$, yielding a finite {\it local} complexity.
However, for large scales $l > \xi$, metastability is destroyed and only the lowest free energy
states are stable. For $T \to T_K^+$, $\xi \to \io$, so below $T_K$ a stable {\it ideal} glass
phase is possible. This idea leads naturally to the identification of the configurational 
entropy $S_c(T)$ with the {\it local} complexity $\Si(T)$, and to a derivation of an
Adam-Gibbs--like relation between the relaxation time $\t_\a$ and $\Si(T)$.

These considerations have been formalized in~\cite{BB04} where a concrete procedure to define and measure
$\xi$ has been proposed, and later generalized to improved geometries~\cite{CB12}. 
Following this work, numerical measurements of $\xi$ have been performed and 
have revealed a good agreement with the theoretical predictions~\cite{BBCGV08,CCGGGPV10,CGV12,BK12}.
Analytic calculations in simplified models with Kac--like long range interactions~\cite{Fr05,DSW05,FM07} and renormalization group
calculations~\cite{CDFMP10,CBTT11}
have been performed confirming this scenario.

\subsection{Summary}

To summarize, the RFOT theory of the glass transition is constructed by following closely the path that led to the
standard theory of critical phenomena, as sketched in Tab.~\ref{tab1:RFOT}.
The theory was started by the pioneering works of Parisi in 1979~\cite{Pa79}, 
who developed the concept of spontaneous replica symmetry breaking and
identified its consequences in terms of proliferation of metastable states~\cite{MPV87},
and of Kirkpatrick-Thirumalai-Wolynes~\cite{KT87,KT87b,KT88,KT89,KTW89,KW87,KW87b}, 
who identified a concrete class of simple mean field models that reproduced many basic facts
of the phenomenology of glasses. At the same time, early attempts to construct approximate quantitative theories of the glass transition,
such as density functional theory~\cite{SW84}, mode-coupling theory~\cite{BGS84}, and replica theory~\cite{MP96}, 
fit well under the RFOT umbrella, and can be thought as different
implementations of RFOT. 
Early consistency checks~\cite{KW87} corroborated this interpretation, even if many details of it remain open and a complete RFOT
theory of infinite dimensional liquid is still lacking (Sec.~\ref{sec:highd}).
Finally, RFOT provides the framework to construct a mean field theory of the jamming transition that happens inside the glass phase
(Sec.~\ref{sec:defjamming})~\cite{BM01,PZ05,MKK08,PZ10,BJZ11}. This will be the subject of Sec.~\ref{sec:finited}.

Nowadays, most of the research on RFOT theory is focused on a detailed understanding of the corrections around mean field theory.
There are at least two different sources of corrections.
Around $T_d$, corrections are due to standard critical fluctuations and can be treated via loop expansions 
both in the statics and in the dynamics,
leading to a Ginzburg criterion that states that the upper critical dimension is $d_u=8$, 
and that in $d=3$ non--mean-field effects (of this kind) should be
barely observable~\cite{BB07,BB09,FPRR11,FJPUZ12}.
Mean field theory predicts in this region the existence of large dynamical heterogeneities
and an increasing dynamical correlation length, that are indeed observed~\cite{BBBCS11}. The associated critical exponents should be 
the mean field ones for $d\geq 8$~\cite{BBMR06,BB07,BBBKMR07a,BBBKMR07b} and close to the mean-field ones for $d<8$, 
but this for the moment remains controversial~\cite{SA08,KDS09,SF10}.
Below $T_d$ and in low dimensions, activation is the main source of corrections around mean field, and
a nucleation theory has to be developed to take it into account~\cite{KTW89}. 
Doing this results in a number of predictions about the existence of a static correlation
length~\cite{XW00,BB04,Fr05,DSW05,MS06,LW07,CB12} that are being extensively tested in simulation~\cite{BBCGV08,CCGGGPV10,CGV12,BK12}.

A crucial issue is of course how much of the mean field picture survives in physical dimensions $d=2,3$.
It is very likely that the glass transition does not exist in two dimensions~\cite{SK00,DTS06}, 
suggesting a {\it lower critical dimension} $d_l > 2$. Whether $d_l < 3$ or not, and what is the important physical mechanism
that determines $d_l$, remain a largely open problem. Preliminary renormalization group 
investigations~\cite{CBTT11,CDFMP10,YM12a,YM12b} give contradictory results. The crucial difficulty,
which has been extensively discussed in~\cite{CBTT12}, is the following.
There are many simple lattice models that display a RFOT at the fully-connected mean field level. 
However, for most of those, the transition either disappears or becomes a continuous spin glass transition as soon as some spatial
fluctuations are included, \eg by considering the model on a Bethe lattice. This is a well known effect: the inclusion of fluctuations
lowers the barriers between states (for example, it lowers considerably the critical temperature in the ferromagnetic Ising model),
in such a way that models that display the transition at the mean field level can become paramagnetic once fluctuations are included.
Of course, there are also lattice models that have a robust RFOT, such as the lattice glass model of~\cite{BM01}, however they are
more complicated to study using renormalization group methods. Hence, the problem of finding a simple enough model 
for which the RFOT is robust enough has not been solved for the moment~\cite{CBTT12}.

It is important to stress, however, that even if the lower critical dimension is bigger than $3$, still
the non-mean-field physics might only be visible close enough to the critical point (or more precisely, when the correlation length
is large enough). In that case, mean field theory would still provide a rather good description of experimental and numerical systems,
whose correlation lengths are never very large. The fact that many predictions of mean field theory compare well with numerical
simulations, as will be shown in Chap.~\ref{chap2}, seems to support this point of view.

\chapter{Glass and jamming transitions of hard spheres}
\label{chap2}

Here the application of the RFOT theory to the glass and jamming transition of hard spheres
will be presented. This line of research started in 2005 with the paper~\cite{PZ05}, where the 
small cage expansion within replica theory previously
developed in~\cite{MP99} was adapted to hard spheres. Technically, this was a non-trivial task because
the small cage expansion of~\cite{MP99} is based on a harmonic expansion and therefore fails badly for singular potentials.
After this extension was achieved, the work evolved in 
two main directions.
The first is the study of the glass transition of hard spheres in dimension larger than 3, motivated by the desire to achieve a better
understanding of how the mean field limit $d\to\io$ is approached.
The second is the study of the jamming transition that happens deep in the glass phase.
As discussed in Sec.~\ref{sec:defjamming}, the jamming transition is characterized by marked structural changes, that are clearly detected in simulation
and experiments, and therefore --besides being an interesting phenomenon in itself-- it provides a unique opportunity to test any theory of the glass state.

All the technical details of this work are contained in~\cite{PZ10,BJZ11}. Here only the main results will be discussed.
Note that, in the following, $v(x)$ is the harmonic sphere potential, using the same notations as in Sec.~\ref{sec:defjamming}.

\section{The glass transition of hard spheres in high dimension}
\label{sec:highd}

Hard spheres in the limit of large spatial dimensions provide an excellent opportunity for an analytic solution covering many aspects of the 
liquid and glass physics~\cite{FRW85,KW87,FP99,PS00}. The reason why this limit is solvable is geometric: consider three spheres A, B and C, with AB and BC in contact.
What are the chances that A and C will themselves also be in contact? In high dimensions, vanishingly small. 
This led to the realization~\cite{FRW85} that all terms in the virial expansion above the second could be neglected in high dimensions, as they involve 
geometrically heavily suppressed ``coincidences'', leaving one with only the two first terms of the series for the entropy 
 \beq\label{1SS}
\SS[\r( x)] = \int d^d x \, \r( x) [ 1 - \log \r( x) ] + \frac12 
\int d^d x d^d y \,  \r(x) \r( y) F(x -  y) \ .
\eeq
Here $F(x) = e^{-v(x)} - 1 = -\theta(\s - |x|)$ is the Mayer function for $\ee\to\io$, \ie for hard spheres, to which we restrict in this section.
For the liquid phase, one has $\r(x)=\r$, a constant, 
and the above expression gives the liquid entropy and coincides with the Van der Waals equation.
Non-uniform phases are described by solutions
of the stationarity equation $\partial \SS/\partial \r(x) = 0$ that are not constant. Indeed, Eq.~\eqref{1SS} provides the exact expression of the density
functional for hard spheres when $d\to\io$, which has basically the form of Eq.~\eqref{eq:DFT}.
Actually, at very high densities Eq.~(\ref{1SS}) would have corrections, 
however these corrections are exponentially small in the density region that will be relevant in the following~\cite{PS00}.
The liquid phase stays metastable at higher densities,
where one expects the thermodynamics to be dominated by a modulated, crystalline phase, which is however only known in small dimensions~\cite{ConwaySloane}.
Despite these simplifications, the problem remains much more difficult than the study of simple mean-field models such as the $p$-spin model.
The key difficulty is that, while for spin glass models the order parameter for the glass transition is the local Edwards-Anderson overlap,
for particle systems the caging order parameter is a non-trivial function of space, the non-ergodic parameter defined in Eq.~\eqref{nonergdyn},
or its Fourier transform, the long time limit of the van Hove function $G_\mathrm{s}(r,t)$~\cite{Go09}.

\subsection{Early theoretical results}

An early attempt to solve the amorphous DFT equations based on Eq.~\eqref{1SS} was performed in~\cite{KW87}. 
First, a Gaussian density profile as
in Eq.~\eqref{eq:DFTrhoG} was assumed.
Next, it was assumed that the equilibrium positions have the structure of a typical liquid configuration, hence
$S_0(q) = N^{-1} \sum_{jk} e^{i q (R_j - R_k)} \sim S(q) = 1/(1 - \rho F(q))$ where $F(q)$ is the Fourier transform of the Mayer function.
Finally, a small $\r$ development was made to obtain $S_0(q) \sim 1 + \rho F(q)$. Under these assumptions, a closed equation for the
cage radius $A$ was obtained. This equation has a finite solution when $\f > \f_d \sim 4.1 \, d \, 2^{-d}$, which defines the dynamical transition.
In the same paper~\cite{KW87}, the behavior of the MCT equations~\eqref{eq:MCT} in large $d$ was considered. Under a Gaussian assumption 
for the non-ergodic parameter, $f(q) \sim e^{-A q^2}$, it was shown that the MCT equations lead to the same equations for $A$ of DFT.
Hence, it was concluded that static and dynamical treatments lead to consistent results for $d\to\io$ like in the $p$-spin model, and consistently with the RFOT theory. 

After the replica theory of~\cite{PZ05} was developed, it was natural to apply it in larger dimension~\cite{PZ06a}.
It was realized that a replicated version of Eq.~\eqref{1SS}, in which the coordinate $x$ describing a single atom is replaced by a ``molecular'' coordinate
$\ol{x} = \{x_1\cdots x_m\}$, provides an exact description of the molecular liquid, for the same reasons why Eq.~\eqref{1SS} provides an exact description
for $m=1$. The geometrical argument carries over to the molecules because when $d\to\io$ the molecules become extremely compact, with a cage radius
$A \propto 1/d^2$.
Hence $\r(x)$ is replaced by $\r(\ol{x})$, with the advantage that the system of molecules is translationally invariant. A Gaussian
assumption for $\r(\ol{x})$, of the form
\beq
\r(\ol{x})=\r \int dX \prod_{a=1}^m \frac{e^{-\frac{1}{2A} (x_a-X)^2}}{ (2\pi A)^{d/2} } \ ,
\eeq
allows one to obtain an exact expression of the replicated entropy as a function of $A$. This procedure leads however to an equation for $A$ that is {\it different}
from the ones derived in~\cite{KW87} from DFT and MCT. In particular, it is found that $\f_d \sim 4.8 \, d \, 2^{-d}$.
A comparison of these equations seems to indicate that the RT equation contains more ``diagrams'' than the DFT and MCT ones. Indeed, an expansion
of the RT equation in powers of $F(q)$ gives back the DFT--MCT equation of~\cite{KW87} at the lowest order.

A few years later, two groups independently found an asymptotically exact solution of the MCT equations~\eqref{eq:MCT} when $d\to\io$~\cite{SS10,IM10}.
The main features of this solution are that the $f(q)$ is markedly non-Gaussian and it remains so when $d\to\io$,
and the dynamical transition is found to scale as $\f_d \propto d^2 \, 2^{-d}$, which is much larger than the previous results obtained from the Gaussian
assumption. 
An inconsistency was pointed out in~\cite{IM10}, namely that MCT gives a negative Van Hove function $G(r,t)$ when $d$ is large enough. 

At this point, one is faced with an unexpected puzzle: why in the limit $d\to\io$, where mean field theory should be exact, the three theories that 
are thought to be different facets of the unified RFOT theory give such different predictions?
Looking carefully at the three theories, it is important to keep in mind that:
\begin{itemize}
\item The DFT result is based --besides the Gaussian assumption-- on a strong assumption, that $S_0(q)\sim S(q)$. This is far from being obvious.
In fact, when the pressure of the glass goes to infinity and $A\to 0$, the particles are stuck in their equilibrium positions $R_j$. But these have then to
be strongly correlated, because of the requirement of mechanical stability of the packing. This will be further discussed in Sec.~\ref{sec:finited}.
In any case it is possible that the assumption $S_0(q) \sim S(q)$ is too strong.
\item MCT has not yet been derived from a systematic large $d$ expansion, therefore it is hard to tell whether the corrections to Eq.~\eqref{eq:MCT}
become small when $d\to\io$. One could add terms to the MCT equations that preserve the ``universal'' MCT phenomenology (\ie the power-law scalings
and the associated critical exponents), 
but would change the ``non-universal'' predictions for the form of $f(q)$ and the value of $\f_d$~\cite{ABB09}.
At present, there is no reason to believe that the MCT equations~\eqref{eq:MCT} should be exact when $d\to\io$.
\item The RT results are obtained under one single assumption, that of the Gaussian form of the molecule distribution. They could therefore only fail if the Gaussian
assumption were not verified.
\end{itemize}

\subsection{Numerical simulations}
\label{sec:nonGaussian}

In absence of any other theoretical indication of a way to resolve this inconsistency, one needs some numerical results to identify the correct path to follow.
This was the main motivation to start a series of simulations trying to reach the largest possible dimensions in order to obtain indications on the scaling
of the transition density and the shape of the non-ergodic parameter~\cite{CIPZ11,CIPZ12}.
Previous investigations had shown that crystallization is strongly suppressed when dimension is increased above $3$ and is basically impossible to
observe on simulation time scales for any $d\geq 4$~\cite{SDST06,VCFC09}. Hence, for these studies one can safely use monodisperse systems without having
to worry about crystallization. 
Moreover, there was indication that the MCT predictions are very accurate in $d=4$~\cite{CIMM10}.

\subsubsection{Compression at constant rate~\cite{CIPZ11}}

In a first set of simulations~\cite{CIPZ11}, we performed event-driven molecular dynamics simulations~\cite{Krauth} of $N$ identical hard spheres of diameter $\sigma$,
enclosed in a periodic box of volume $V$.
A modified Lubachevsky-Stillinger
algorithm was employed, which starts from a low-density gas (small $\f$) and densify it by growing uniformly the particles
at a constant rate $\dot{\sigma}$, reported here in standard reduced units such that the particle mass is $m=1$, the temperature is $k_B T=1$, and $\sigma=1$.
Time evolution stops when the system reaches a high reduced pressure $p\equiv\beta P/\rho=10^3$. 
Systems with $N=8000$ are simulated for $d\leq9$ and larger ones for $d=10$--$13$.
These sizes ensure that even when the system is in its densest state the box edge remains larger than $2\sigma$, 
which prevents a particle from ever having two direct contacts with another one. 
There are strong reasons to believe that although relatively small these sizes nonetheless provide a
reliable approximation of bulk behavior. First, with increasing $d$ the box edge becomes less representative of the
overall box size. The largest diagonals are $\sqrt{d}$ larger and there are many more diagonals than edges.
Second, by analogy to spin systems, mean-field arguments indicate that for $d\to\infty$, a hypercube of side two is
sufficient to capture the full thermodynamic behavior.
Even at the critical point, finite-size corrections are proportional
to $1/N^{\delta}$, where
the exponent $\delta$ is model dependent (e.g. 1/2 at the ferromagnetic transition), 
and do not directly involve the edge length $L$~\cite{ParisiBook}. 
Similar results hold for dimensions greater than the upper critical dimension, where the exponents coincide with the mean field ones.
Third, the fluid structure is expected to become uniform at ever smaller distances with increasing
$d$~\cite{FP99,PS00,SDST06}. Nearest-neighbor ordering should thus mainly be influenced by particles in contact or nearly so,
 with the rest of the fluid acting as a continuum.
Indeed, in the fluid phase, finite volume corrections are proportional to the pair correlation $h(L)$,
and at fixed $L$, $h(L)$ goes to zero exponentially with $d$~\cite{PS00}. The validity of these rationalizations
has been satisfactorily tested by simulations in $d=8$, where no finite size effect has been detected~\cite{CIPZ11}.

\begin{figure}[t]
\begin{center}
\includegraphics[width=.45\columnwidth]{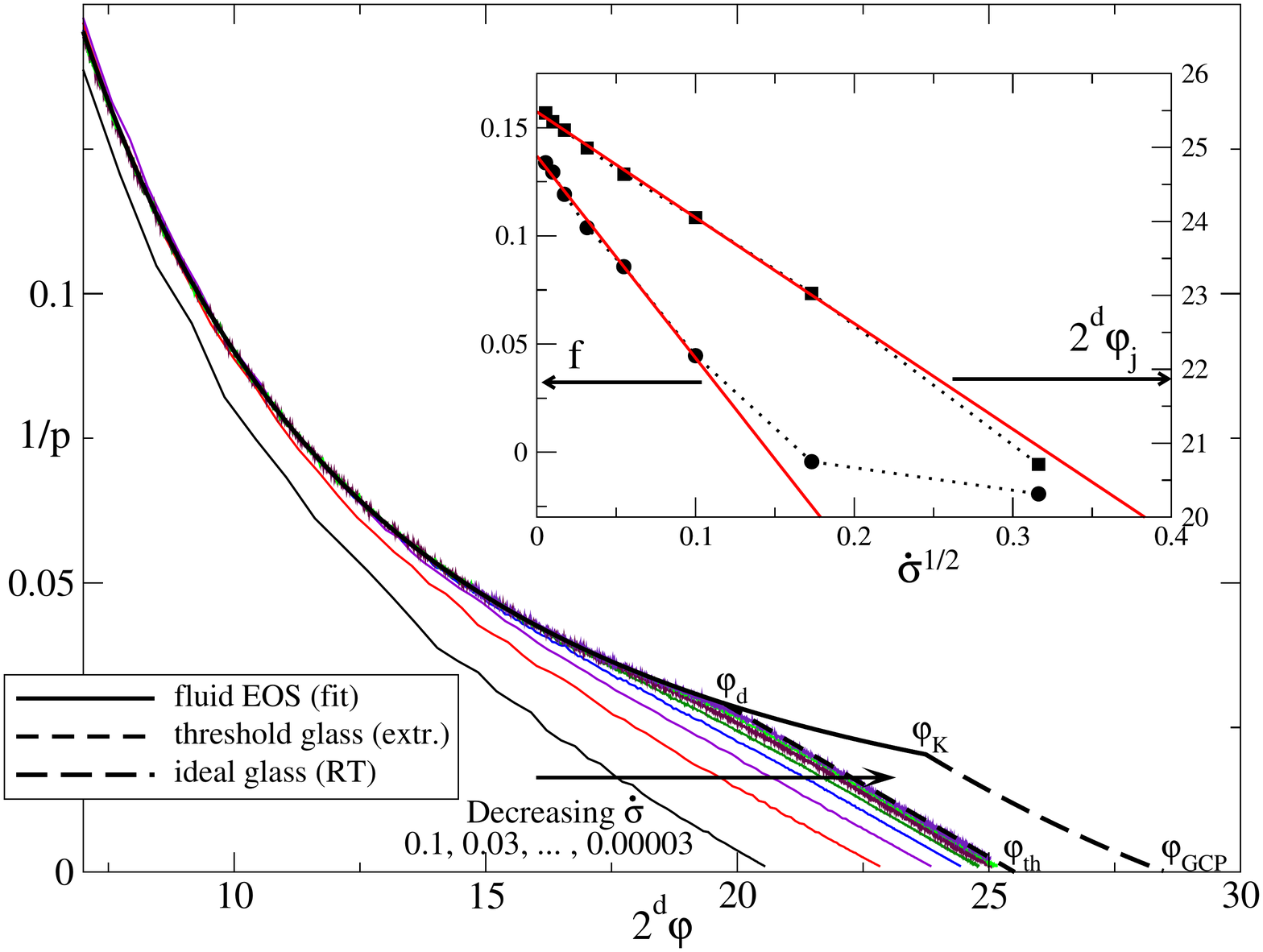}
\includegraphics[width=.49\columnwidth]{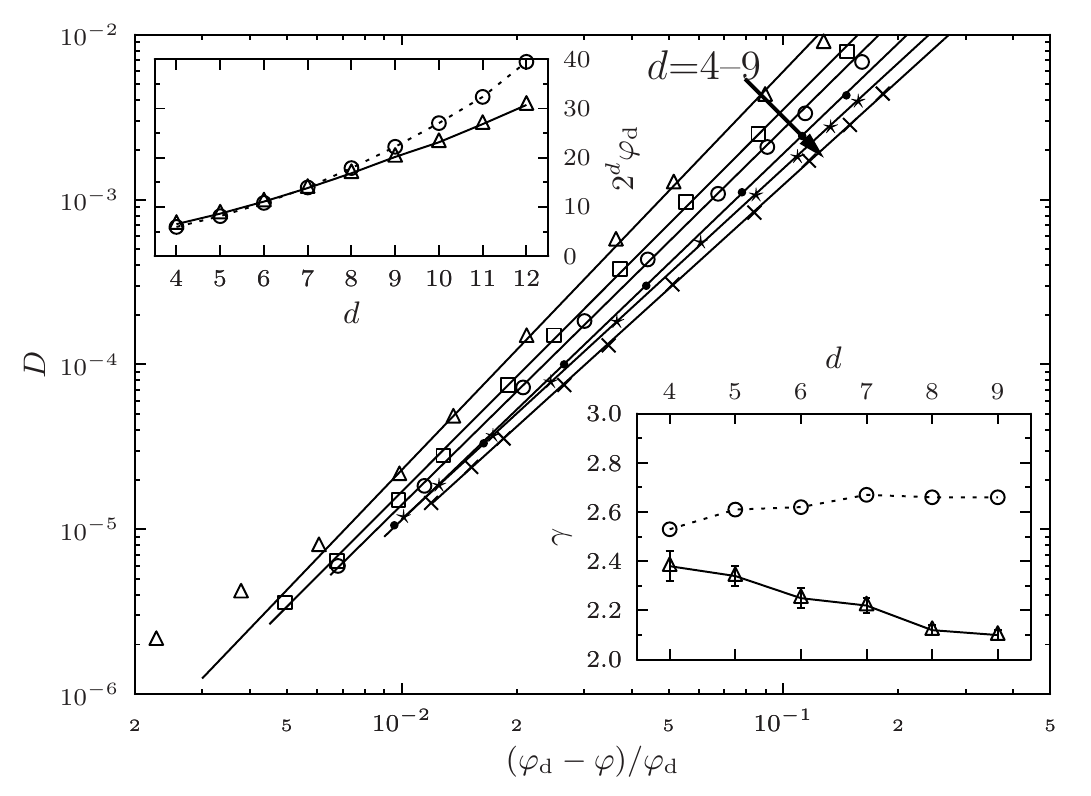}
\caption{
(Left, from~\cite{CIPZ11})
Different compactions of $N=8000$ particles in $d=9$.
With growing $\f$, the pressure first evolves like the fluid EOS then like a free volume EOS. 
In the inset, the fitted glass EOS parameters are reported as functions of $\dot{\s}$ and the
power-law extrapolation to $\dot{\s}\to 0$ is shown. The resulting EOS for the threshold glass
is reported in the main panel. It crosses the liquid EOS at $\f_d$, and its pressure diverges at
$\f_{th}$. The prediction of replica theory for the ideal glass EOS is also reported for comparison:
it leaves the liquid EOS at the Kauzmann transition $\f_K$, and its pressure diverges at the glass
close packing point $\f_{GCP}$, see Sec.~\ref{sec:finited}. \newline
(Right, from~\cite{CIPZ12})
Power-law fits (lines) to the vanishing diffusivity improve with $d$=4--9 (different symbols), spanning over three decades of $D$. 
In the top inset, the resulting numerical $\f_{d}$ values, which coincide with the compression results from~\cite{CIPZ11},
are reported as triangles.
The MCT prediction for $\f_d$ is reported with circles for comparison.
In the bottom inset, the value of the exponent $\gamma$ from numerics (triangles) and from MCT (circles) is reported.
}
\label{fig:comp}
\end{center}
\end{figure}

The compression results for $d=9$ shown in Fig.~\ref{fig:comp}
are representative of the behavior observed for all $d>3$.
The system first follows the equilibrium fluid equation of state (EOS) at low density and falls out of
equilibrium at high density. Beyond this point, the pressure increases faster than in the equilibrium fluid and ultimately
diverges at packing fraction $\f_{\rm j}(\dot{\s})$.
A Carnahan-Starling form
$p_{\rm fluid}(\f)=1+ 2^{d-1} \f \, \frac{1-A_{d} \, \f}{(1-\f)^d}$
captures well the pressure growth with $\f$
in the fluid regime (Fig.~\ref{fig:comp}), provided, for each $d$,
that one fits $A_{d}$ to the data from the slowest compression
rate available.
Note that the
contribution of the $A_d$ term is found to decrease with increasing $d$, consistently with the asymptotic result of
Eq.~\eqref{1SS}.
In the high-density non-equilibrium regime, compaction runs with different $\dot{\s}$ follow separate branches
along which the pressure evolution is dominated by the expulsion of free volume~\cite{DTS05}.
Upon approaching jamming, the pressure is well approximated by the free volume form in which $1/p$ is linear in $\f$:
$p_{\rm fv}(\dot{\s},\f)=\frac{d \, \f_{\mathrm{j}}(\dot{\s}) [1-f(\dot{\s})]}{\f_\mathrm{j}(\dot{\s})-\f}$.
Both $f(\dot{\s})$ and $\f_\mathrm{j}(\dot{\s})$ are extracted from fitting the simulation data for
$p\geq p_\mathrm{min} \sim 50$.

These numerical results qualitatively agree with the RFOT scenario~\cite{PZ10}.
According to the theory, 
the compression curves for moderately small $\dot{\s}$ should converge as a power law $\dot{\s}^\alpha$ to a ``threshold'' glass.
The dynamical transition density $\f_{d}$
separates the equilibrium fluid from this glass.
Equilibrating the liquid at $\f > \f_d$ requires
much slower compaction rates,
 $\dot{\s} \lesssim\exp(-d)$,
such that activated events allow the system to remain in
equilibrium. In this regime one would expect a crossover to a logarithmic dependence of the glass EOS on $\dot{\sigma}$, e.g., 
of the form $1/|\log(\dot{\sigma})|$, but it turns out that this regime is out of reach of present time numerical simulations.
Indeed, the inset of left panel of Fig.~\ref{fig:comp} shows that 
the reported $f(\dot{\sigma})$ and $\f_\mathrm{j}(\dot{\sigma})$ are both linear as
functions of
$\dot{\sigma}^{\alpha}$ with $\alpha\approx0.5$.
The exponent is expected to weakly depend on dimensionality, especially at low $d$, but this value is within the
numerically reasonable range for all systems studied. Because the subsequent analysis is rather insensitive to the
precise value of $\alpha$, for simplicity it is kept constant. Extrapolating the results to $\dot{\sigma}=0$ gives
the parameters $\f_{\rm th}$ and $f_{\rm th}$ for the threshold glass free volume EOS, also reported in Fig.~\ref{fig:comp}.
Its intersection with the fluid EOS then provides a numerical estimate for $\f_{\rm d}$.
These values are very close to previous numerical estimates in
$d=3$~\cite{BEPPSBC08} and $d=4$~\cite{CIMM10},
which demonstrates the coherence of this analysis in low dimensions.
The results for $\f_d$ reported in Fig.~\ref{fig:comp}
show that its dimensional evolution is smooth.
Interestingly, dimensions where the crystal structure is singularly dense,
such as $d=8$ and $d=12$~\cite{ConwaySloane}, do not present any echo of that singularity, 
which illustrates
the smooth $d$ dependence of the fluid structure and its independence from the crystal.

\subsubsection{Equilibrium dynamics in the liquid phase~\cite{CIPZ12}}

Having obtained an estimate of $\f_d$ as a function of dimension, one can proceed to investigate the equilibrium dynamics
in the liquid phase on approaching $\f_d$. This has been done with the same algorithm as before, but stopping the compression
at the desired $\f<\f_d$, letting the system equilibrate, and then performing equilibrium measurements of dynamical quantities.
In particular, the average mean-square displacement (MSD) defined in Eq.~\eqref{eq:MSDdef}
was obtained. 
At times shorter than the collision time, MSD displays a ballistic regime $\langle \Delta r^2(t)\rangle=d \, t^2$, 
and at long times it has a diffusive regime $\langle \Delta r^2(t)\rangle=2d \, D \,t$.
Fitting these numerically determined $D$ to the MCT
power-law form $D\sim(\f-\f_{d})^{\gamma}$
is reasonably good for $D<0.005$, and improves with increasing $d$. The resulting values of $\gamma$ and $\f_{d}$ are reported in Fig.~\ref{fig:comp}.
For $d\geq5$, 
the full accessible dynamical range studied is used, spanning up to three $D$ decades. 
The values of $\f_d$ obtained in this way up to $d=9$ are consistent with the ones obtained from the constant compression data.
The values of $\g$ are reported in Fig.~\ref{fig:comp}.

\begin{figure}[t]
\begin{center}
\includegraphics[width=.7\columnwidth]{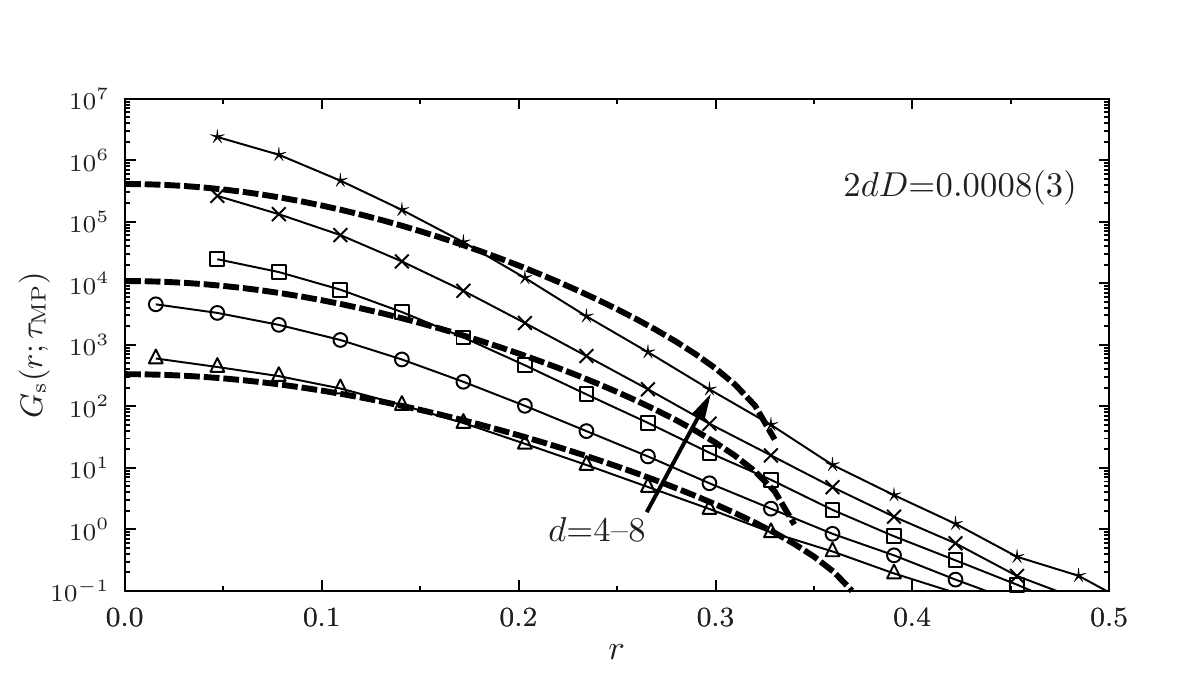}
\caption{
(From~\cite{CIPZ12})
The self Van Hove function, computed at the mid-plateau point $\t_{\rm MP}$ and at a density such that $2 d D = 0.0008$, hence close to $\f_d$,
for several dimensions $d=4, 5, 6, 7, 8$. Dashed lines are the MCT predictions for $d=4, 6, 8$. A fat non-Gaussian tail is clearly visible and is
not reduced upon increasing dimension.
}
\label{fig:cage}
\end{center}
\end{figure}

The cage form is described by the self part of the van Hove function 
\begin{equation}
G_\mathrm{s}(r;t)=\frac{1}{N}\sum_{i=1}^N\langle\delta(x_i(t)-x_i(0)-r)\rangle \ ,
\end{equation}
which in the ballistic and diffusive regimes is well-approximated by a pure Gaussian.  The logarithmic mid-point of the caging plateau 
$\tau_{\mathrm{MP}}$ is chosen at the mid-time on a logarithmic scale, intermediate between the ballistic and the diffusive 
extrapolations of the MSD, and provides an estimate of the {\it plateau} regime discussed in Fig.~\ref{fig1:Fs} (see~\cite{CIPZ12} for details). 
The results close to $\f_d$, reported in Fig.~\ref{fig:cage}, clearly shows that the cage is not Gaussian, with a fat tail at large $r$ that does
not shrink upon increasing dimension. By contrast, a comparison with the D5 crystal in $d=5$ and the E8 crystal in $d=8$ (unpublished) 
shows that the cage is there very well described by a Gaussian form.

\subsubsection{Discussion}

In summary, numerical results indicate that:
\begin{itemize}
\item
The MCT qualitative predictions are very well obeyed by the data, as previously observed in $d=4$~\cite{CIMM10}.
The diffusion constant vanishes as $(\f-\f_d)^\g$. The vanishing of the diffusion is accompanied by a growing caging plateau.
It remains to be checked whether
the approach and departure from the plateau define power-law regimes in time with exponents $a$ and $b$, and whether the relation
between $a, b, \g$ is consistent with MCT predictions.
\item
Dynamical heterogeneity is reduced with increasing dimension~\cite{CIPZ12}. Together with the previous observation,
this suggests that indeed when $d$ increases the system approaches a mean field regime well described by the RFOT scenario.
\item
Yet, the quantitative MCT predictions for $\f_d$, $\g$ and the cage radius become increasingly poor with increasing $d$.
This is not a problem for the RFOT scenario: it is reasonable to expect that 
a different theory with a similar structure to MCT but a different kernel
exist. Such a theory would provide the same qualitative predictions as MCT~\cite{ABB09}, and exact quantitative results
for $\f_d$, $\g$ and the cage structure. But this theory has yet to be found.
\item
The Gaussian assumption for the cage form done by RT and DFT is invalidated by the data, calling for a revision of these theories.
Unfortunately, the predictions for $\f_d$ of RT in low $d$ are not reliable because of the inaccuracy of the small cage expansion
(the replicated HNC calculation has not been performed yet).
However, the result of RT for $\f_K$ should provide an upper bound for $\f_d$, and this is consistent with the data.
Also the predictions of RT for the ideal glass EOS (Fig.~\ref{fig:comp}) are consistent with the data, as the ideal glass EOS should be
similar to the threshold glass EOS but shifted in density.
\end{itemize}

\subsection{New theoretical results}

The numerical results obtained in~\cite{CIPZ11,CIPZ12} give enough suggestions on how to proceed to improve the theory.

On the side of replica theory, a first step has been obtained in~\cite{KPZ12}, where the exact description of the molecular liquid
in the high-dimensional limit, without further assumptions, was found. 
The strategy is the following: a generic replica problem is written in terms of the order parameter $\r(\bar x)$. 
The $x_a$ are vectors in $d$-dimensional space, and one is looking for a solution that is statistically translationally invariant and isotropic.
 The only possibility with these properties is that $\r(\bar x)$ depends exclusively on the $|x_a-x_b|^2 = q_{aa} + q_{bb} -2 q_{ab} $, where $q_{ab}=x_a \cdot x_b$.
 All $d$-dimensional integrals may be expressed as {\em low dimensional} integrals in terms of the $q_{ab}$, with volume factors,
  a simple generalization of spherical coordinates.
Then, the dimension $d$ appears explicitly, and the limit of large $d$ may be taken in a straightforward way, by saddle point evaluation of the integrals.
The result is that 
 the Gaussian ansatz turns out to be, in a sense, exact: it gives for large dimensions the exact values for the thermodynamic quantities. 
 The reason is that in the large $d$ limit all the saddle points are dominated by a typical distance between atoms in a molecule, and therefore
 the exact shape of the cage function is irrelevant. Any reasonable ansatz would give the same results. This explains why the Gaussian results for the
 glass EOS, as well as the transition densities, are consistent with the numerical data, as shown in Fig.~\ref{fig:comp}.
 The next step in this direction is to work out the prediction of RT for the cage shape and compare it with the results in Fig.~\ref{fig:cage}.
 This work is currently in progress.

On the side of dynamics, it might be possible to obtain an exact solution 
starting from the approach of~\cite{MK11} and following the same lines of the replica calculation.
Also this work is currently in progress.

In summary, although the physics in large $d$ seems quite well understood, its mathematical description is surprisingly complex.
Yet a lot of progress has been made and it seems that we are on the right path towards a complete RFOT theory in this limit.

\section{Mean field theory of the jamming transition}
\label{sec:finited}

The main features of the jamming transition have been introduced in Sec.~\ref{sec:defjamming}. Consider the simple model of harmonic spheres
described by Eq.~\eqref{eq1:pot}. The two control parameters are the packing fraction $\f$ and the ratio of thermal energy to potential energy $T/\ee$
(we always use $k_B=1$).
As discussed in Sec.~\ref{sec:defjamming}, the jamming transition happens out of equilibrium at a packing fraction $\f_j$, 
when the system is inside its glassy phase.
At this transition, the origin of the rigidity of the glass changes in nature~\cite{IBS12}: 
below $\f_j$, particles do not overlap and rigidity is {\it entropic} in nature,
in such a way that pressure and elastic moduli are proportional to $T$; above $\f_j$, particles overlap and rigidity is {\it mechanical} in nature,
hence pressure and elastic moduli have a finite limit at $T\to 0$.
Hard spheres have $\ee\to\io$, hence $T/\ee \to 0$, and they lie on the horizontal axis in the phase diagram of Fig.~\ref{fig:PDsch}, but they
exist only for $\f < \f_j$. At $\f_j$, the reduced pressure of hard spheres, defined by $p = P / (\r T)$, diverges, see Fig.~\ref{fig:comp}. 
This can be understood in two equivalent
ways. From the hard sphere perspective, $\f_j$ is the point where the spheres get in contact with their neighbors. The cage radius goes to zero, and
collisions with the neighbors become infinitely frequent. If $T$ is fixed, the momentum transfer per unit time between neighbors, which is proportional to $p$, 
diverges. From the soft sphere perspective (\ie at finite $\ee$), the pressure becomes non-zero at $\f_j$ even at $T=0$, hence $p \propto P/T = \io$ in the whole
region above $\f_j$. In other words, $\f_j$ is identified with the {\it close packing} point~\cite{BM60,BM62}, the maximal density at which the spheres can be packed avoiding overlaps.
This has been called {\it random close packing}, $\f_{RCP}$, in the literature. In present notations, $\f_j$ and $\f_{RCP}$ coincide.

Independently from these motivations originating in the physics of amorphous materials, the problem of packing spheres in $d$ dimensions
is intimately related to several important mathematical problems, notably in the
context of signal digitalization and of error correcting codes, especially when $d$ is large~\cite{ConwaySloane}.
Although the densest packings are known (either rigorously or at the level of good conjectures) in low dimensions,
as soon as $d$ becomes large, almost no results are available, see~\cite{ConwaySloane} and the very nice review paper~\cite{Co10}.
Providing a ``physical'' estimate of the closest packing density in large dimensions would be very interesting for the mathematical community.
Attempts in this direction have been performed, and even if intuition suggests that lattice structures should fill space better than amorphous ones,
some have suggested that for large $d$ disordered packings may be at least as
efficient as lattice-based versions~\cite{TS06}.
Although the density of lattice geometries is strongly $d$ dependent~\cite{ConwaySloane}, the fluid structure and its
properties are much less sensitive to $d$, once $d \geq 3$~\cite{VCFC09,CIPZ11,CIPZ12}.
A general understanding of the jamming transition in arbitrary $d$ would thus also clarify
the density scaling of amorphous packings and their potential mathematical role.

In summary, the jamming transition is interesting for wildly different problems, ranging from the physics of foams, pastes, emulsions, powders and other
granulars, to the physics of thermal glasses in the limit of low temperature, and to mathematics and information theory~\cite{PZ10,TS10}. 
Despite all this interest, and despite
the fact that the random close packing concept has been proposed long ago~\cite{BM60,BM62}, a first principle theory of jamming has
not been developed. ``First principle" here means a theory that is able to {\it predict} the existence of the transition, its location in density, and the associated
structural changes, from the sole knowledge of the particles' interaction potential.
A first step in this direction was obtained --unexpectedly-- as a side product of the replica theory of hard sphere glasses in~\cite{PZ05}. Subsequent improvements
have been made in~\cite{PZ10,BJZ11}. This theory will be reviewed in the rest of this section, highlighting its successes and the many problems that are still
unresolved.

\subsection{Is random close packing well defined?}
\label{sec:RCP}

\subsubsection{Predictions of replica theory}

How does the jamming transition appear in the replica theory? 
Recall from the discussion of Sec.~\ref{sec:MFTqual} that one has to introduce $m$ replicas as a tool to take into account the existence
of multiple amorphous glassy states. $m$ is a free parameter and in the end one has to optimize the free energy with respect to it,
the optimal value for a given state point $(\f,T)$ being called $m^*(\f,T)$. It is expected that $m^*=1$ in the liquid phase, and
$m^*$ continuously becomes smaller than 1 at the glass transition~\cite{MP99}.
What came as a surprise in~\cite{PZ05} is that, for the hard sphere system, $m^*(\f,T=0)$ continuously decreases starting from 1
at $\f=\f_K$, and eventually vanishes linearly at some ``(ideal) glass close packing'' (GCP) density $\f_{GCP}$, $m^*(\f_{GCP},T=0) \sim | \f - \f_{GCP} |$.
This is crucial, because within replica theory the reduced pressure $p$ of the (ideal) glass is proportional to $1/m^*$, hence the vanishing of
$m^*$ drives a divergence of $p$ and signals the jamming transition! 
Note that this result was not put by hand in the theory: the calculation of the replicated free energy can be done starting from the interaction potential, 
as a function of $(\f,T)$ and $m$, and it is the process of optimization with respect to $m$ and the vanishing of $m^*$ that is behind the appearance
of the transition. A posteriori, the fact that the jamming transition is related to the vanishing of $m$ is obvious thanks to the connection with
combinatorial optimization problems where a similar transition happens 
(and is called the SAT--UNSAT transition in that context~\cite{MM09}). This connection was explored in details in~\cite{MPR05,KK07}. Note also that the lattice glass
model of~\cite{BM01,RBMM04} is a perfect example of this connection.

However, what is the relation between $\f_{GCP}$ and the random close packing of Bernal~\cite{BM60,BM62}? The GCP is the point where the pressure
of the ideal glass diverges. However, as discussed in Sec.~\ref{sec:viscosity}, even if one accepts the mean field prediction that such ideal glass really exists,
producing it in a reasonable amount of time is impossible both in numerical simulations and experiments. The liquid will always fall out of equilibrium much before $\f_K$ is reached, and 
a non-equilibrium glass of lower density will be obtained. This is well illustrated in Fig.~\ref{fig:comp}. Hence, in real-life experiments and numerical simulation,
one will be stuck in out of equilibrium states, and in general the random close packing point obtained in a given experiment will be at lower density,
$\f_j = \f_{RCP} < \f_{GCP}$, corresponding to a system falling out of equilibrium at some $\f_g < \f_K$.

It might seem that the replica description of the GCP is therefore useless. Fortunately, this is not the case. Indeed, as discussed in Sec.~\ref{sec:MFTqual},
RFOT theory predicts the existence of (exponentially) many metastable states, each corresponding to a different glass, with different free energies.
And replica theory allows one to compute their properties. Performing the computation~\cite{PZ10}, one discovers that each different glass 
will jam at a {\it different} $\f_j$. Therefore, replica theory {\it predicts} the existence of a {\it range} of densities $\f_j \in [\f_{j}^{\rm min}, \f_{GCP}]$ in which
amorphous jammed packings exist. Each of them corresponds to jamming of a different non-equilibrium glass.
Therefore, replica theory predicts that the jamming (or RCP) point is not well defined: it depends on the protocol that one uses in the experiment.
Actually, the whole phase diagram in Fig.~\ref{fig:PDsch} (\ie the transition line $T_g(\f)$ and the jamming point $\f_j$) depends on the particular metastable state (as parametrized
for instance by the equilibrium positions $R_j$) that is selected in the experiment under consideration.
So what is left of the predictive power of replica theory, besides setting upper and lower bounds for $\f_j$?
Here comes another crucial observation: 
replica theory predicts that {\it all amorphous packings that correspond to jamming of a metastable glass
have equivalent structural properties}~\cite{PZ10}. This means that even if the jamming density $\f_j$ is protocol-dependent,
the structural signatures of jamming are universal.

This prediction of replica theory should not be confused with the idea that RCP is not well defined because one can construct
packings with arbitrary degree of crystallization~\cite{TTD00,JM10}. Although this is certainly true, the presence of such microcrystallites
is detectable using structural measures (so-called order metrics~\cite{TTD00}). The prediction of RT is that there is an interval of perfectly amorphous packings.
Of course those can be mixed with crystalline regions to produce any intermediate density up to the crystal close packing.

\subsubsection{Numerical tests}

\begin{figure}
\begin{center}
\includegraphics[width=0.5\textwidth]{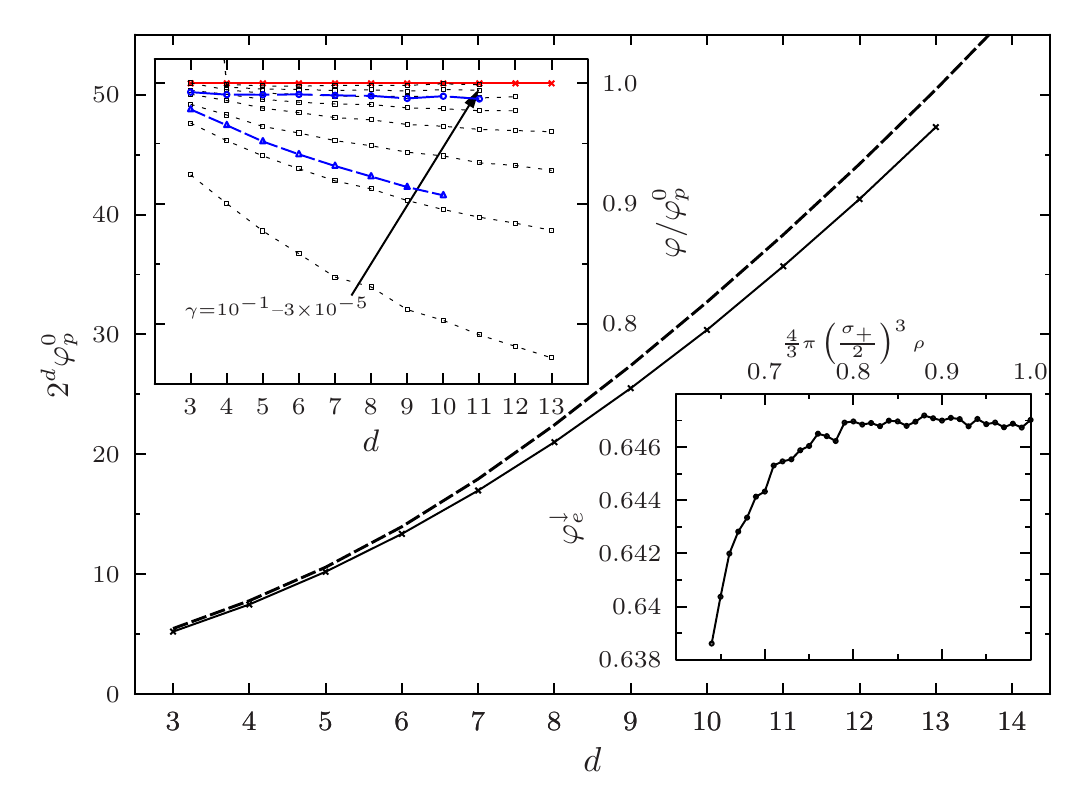}
\caption{
(From~\cite{CCPZ12})
Main panel:
the extrapolated jamming density $\varphi_p^{\dot\sigma\rightarrow 0}$ following the protocol described in Fig.~\ref{fig:comp} (full line and crosses), and
compared with the prediction of RT for $\f_{GCP}$ (dashed line).
Top inset: the range of jamming densities $\f_p^{\dot\s}$ (squares) 
is compared to $\f_e^{\mathrm{max}}$ (circles) and $\f_e^{\mathrm{min}}$ (triangles). 
Note that $\f_e^{\mathrm{max}} \sim \f_p^{\dot\s = 3\times10^{-4}}$ and $\f_e^{\mathrm{min}} \sim \f_p^{\dot\s=3\times10^{-2}}$. 
Bottom inset: the increase of $\varphi_e^\downarrow$ with $\sigma_+$, in terms of the initial effective packing fraction, in $d=3$. 
}\label{fig:phiJ}
\end{center}
\end{figure}

This prediction has been successfully tested in a number of independent studies~\cite{BW09,CBS09,HD09}.
However, these studies were limited to $d=3$, hence subject to the criticism that crystallization might play a hidden 
role in the density variation~\cite{TTD00}. Moreover, the density range of amorphous packings in $d=3$ is quite small~\cite{CBS09}.
Hence, more recently~\cite{CCPZ12}, a systematic study as a function of dimension was performed,
as part of the previously mentioned effort towards a systematic understanding of the RFOT theory in large $d$~\cite{CIPZ11,CIPZ12}.
Amorphous packings were prepared using two different numerical protocols.
\begin{itemize}
\item[\emph{(i)}] Approaching jamming from densities below it by Lubachevsky-Stillinger (LS) compressions of hard spheres undergoing Newtonian dynamics while $\sigma$ 
grows at a fixed rate $\dot{\sigma}$, as previously discussed.
The compression stops when particles are in contact, 
defining the packing fraction $\f_p^{\dot{\s}}$, at which the reduced hard spheres pressure becomes infinite. 
\item[\emph{(ii)}] Approaching jamming from densities above or below it by minimizing the energy $E(X) = \sum_{i < j} v(|x_i - x_j|)$ of a 
soft sphere system with interaction potential $v(r)$ in Eq.~\eqref{eq1:pot}~\cite{OLLN02}. 
One starts from random sphere configurations $X_+ = X_-$ and uses $\s_-$ and $\sigma_+$ that bracket the jamming point,
in such a way that $E(X_-)=0$ at $\s_-$, while $E(X_+)>0$ at $\sigma_+$.
The onset of non-zero energy is found using an iterative bisection method.
At each iteration, an intermediate value $\s_{\rm m}$ is chosen, and
the energy at $\s_{\rm m}$ is minimized starting from either $X_+$ or $X_-$. This choice defines the procedure
to approach jamming
from \emph{above} or from \emph{below}, respectively.
Calling $X_{\rm m}$ the configuration obtained after minimization, one substitutes $\s_+ = \s_{\rm m}$ and $X_+ = X_{\rm m}$ if 
$E(X_{\rm m})>0$, while $\s_- = \s_{\rm m}$ and $X_- = X_{\rm m}$ if 
$E(X_{\rm m})=0$. 
The procedure stops when $E(X_+)$ is below a given threshold (of the order of numerical accuracy), and we call the final density
$\varphi_e^{\downarrow}$ in the procedure from above, and $\varphi_e^{\uparrow}$ in the procedure from below.
From above, the energy vanishes
with $e=E/N \sim \Delta\f^2$ and the static pressure $P\sim \Delta\f$, where
$\D\f$ is the distance from jamming~\cite{OLLN02}
\end{itemize}

For protocol \emph{(ii)}, both $\sigma_\pm$ have no measurable effect on $\varphi_e^{\uparrow}$.
One can formally define $\f_{e}^{\rm min}=\min_{\s_\pm} \f^\uparrow_e(\s_\pm)$, but any reasonable $\s_\pm$ results in the same final density.
By contrast, $\varphi_e^{\downarrow}$  strongly depends on $\s_+$ (Fig.~\ref{fig:phiJ}), but is also independent of $\s_-$. 
One can therefore define $\f_{e}^{\rm max} = \max_{\s_\pm} \f^\downarrow_e(\s_\pm)$. 
A practical way of constructing both $\f_{e}^{\rm min}$ and $\f_{e}^{\rm max}$ is to run the energy minimization (respectively from below and from above) 
starting from $\s_-=0$ and a $\s_+$ large enough that $\varphi_e^{\downarrow}$ saturates to its maximum;
intermediate packing fractions can then be obtained by reducing $\sigma_+$ (Fig.~\ref{fig:phiJ}).
It is found that
$[\varphi_e^{\rm min},\varphi_e^{\rm max}]$ roughly corresponds to $[\varphi_{p}^{\dot\s_-},\varphi_p^{\dot\s_+}]$ 
with $\dot\s_-\approx3\times10^{-2}$ and $\dot\s_+ \approx 3\times10^{-4}$, and that large $\dot\s$ generate mechanically unstable packings.
In other words, although the jamming density sensitively depends on the protocol's specific parameters, the jamming density range is much less sensitive to the protocol's class. 
The resulting density range is found to grow steadily from about $2\%$ 
in $d=3$ to nearly $10\%$ in $d=11$. 
Therefore, the similar observations made for $d=3$ binary mixtures~\cite{BW09,CBS09,HD09} are confirmed, but in larger dimension the effect is much larger and leaves
no room for ambiguity, also because crystallization is totally absent as soon as $d>3$~\cite{VCFC09}.

The value of $\f_e^{\rm min} \sim \f_p^{\dot\s\to 0}$ are both lower than $\f_{GCP}$ predicted by replica theory (Fig.~\ref{fig:phiJ}).
This is expected because, according to the analysis of the glass problem (Sec.~\ref{sec:viscosity}) and of analogous combinatorial optimization problems~\cite{KK07,MPR05},
achieving a packing density of $\f_{GCP}$ requires at least to equilibrate the liquid close to $\f_K$, which can only happen through activated processes and can be expected to require
a time increasing exponentially with $d$ (or with $N$ in mean field models), hence being out of reach of numerical simulations.
Unfortunately, a reliable prediction from RT for the minimal value of the jamming density cannot be obtained, for similar reasons as for $\f_d$.

\subsection{Finite temperature scaling relations}

As shown in Fig.~\ref{fig:PDsch},
the jamming transition exists only at strictly zero temperature (here we consider soft spheres with a finite $\ee$). 
In fact, as soon as temperature is non-zero, overlaps between particles are allowed
by thermal fluctuations, and a strict distinction between zero energy non-overlapping configurations and positive energy overlapping configurations
is not possible. However, if temperature is low enough, the proximity of the jamming transition induces scaling relations for all thermodynamic quantities.
Here these scaling relations are discussed. For this discussion, one can restrict to a given non-equilibrium glass with given jamming density $\f_j$
(for instance, the ideal glass with jamming density $\f_{GCP}$) because the results are independent of the choice of glass state and the corresponding $\f_j$. 

\begin{figure}
\begin{center}
\includegraphics[width=.49\textwidth]{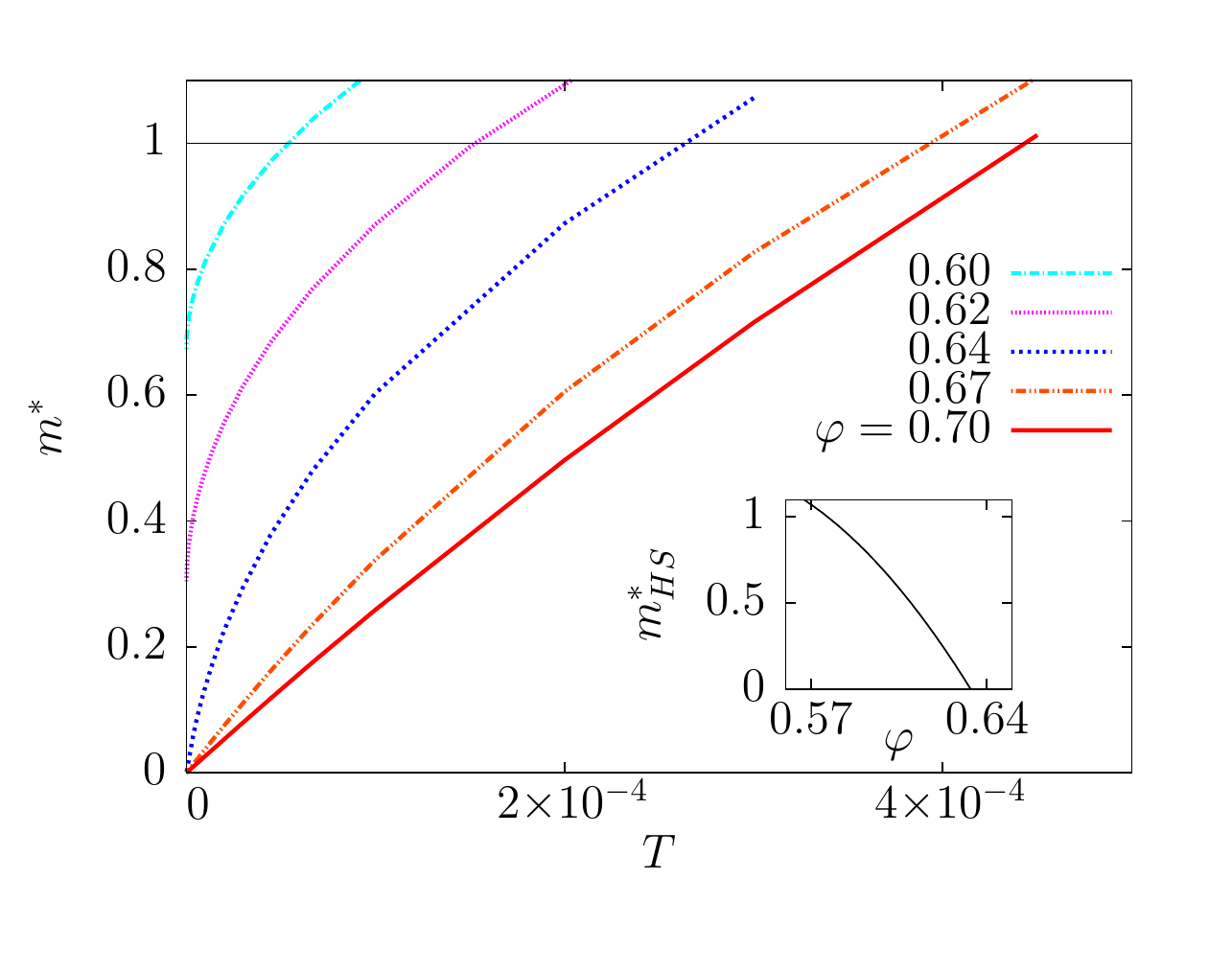}
\includegraphics[width=.49\textwidth]{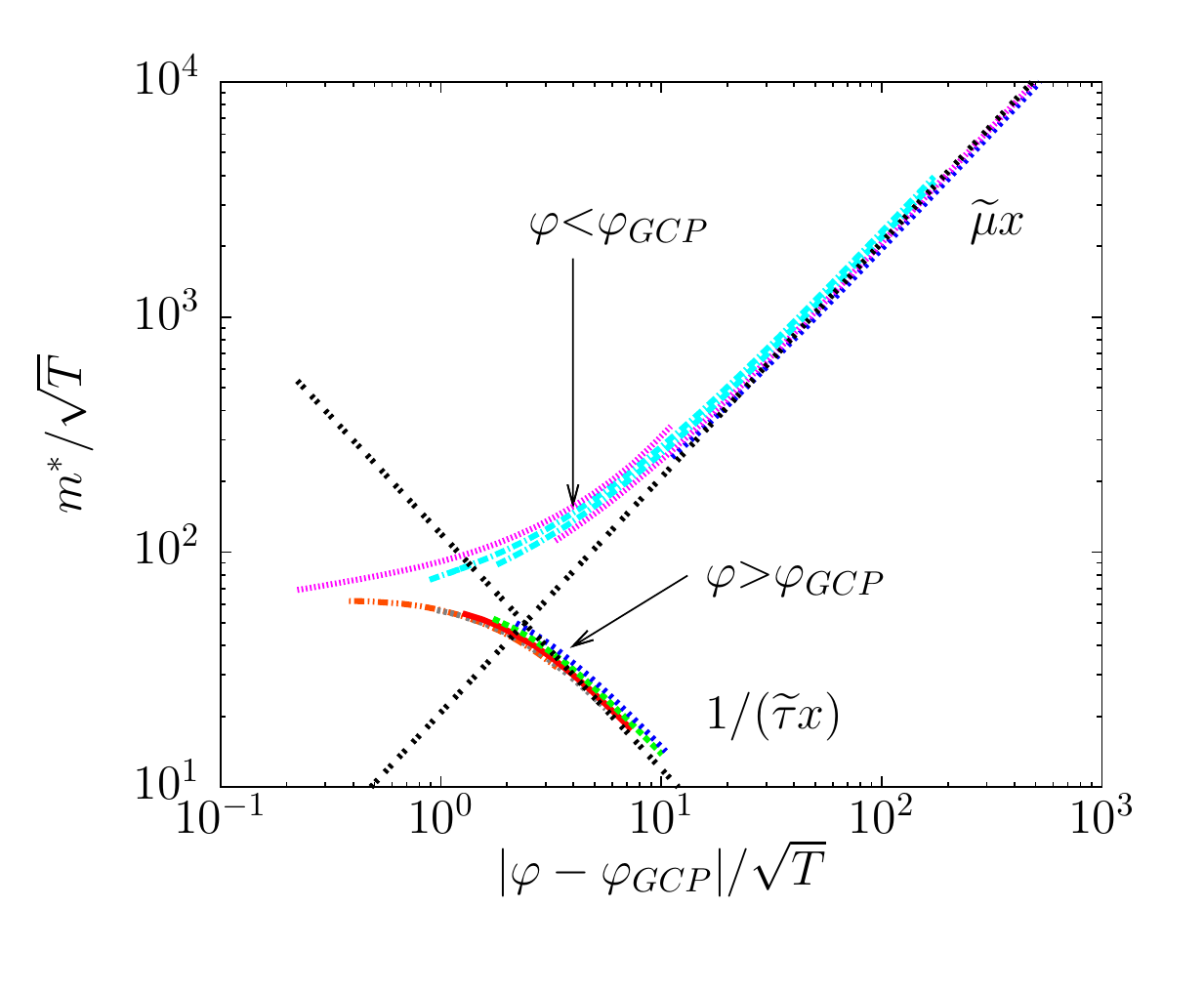}
\caption{(From~\cite{BJZ11}) Results for soft spheres in $d=3$.
(Left) 
Optimal number of replicas $m^*$ as a function of $T$, for several volume fractions. The inset shows the $T \to 0$ limit of $m^*$ as a function of the volume fraction.
(Right)
Scaling of $m^*(T,\ph)/\sqrt{T}$ as a function of the rescaled temperature 
$|\ph-\ph_{GCP}|/\sqrt{T}$ for $\varphi=0.58 , 0.59 , \ldots , 0.70$. The asymptotic 
forms corresponding to Eq.~(\ref{eq:mstarscaling}) are also reported.
}
\label{fig:scaling_m}
\end{center}
\end{figure}

It turns out from the solution of the replica equations~\cite{BJZ11}
that all the thermodynamic quantities are singular at $\ph_{j}$
and $T=0$;
for instance $m^*$ is finite below $\ph_{j}$ while it vanishes proportionally to $T$ above $\ph_{j}$.
The reduced pressure is also finite below $\ph_{j}$, while it diverges at $\ph_{j}$ and is formally infinite
above $\ph_{j}$.
Both the energy and the pressure vanish below $\ph_{j}$ while they are finite above $\ph_{j}$.
From this observation, and because all quantities are analytic at finite $T$, 
it follows that all these quantities must satisfy scaling relations if $T$ is small
enough and $\ph$ is close enough to $\ph_{j}$~\cite{OT07,BW08,BW09}. 
In the replica theory, all the scalings are driven by the one of $m^*(\f,T)$, for which one finds
\beq\label{eq:mstarscaling}
\begin{split}
& m^*(T,\ph) = \sqrt{T} \, \widetilde m_\pm\left( \frac{|\ph - \ph_{j}|}{\sqrt{T}} \right) \ , \\
& \wt m_-(x \to \io) = \wt \mu x \ , \\
& \wt m_+(x \to \io) = \frac{1}{\wt\t x} \ .
\end{split}\eeq
This ensures that
in the hard sphere limit $T\to 0$ and $\ph < \ph_{j}$, one recovers $m^* \sim |\f-\f_{j}|$,
while for $\ph > \ph_{j}$,
one finds that 
$m^* = T/|\f-\f_{j}|$. Note that $\tilde\mu$ and $\tilde\tau$ can be computed explicitly from the theory and
they slightly depend on $\f_j$~\cite{BJZ11}.
This scaling is shown in Fig.~\ref{fig:scaling_m}.

\begin{figure}[t]
\begin{center}
\includegraphics[width=.49\textwidth]{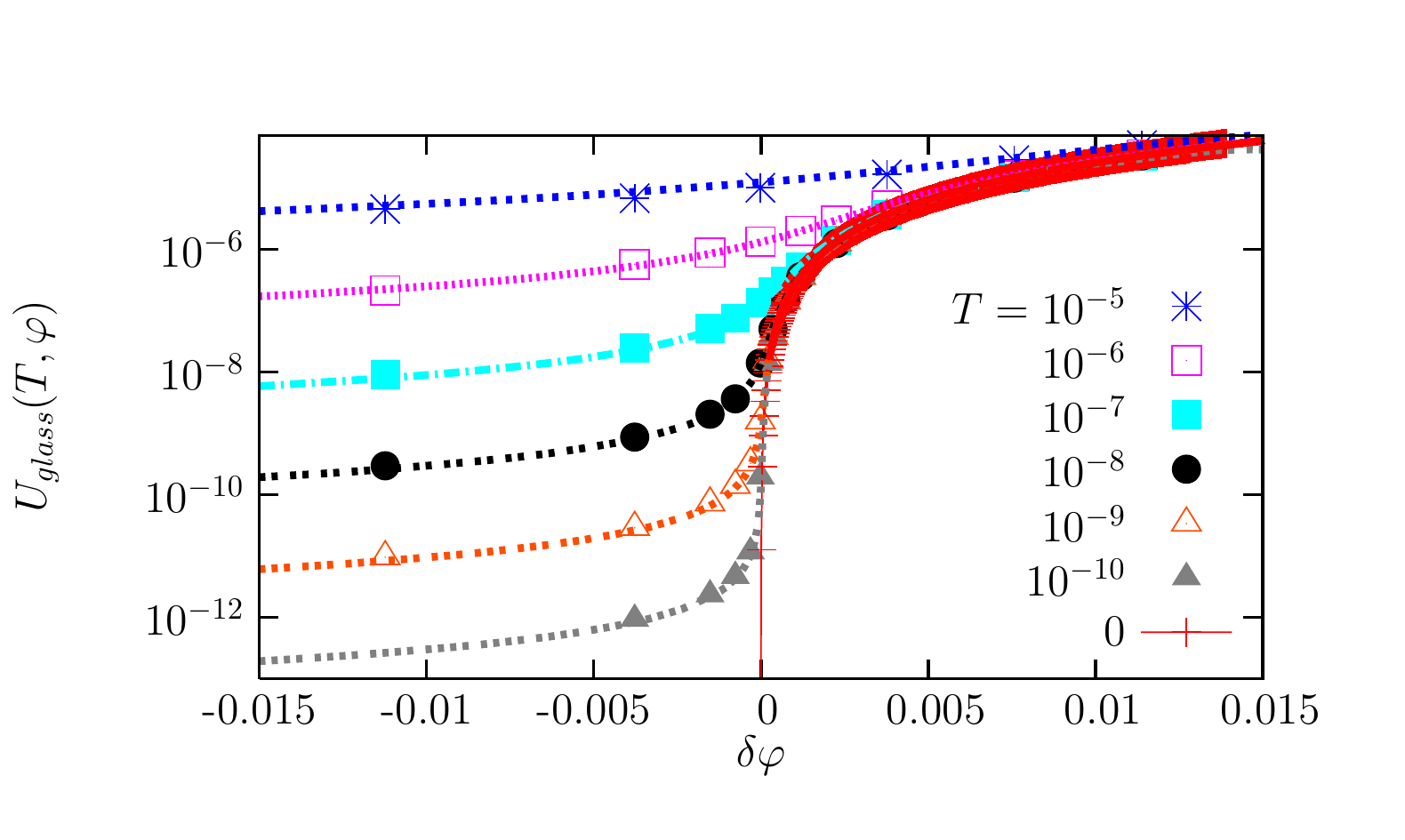}
\includegraphics[width=.49\textwidth]{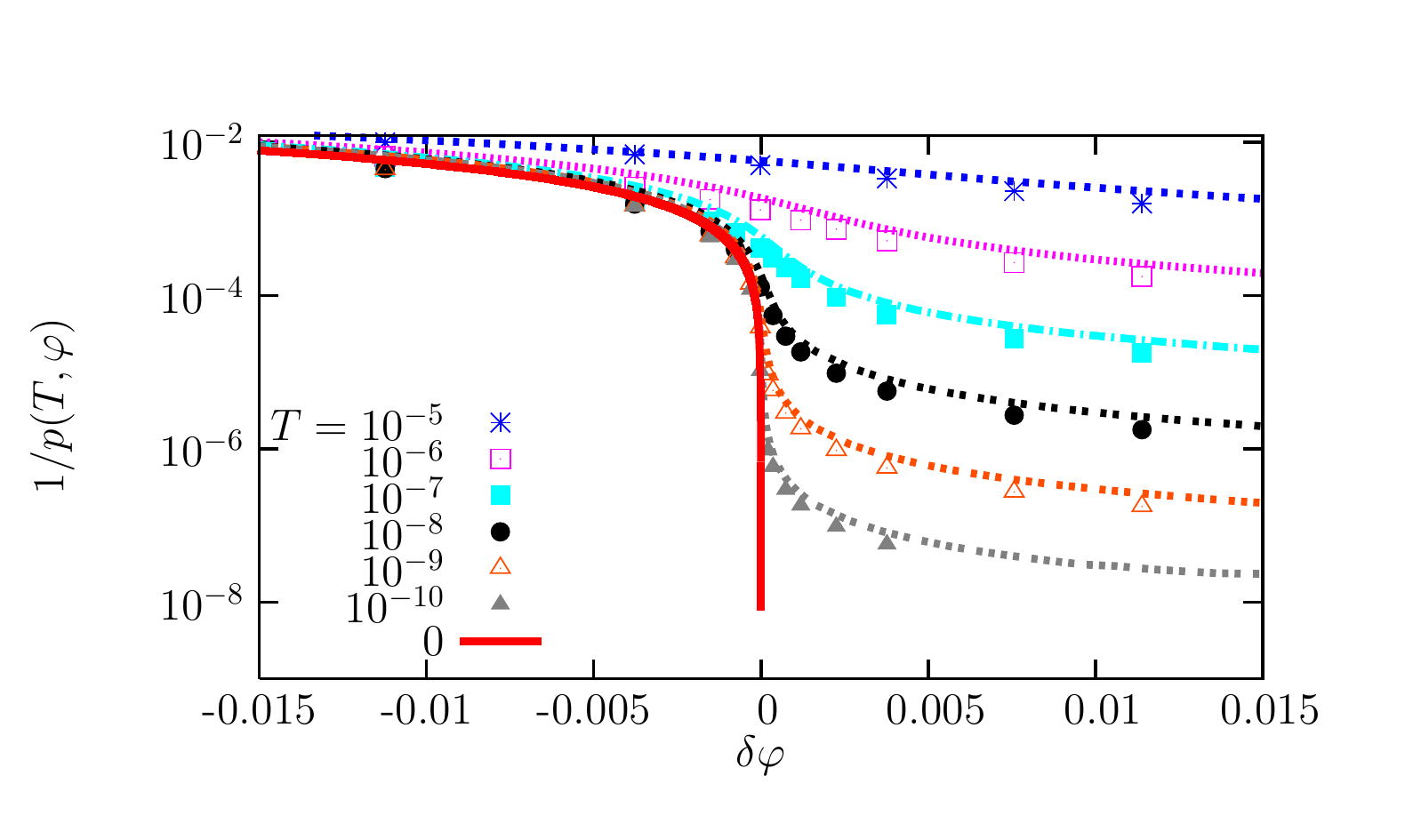}
\caption{(From~\cite{BJZ11})
Energy $U_{glass}(T,\ph)$ (left panel) and inverse reduced pressure $1/p_{glass}(T,\ph)$ (right panel)
as functions of distance from jamming $\d \ph$, for several temperatures.
Here $\d \ph = \ph -\ph_{GCP}$ for the theory (lines),
and $\d \ph = 1.46562 (\ph - \ph_j)$ for the numerical data (points).
}
\label{fig:ep}
\end{center}
\end{figure}

Qualitatively similar scaling forms (in some cases with different exponents) 
apply to thermodynamic quantities such 
as the energy, the pressure, or the complexity. 
Although the corresponding scaling plots are not shown explicitly, 
the energy and the pressure of the glass are plotted and
compared with numerical data in Fig.~\ref{fig:ep}.
The latter have been obtained by preparing an amorphous packing through the energy minimization protocol
discussed above, and then performing standard Monte Carlo simulations at finite temperature in the vicinity of $\f_j \sim 0.643$,
and checking carefully that no structural changes are observed, because in the simulation one is always in the regime $T \ll T_g(\f)$~\cite{BJZ11}.
The agreement between theory and numerical data, with a rescaling of $\d\ph$ discussed in~\cite{BJZ11} that is needed
to correct for a thermodynamic inconsistency of the theory,
is nearly perfect. The scaling around jamming is clearly visible in the figures. For instance, the average energy $U_{glass}(T,\ph)$ tends
to a finite value for $\d\ph>0$, while for $\d\ph < 0$ it goes to zero as a power law, because in this case the system becomes a hard
sphere glass. Similarly, the reduced pressure is finite for $\d\ph<0$, while it diverges proportionally to $\b$ for $\d\ph>0$, because
in this case the pressure is finite at zero temperature. 
At finite temperature, the curves interpolate between the two regimes. Similar scalings have been discussed in~\cite{OH11}.

\begin{figure}
\begin{center}
\includegraphics[width=.49\textwidth]{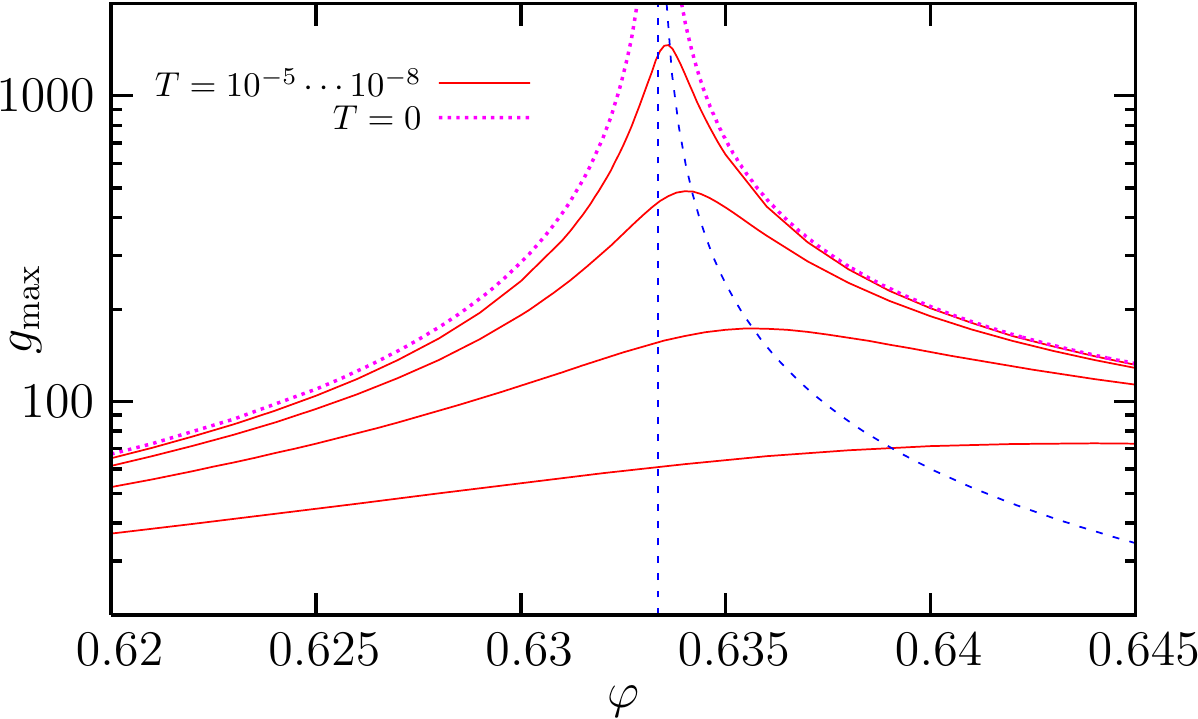}
\includegraphics[width=.49\textwidth]{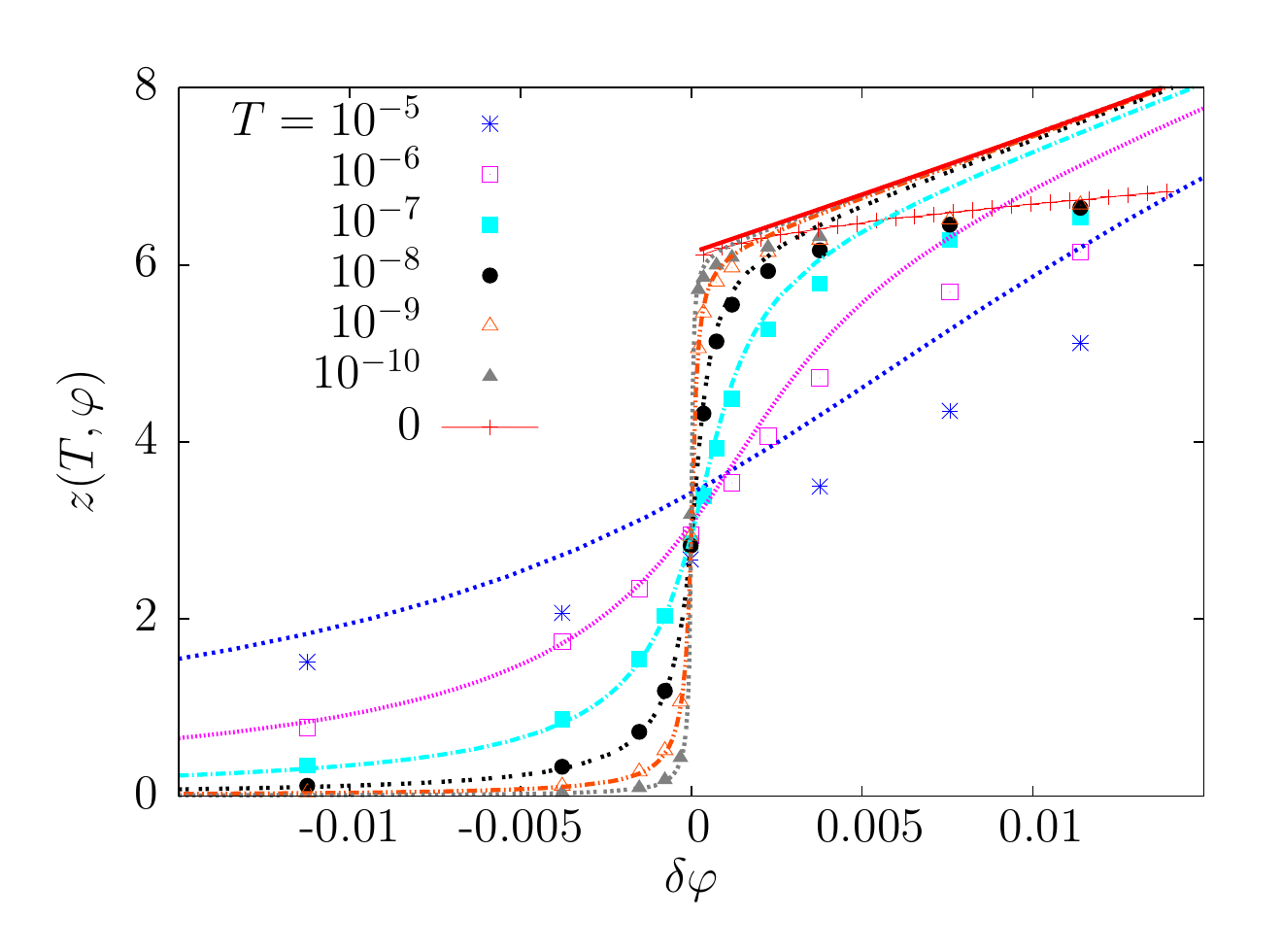}
\caption{(From~\cite{BJZ11})
Left panel:
evolution of the maximum of the glass pair correlation function with $T$ and $\ph$. Right panel: average number of contacts $z$ as a function of distance from jamming $\d \ph$, compared with numerical data (as in Fig.~\ref{fig:ep}), for several temperatures.
}
\label{fig:gmax}
\end{center}
\end{figure}

\subsection{Scaling relation for the structure}

At the jamming transition, a marked structural change happens, as illustrated in Fig.~\ref{fig:PDsch}.
Particles are not in contact below $\f_j$, they just touch at $\f_j$, and they overlap above $\f_j$.
This is clearly revealed in the pair correlation function
$g(r)=(\rho N)^{-1}\langle\sum_{i\neq j}\delta( r+ x_i- x_j)\rangle$
--the Fourier transform of the structure factor $S(k)$--,
and
in the cumulative structure function
\beq\label{Zr}
Z(r) = \rho \, \Omega_{d}  \int_0^r ds s^{d-1} g(s) \ ,
\eeq
where $\Omega_{d}$ is the $d$-dimensional solid angle. 
The function $g(r)$ is proportional to the average number of particles at distance $r$ from a given particle, while
$Z(r)$ counts the average number of particle centers in a shell of radius $r$ around a given particle.

\subsubsection{Number of contacts}

In fact, $g(r)$ develops a delta peak in $r=\s$ when $T=0$ and $\f=\f_j$, corresponding to the touching neighbors.
The coefficient of the delta peak, which is also equal to $\ol{z}=Z(r=\s)$, gives the average number of neighbors.
The maximum of $g(r)$ as well as $\ol{z}$ are reported in Fig.~\ref{fig:gmax} as a function of $(T,\f)$~\cite{BJZ11}. 
It is indeed observed in the theory that $g_{max}$ diverges on both sides of the transition at $T = 0$ as 
$g_{max} \sim |\ph - \ph_{j}|^{-1}$, 
while this divergence becomes a smooth maximum at finite $T$ near the transition, whose position shifts with temperature. 
This behavior compares very well with numerical \cite{DTS05,SLN06} and experimental~\cite{ZXCYAAHLNY09,Ch10}
observations.

Similarly, $\ol{z}$ is a smooth function of density at $T>0$, but it develops a jump
at $T=0$ and $\f=\f_j$. The limit $\ol{z}(T=0,\f\to\f_j^+) = 2d$ corresponds to the so-called {\it isostatic} condition.
This is a crucial property of jammed amorphous packings~\cite{OLLN02,Al98,LNSW10,He10}. Indeed, it is known since
the work of Maxwell that isostatic systems are at the verge of mechanical instability: removing one contact 
creates one ``soft'' mode that allows particles to slide without paying any energy cost.
This is believed to be the origin of all the peculiar mechanical properties of nearly jammed systems~\cite{OLLN02,Wyart,LNSW10,He10}.
It is therefore quite surprising that a mean field theory, that is a priori completely unrelated to the detailed mechanical structure
of the system, is able to predict that $\ol{z}=2d$ at the transition.
However, an important discrepancy between theory and numerical data is revealed in Fig.~\ref{fig:gmax}. In fact, while the theory
predicts $\ol{z} - 2d \propto  \d\f$ for $T=0$ and $\d\f>0$, numerical data show that $\ol{z} - 2d \propto \sqrt{\d\f}$
(we will discuss this point later).

\subsubsection{Scaling of the pair correlation around the delta peak}

\begin{figure}[t]
\begin{center}
\includegraphics[width=.49\textwidth]{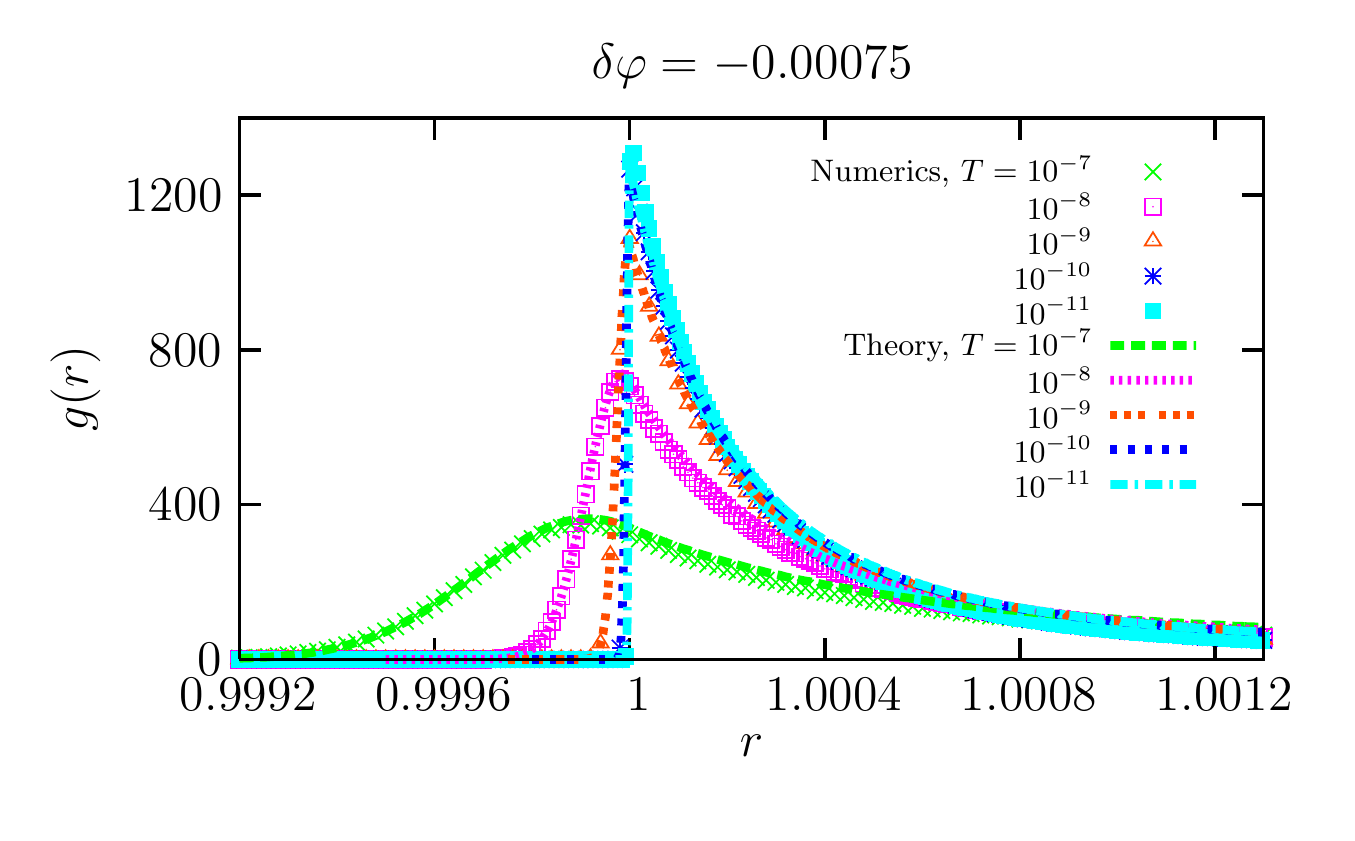}
\includegraphics[width=.49\textwidth]{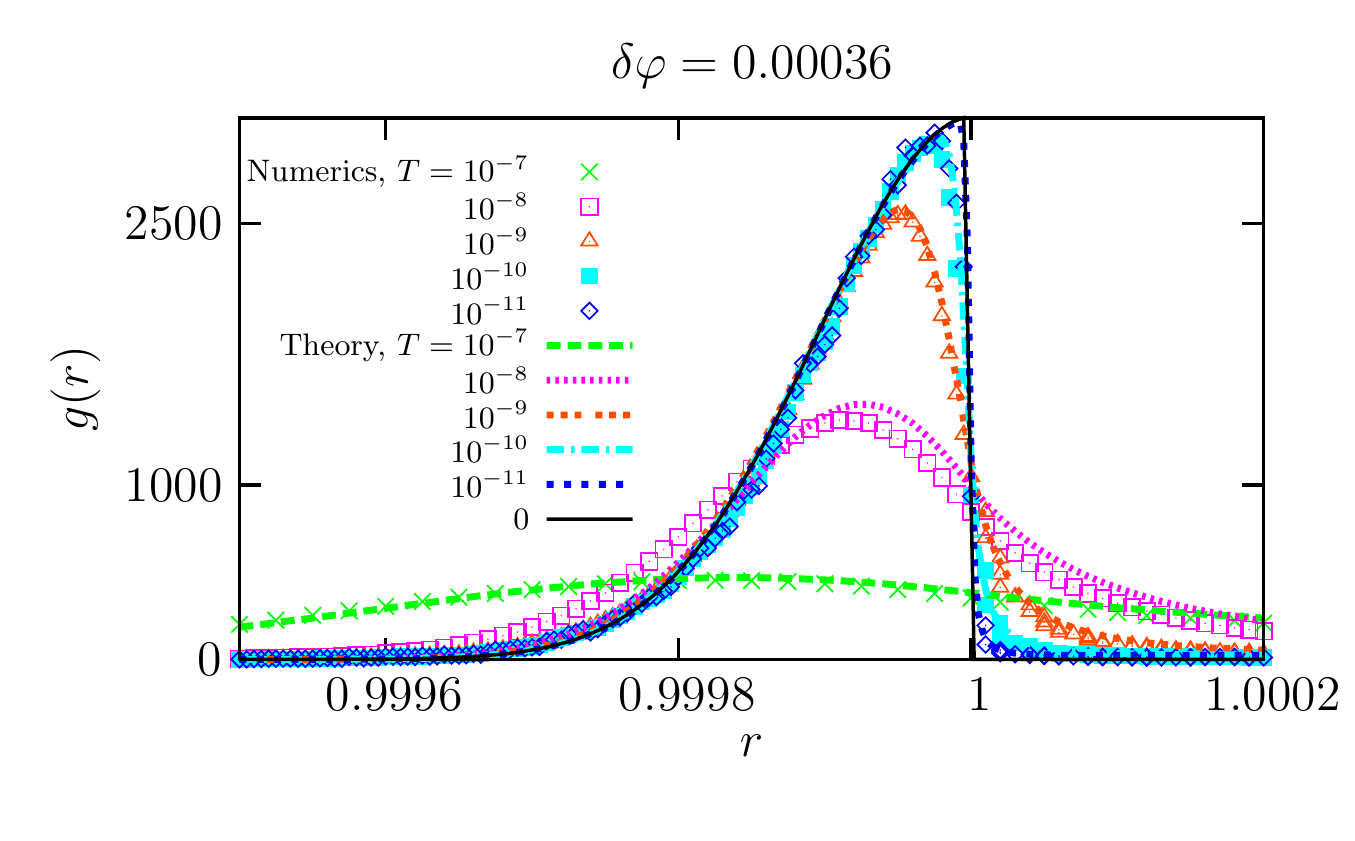}
\caption{(From~\cite{BJZ11})
Pair correlation $g(r)$ in the glass phase just below (left panel) and just above (right panel) 
jamming, predicted by theory 
(full lines) and measured in numerical simulations (symbols).
}
\label{fig:grbelow}
\end{center}
\end{figure}

The shape of $g(r)$ is deeply modified around jamming with respect to the liquid phase, 
due to the development of the contact structure. The most spectacular feature is the emergence
of the delta peak at $T\to 0$ and $\f \to \f_j$. 
Replica theory gives a prediction for the shape of the delta peak in the whole region for $T \gtrsim 0$ and $\f \sim \f_j$,
which is tested against numerical data in Fig.~\ref{fig:grbelow} and is shown to work quite well~\cite{BJZ11}.

\begin{figure}
\begin{center}
\includegraphics[width=0.49\textwidth]{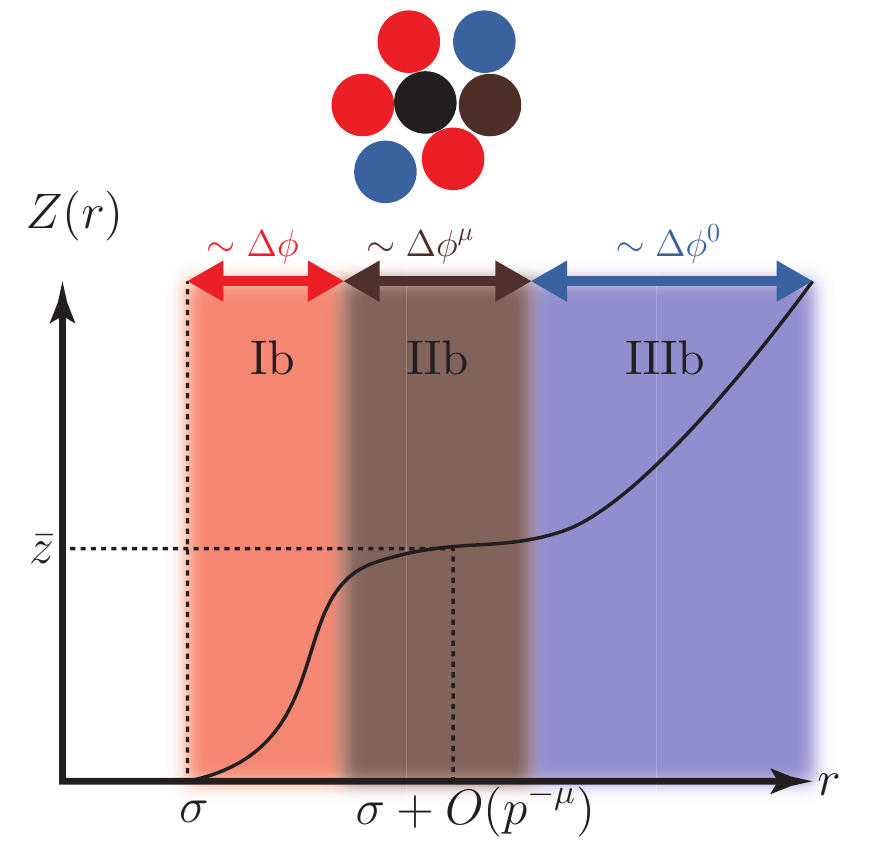}
\includegraphics[width=0.49\textwidth]{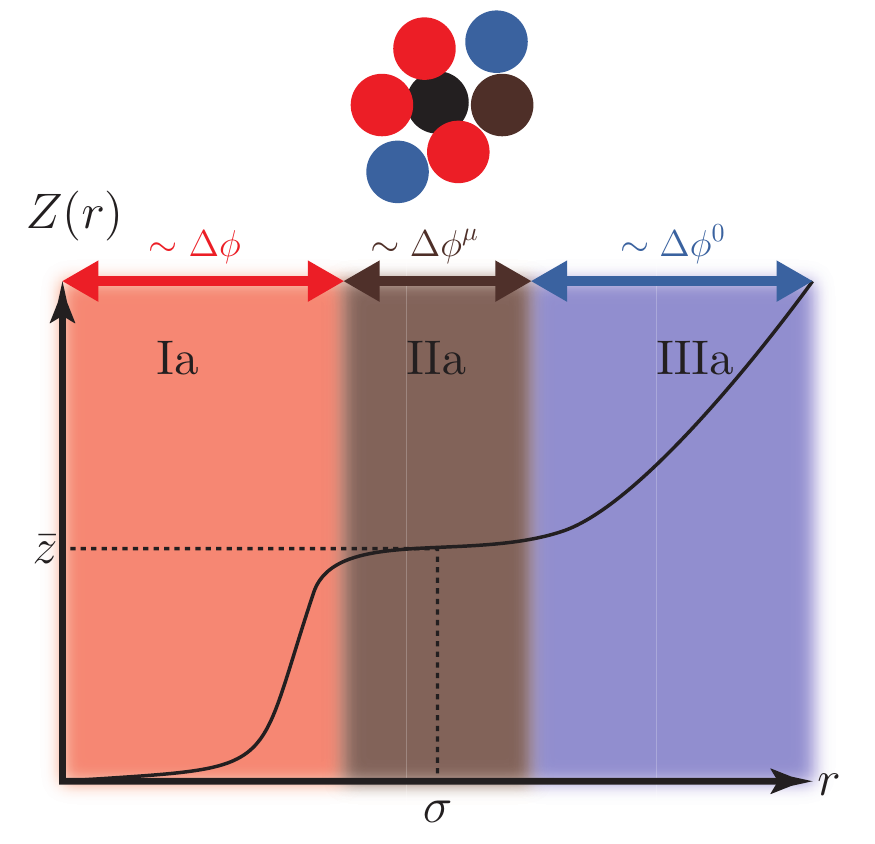}
\caption{(From~\cite{CCPZ12})
Schematic of $Z(r)$ when approaching jamming {\bf (b)} from below  and {\bf (a)} from above.
Three distinct scaling regimes can be identified.
The first regime is related to the growth of $Z(r)$ from 0 to $\ol{z}$. It corresponds
to interparticle gaps $h = |r-\s| \sim \D\f$, and hence to
particles that are in contact when $\D\f\to 0$. The last regime corresponds to gaps $h$ that remain finite for $\D \f\to 0$,
and therefore these particles remain separated at jamming.  These small gaps, $Z(r) -\ol{z} \propto h^{1-\a}$, nonetheless result in a ``quasi-contact'' regime.
The intermediate regime corresponds to gaps $h  \sim \D\f^\mu$. It matches
the two other regimes and disappears when $\D\f\to 0$. 
}
\label{fig:scheme}
\end{center}
\end{figure}

A more detailed investigation can be performed at $T=0$, approaching $\f_j$ from above and from below
using protocols {\it (i)} and {\it (ii)} described in Sec.~\ref{sec:RCP}~\cite{CCPZ12}.
For both protocols, at fixed $T=0$ and $\f\sim \f_j$ with $\D\f = | \d\f|= | \f - \f_j |$, it is found that
$Z(r)$ jumps from 0 to the isostatic average number of contacts $\ol{z}\approx2d$ around $r = \s$, on a scale $r-\s \propto \D \f$.
This regime corresponds to the contact delta peak of $g(r)$.
The plateau is extended by a second power-law regime that corresponds to particles in ``quasi-contact'', carrying no force at jamming.
In this regime the scaling is the same for both protocols and $Z(r)$ displays the same growth
until it reaches the trivial large $r$ regime. 
A third intermediate regime is required to match the first two around the plateau, as illustrated in Fig.~\ref{fig:scheme}.

Consistency sets clear scaling requirements for the different regimes.
When approaching jamming with protocol \emph{(i)}, $Z(r)=0$ for $r<\s$, and for $r\geq \sigma$, the pressure $p \propto \D\f^{-1}$ 
parametrizes the scaling function for $Z(r\geq\sigma)$.
A first scaling regime $r-\s \sim p^{-1}$ sees $Z(r)$ grow from 0 to the average number
of ``contacts'' $\ol{z}$ as 
\begin{equation}
Z(r) = \ol{z} \ZZ_-[(r-\s) p /\s ]
\end{equation}
with $\ZZ_-(x) \sim 1 - C x^{-1-\theta}$ when $x\to\io$ for a constant $C$~\cite{DTS05}. 
Force-bearing contacts are only observed at jamming proper, but their signature develops asymptotically.
A second regime for finite $r-\s$ has 
\begin{equation}
Z(r) = \ol{z} + C' (r-\s)^{1-\a}\ , 
\end{equation}
where $C'$ is a constant.  At jamming, these nearly touching ``quasi-contacts'' carry no force. 
For large $r$, a trivial regime develops independently of $|\Delta\varphi|$. 
Matching the first two scaling regimes implies the existence of an additional intermediate regime for
 $r-\s \sim p^{-\mu}$ where
\begin{equation}
Z(r) = \ol{z} + p^{\nu-\mu} \HH_-[(r-\s) p^\mu/\s]
\end{equation}
with $\mu<1$ and $\nu<\mu$ (Fig.~\ref{fig:scheme}).
Consistency then requires that $\HH_-(x\to 0) \propto -x^{-1-\theta}$ and $\HH_-(x\to\io) \propto x^{1-\a}$ with 
scaling relations $\nu = \a \mu$ and $\mu = (1+\th)/(2+\th-\a)$.

When approaching jamming with protocol \emph{(ii)} from above, the remaining overlaps provide
a scaling variable $\e = \sqrt{e \, d/(2\ee)}\propto|\Delta\f|$~\cite{BJZ11}. In spite of the very different preparation protocol, similar structural regimes are identified.
In the $r<\s$ contact regime, $Z(r)$ grows from 0 to $\ol{z}$ by a universal scaling function 
\begin{equation}
Z(r) = \ol{z} \ZZ_+[(\s-r) \e^{-1} / \s ]
\end{equation}
with $\ZZ_+(x) \sim 1 - C'' x^{1+\th}$ when $x\to 0$ for a constant $C''$.
For $r>\s$, here again 
\begin{equation}
Z(r) = \ol{z} + C' (r-\s)^{1-\a}\ ,
\end{equation}
hence the two regimes must be matched
by an intermediate scaling function $\HH_+$ for $r - \s \sim \e^{\mu}$
\begin{equation} 
Z(r) = \ol{z} + \e^{\mu-\nu} \HH_+[(r-\s) \e^{-\mu}/\s]
\end{equation}
with  $\mu>1$  and $\nu<1$ (Fig.~\ref{fig:scheme}). Consistency here requires that $\HH_+(x \to -\io) \propto  - |x|^{1+\th}$ and $\HH_+(x\to\io) \propto x^{1-\a}$ with $\nu = \a\mu$ and $\mu = (1+\th)/(\a+\th)$. 

\begin{figure}
\begin{center}
\includegraphics[width=0.49\textwidth]{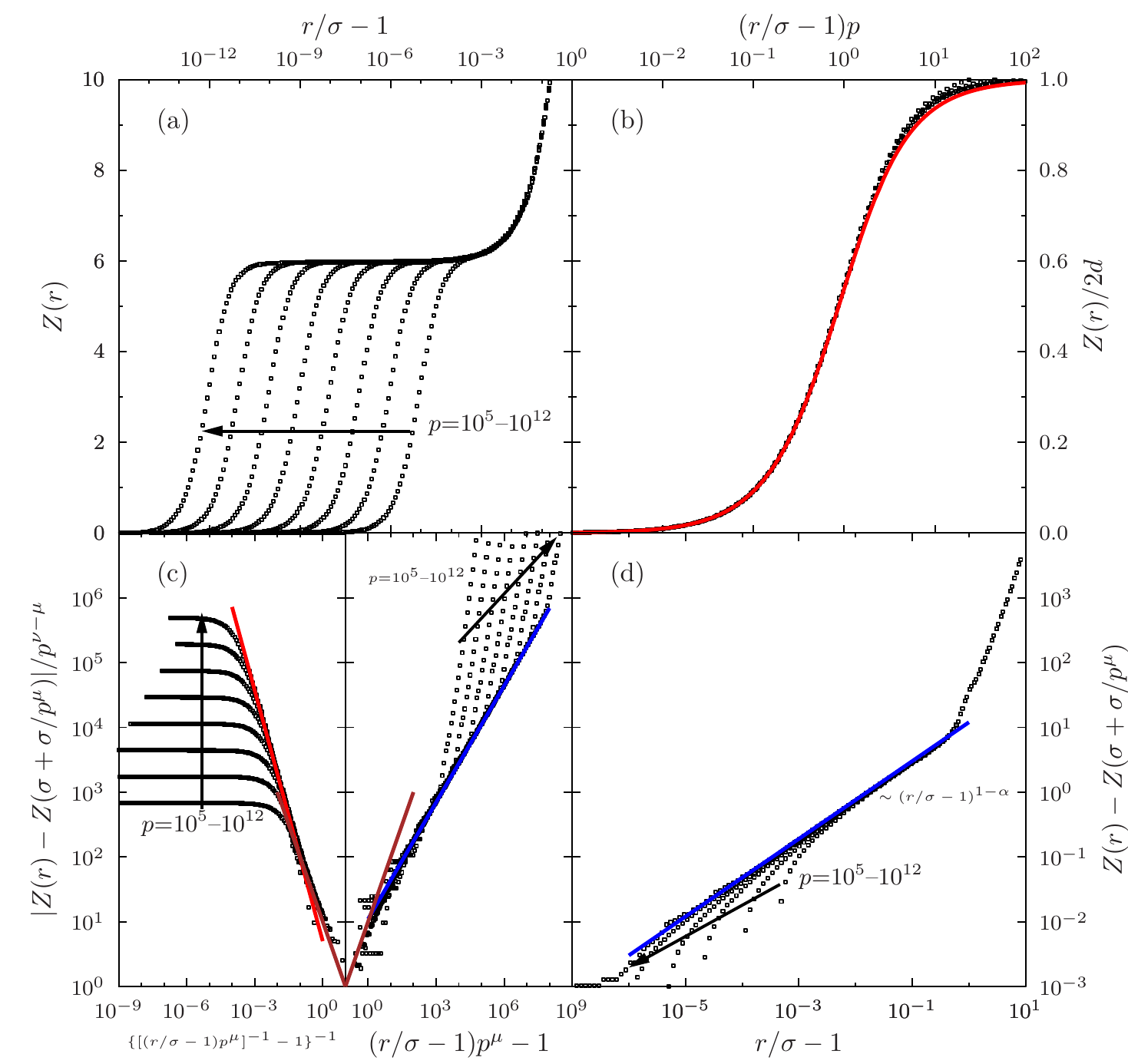}
\includegraphics[width=0.49\textwidth]{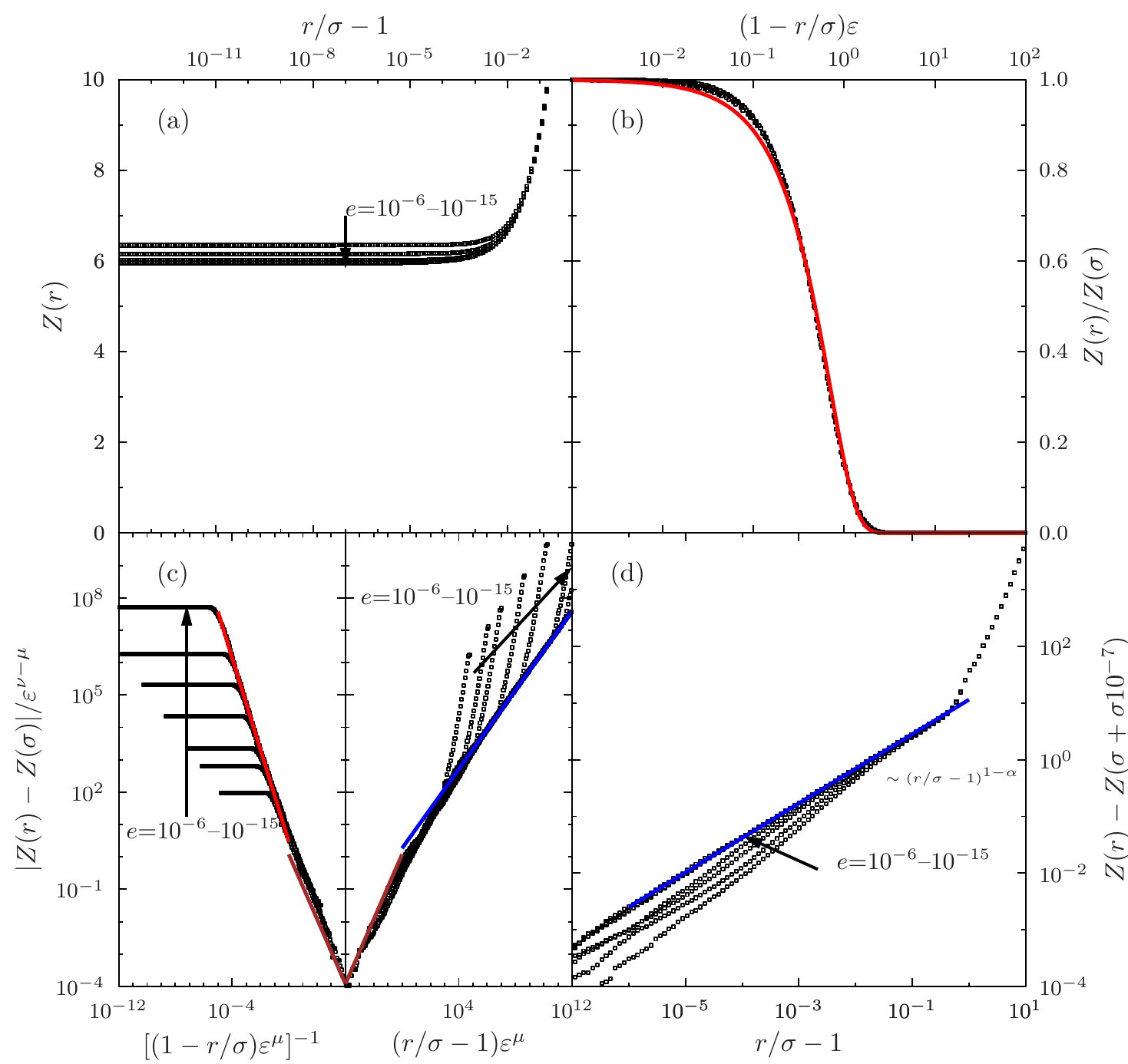}
\caption{
(From~\cite{CCPZ12})
Left panel: scaling of $Z(r)$ upon approaching jamming by hard sphere compressions ($p \propto | \D\f|^{-1} \to\io$) in $d=3$.
{\bf (a)} For different $p$, $Z(r)$ grows on a scale $r-\sigma \sim p^{-1}$ to a plateau at $\ol{z} \sim 6$, the isostatic value.
{\bf (b)} The small $r-\sigma$ regime shows the ``contacts'' scaling function $\ZZ_-(x)$, which agrees well with the G-RT prediction (red line) at small $r-\sigma$. 
{\bf (c)} Rescaling $Z(r)$ using $\mu = (1+\th)/(2+\th-\a)$ and $\nu = \a \mu$ highlights the behavior of the scaling function $|\HH_-(x) - \HH_-(1)|\sim10x$ (brown line) along with the $\theta=0.30(3)$ (red line) and $\alpha=0.40(1)$ (blue line) power-law regimes. 
{\bf (d)} The ``quasi-contact'' regime with $\a= 0.40(1)$ (blue line) eventually gives way to the trivial $r^d$ regime as $r\gg\sigma$~\cite{DTS05}.
\newline
Right panel: scaling of $Z(r)$ upon approaching jamming by soft sphere energy minimization from above in $d=3$.
Recall that the energy $e\sim|\Delta \phi|^2$ and the scaling variable $\e = \sqrt{e \, d/(2\ee)}\propto|\Delta\f|$.
{\bf (a)} For diminishing $e$, the height of the plateau converges to the isostatic value  $Z(\sigma)$=$2d$. 
{\bf (b)} The small $r<\sigma$ regime shows the ``contact'' scaling function $\ZZ_+(x)$, which agrees well with the G-RT prediction (red line) at small $r-\sigma$. 
{\bf (c)} Rescaling $Z(r)$ using $\mu = (1+\th)/(2+\th-\a)$ and $\nu = \a \mu$ highlights the behavior of the scaling function $|\HH_-(x) - \HH_-(1)|\sim1.2x$ (brown line) along with the $\theta=0.42(2)$ (red line) and the $\alpha=0.39(1)$ (blue line) power-law regimes. 
{\bf (d)} The ``quasi-contact'' regime with $\a\sim 0.39(1)$ (blue line) eventually gives way to the trivial $r^d$ regime as $r\gg\sigma$.
}
\label{fig:gr_scaling_below}
\end{center}
\end{figure}

These scalings are verified using numerical data from~\cite{CCPZ12} in Fig.~\ref{fig:gr_scaling_below}.
The Gaussian replica theory provides precise and simple
predictions for the contact regime scaling function $\ZZ_{\pm}(x)$~\cite{PZ10,BJZ11,CCPZ12}:
\beq\label{GRT_scaling}
\begin{split}
\ZZ_-(x) &= 1 - e^{\frac\pi2 x^2} \left[ 1- \erf\left( \frac{\sqrt\pi}2 x \right) \right] \ , \\
\ZZ_+(x) &= 1 -\erf(x) \ . 
\end{split}
\eeq
These predictions are found to be extremely accurate when $x$ is of order 1, 
but they fail to capture the power-law tails at large or small $x$
(Fig.~\ref{fig:gr_scaling_below}). 
Gaussian RT indeed predicts an exponent $\th=0$ for both protocols, and completely misses
the power-law divergence related to $\a$, predicting $\a=0$. 
Note that, although the exponent $\th$ plays an important role as it will be shown below, the deviation
between the RT prediction and numerical data is barely visible in Fig.~\ref{fig:gr_scaling_below}, and it is not visible 
at all in Fig.~\ref{fig:grbelow} because it is hidden in the tails of the delta peak, which are also smoothed by thermal fluctuations.

The origin of these discrepancies is not clear yet.
One possibility is that they are due to the
Gaussian assumption for the cage form, 
which was found to be erroneous as detailed in Sec.~\ref{sec:nonGaussian} and in~\cite{CIPZ12}.
It is possible that the non-Gaussian replica theory of~\cite{KPZ12} will lead to an improved prediction, however this has still to be worked out.
The other possibility is that these discrepancies have their origin in the physics of the low frequency vibrational modes (often called ``soft modes'') 
that appear at the jamming transition due to the approaching mechanical instability~\cite{OLLN02,Wyart,LNSW10,He10}. 
It is worth stressing that the analysis of Fig.~\ref{fig:gr_scaling_below} has been repeated
in~\cite{CIPZ12} in dimensions $d=3 \cdots 10$ and no difference has been found. The fact that these scalings persist up to high dimensions strongly suggest that the mean
field theory should be able to capture them. However, how to include the soft modes physics in the replica theory is not clear yet.

\subsubsection{Force distributions and mechanical stability}

\begin{figure}
\begin{center}
\includegraphics[width=0.7\textwidth]{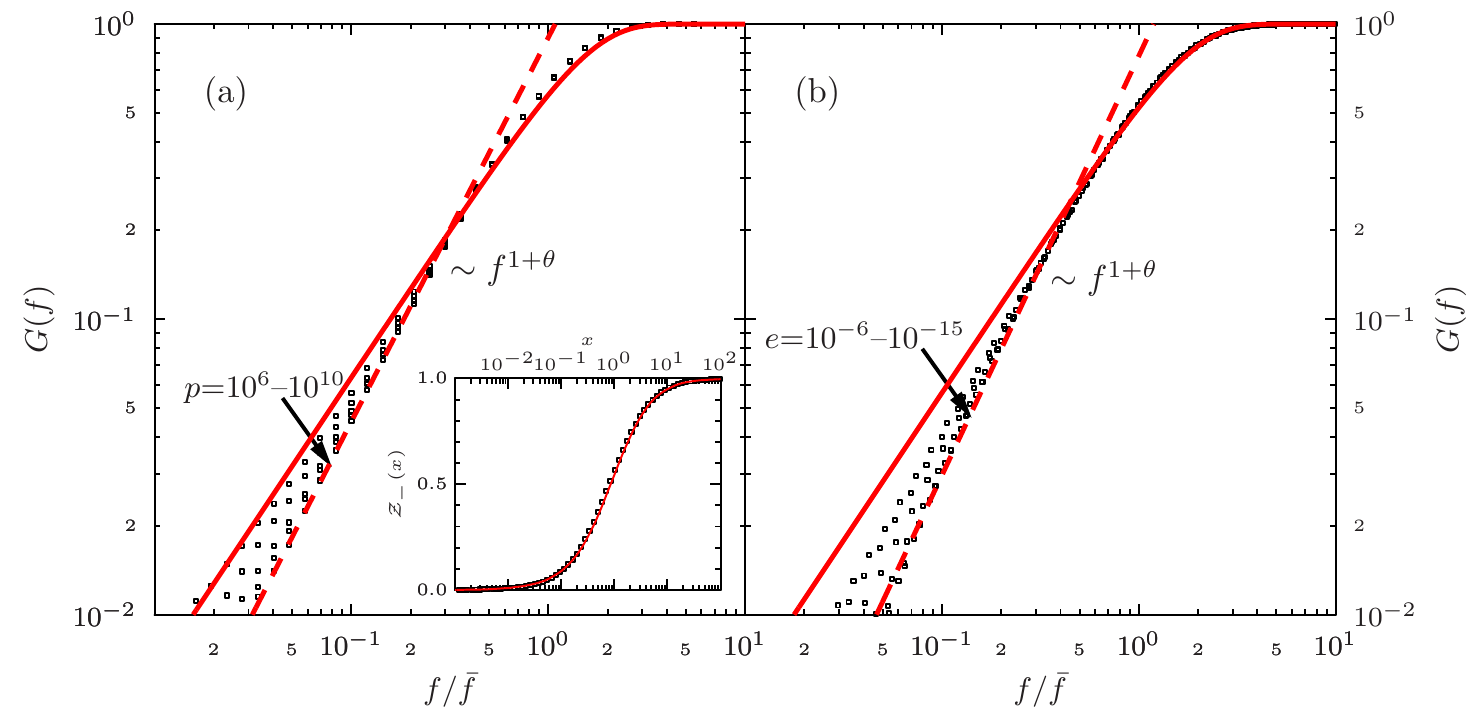}
\caption{(From~\cite{CCPZ12})
Cumulative force distribution $G(f)$ compared with the Gaussian RT prediction (solid line)  {\bf{(a)}} for HS with power-law tail exponent $\theta=0.30(3)$ 
(dashed line) and {\bf (b)} for SS with power-law tail exponent $\theta=0.42(2)$ (dashed line). Inset: test of Eq.~\eqref{eq:FP} in $d$=3.
The left hand side (from distances) is plotted with points, the right hand side (from forces) is plotted with a solid line.
}
\label{fig:forces}
\end{center}
\end{figure}

The scaling functions $\ZZ_\pm(x)$ characterize the contact regime and 
are related to corresponding scaling functions for the probability distribution of inter-particle forces $f$. Here again, it is convenient to consider
the cumulative distribution $G(f)$ rather than the distribution itself.

For hard spheres approaching jamming from below, the average force $\overline{f}\propto P$ and
the cumulative force distribution $G(f)$
approaches a scaling function defined by $G(y \overline{f}) \to \GG_-(y)$.
In the contact regime the force and distance distributions are also related through a Laplace transform
\beq\label{eq:FP}
\ZZ_-(x) = 1 - x \int_0^\io dy \, \GG_-(y) \, e^{-x y} \ ,
\eeq
as suggested in~\cite{DTS05} and tested in Fig.~\ref{fig:forces}. 
It follows that if $\ZZ_-(x) \sim 1 - C x^{-1 - \th}$ for $x\to\io$, then $\GG_-(y) \sim y^{1+\th}$ for $y\to 0$. 
The low-force distribution is indeed consistent with $G(f) \propto f^{1+\th}$ and $\th =0.30(3)$ (Fig.~\ref{fig:forces}). 
For soft spheres approaching jamming from above, the pair potential sets the relation between the force and the pair distributions: 
it gives $f = 2 (\ee/\s)( 1-r/\s )$ for $0\leq r \leq \s$ and zero otherwise, so
$G(f) = 1 - \frac{Z(\s - f \s^2 / (2 \ee))}{Z(\s)}$. In the jamming limit $G(2 y \, \e \, \ee/\s) \to \GG_+(y) = 1 - \ZZ_+( y )$,
and therefore $\GG_+(y) \sim y^{1+\th}$, as in the previous case.
Indeed, here again, the low-force tail is consistent with $\th= 0.42(2)$ (Fig.~\ref{fig:forces}). 
For both protocols, however, the regime intermediate between contacts and quasi-contacts results in deviations from this power-law decay at very weak forces away from jamming.

The exponents $\th$ and $\a$ defined above might have an important role in the mechanical stability of the amorphous packings.
In fact, it has been proposed by Wyart~\cite{Wy12} that $\alpha \geq 1/(2+\theta)$ is required for mechanical stability.
Both the soft spheres values ($\a= 0.39(1)$, $\th=0.42(2)$) and the hard sphere ones ($\a=0.40(1)$, $\th= 0.30(3)$), however, indicate a slight violation of this condition.
A generalized stability condition of the form $\a \geq (1-\d/2)/(2+\th-\d/2)$~\cite{Wy12} is consistent with the numerical findings 
for $\d\gtrsim 0.2$, but a direct test of this extended relation goes beyond the scope of the current analysis because the measurement of the value of $\d$ requires 
a completely different numerical procedure.

\subsection{Binary mixtures}
\label{sec:binary}

The replica theory can be extended to binary mixtures~\cite{BCPZ09}. This is important
because in these systems crystallization can be easily avoided also in $d=3$; this allows one to
study their glass transition in great 
detail, see \eg~\cite{GV03,FGSTV03,FGSTV04,DTS06,SK00,BW08,BW09,HD09}.
Jamming of binary mixtures has also been extensively investigated,
see \eg~\cite{Do75,Do80,OT81,OLLN02}.
Their investigation allows one to test the prediction of the theory concerning the variation
with composition of packing density, structure, etc. 
Here results are reported for a binary mixture of two types of three-dimensional
spheres $\mu=A,B$ in a volume $V$, 
with different diameter $\s_\mu$ and density $\r_\mu = N_\mu/V$. 
Define $r = \s_A / \s_B > 1$ the diameter ratio and $x= N_A/N_B$ the concentration ratio;
$\f = \r_A V_3(\s_A)+\r_B V_3(\s_B)$ the packing fraction;
$\eta = \r_B V_3(\s_B) / \f = 1/(1+ x r^3)$ the volume fraction of the small ($B$) component.
The equation of state used in \cite{BCPZ09} to describe the liquid is a generalization of
the Carnahan-Starling equation. The theoretical results were compared with numerical
results obtained from the Lubachevsky-Stillinger algorithm discussed in 
section~\ref{sec:nonGaussian}.

\begin{figure}
\begin{center}
\includegraphics[width=.6\textwidth]{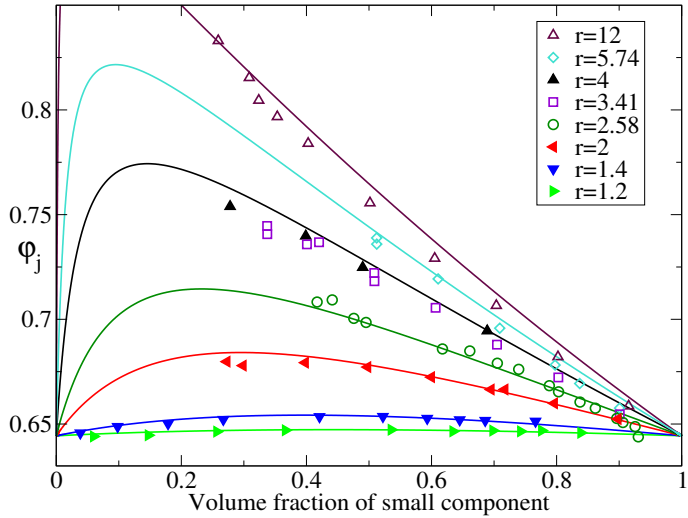}
\caption{
(From~\cite{BCPZ09})
Packing fraction $\f_j$ as a function of 
$\eta = 1/(1+x r^3)$ at fixed $r$.
Full symbols are numerical data from~\cite{BCPZ09}. Open symbols are experimental results from~\cite{YCW65}. 
Lines are predictions from theory. Note that the large $r$-small $\h$ region cannot be explored, because for
such very asymmetric mixtures the large spheres form a rigid structure while
small spheres are able to move through the pores and are not jammed~\cite{Do80,OT81}.
}
\label{fig:phij}
\end{center}
\end{figure}

\begin{figure}
\begin{center}
\includegraphics[width=.4\textwidth]{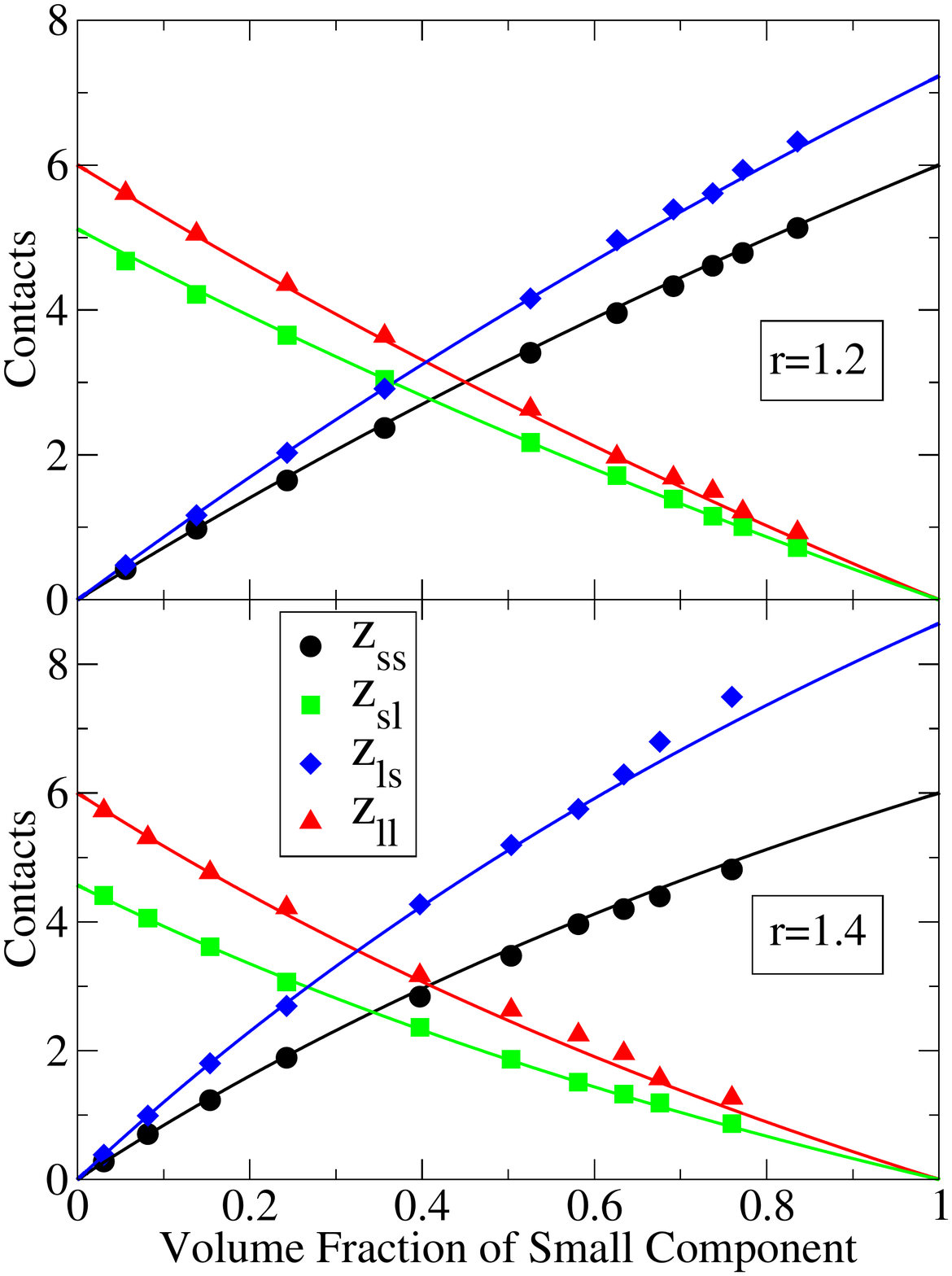}
\includegraphics[width=.4\textwidth]{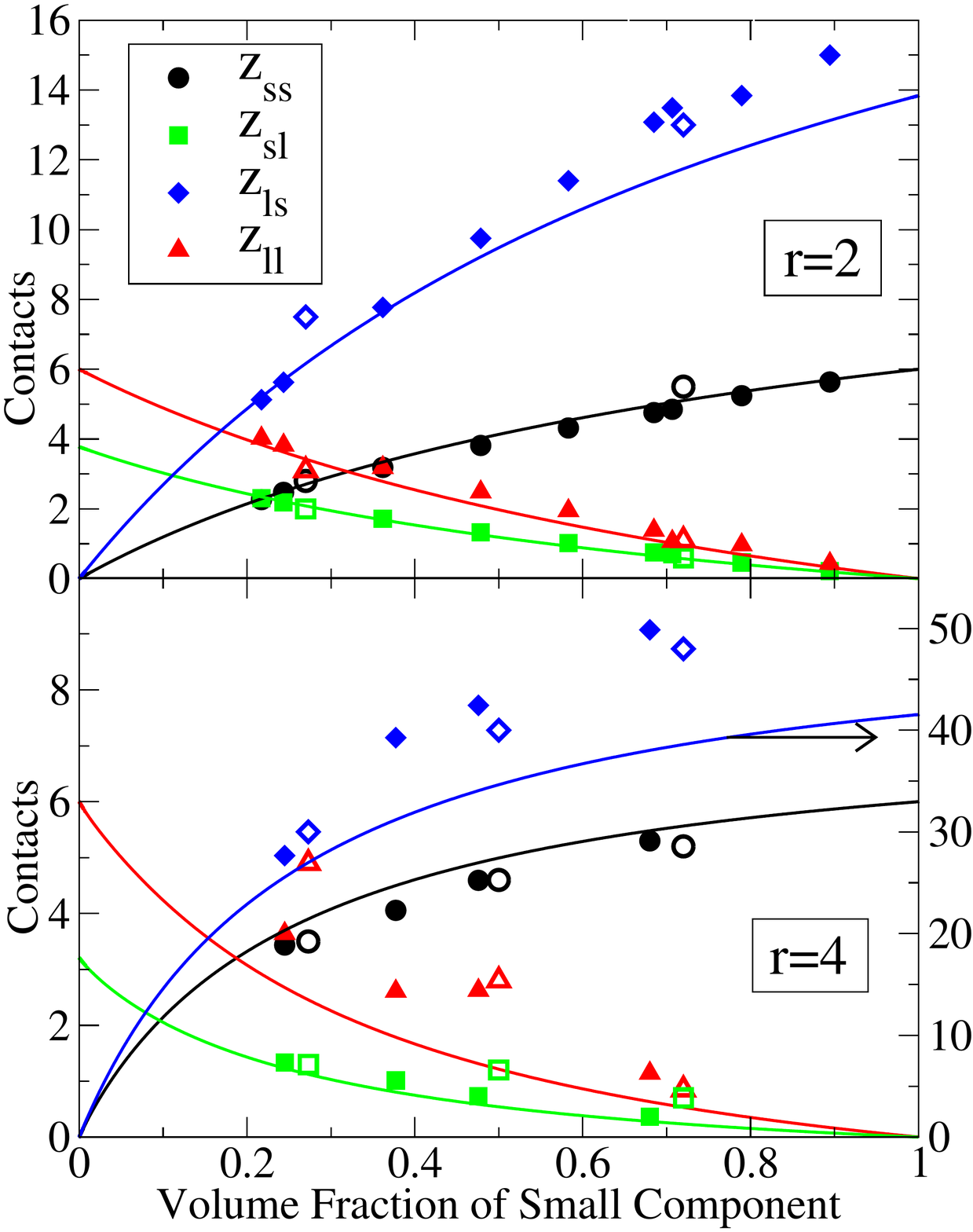}
\caption{(From~\cite{BCPZ09})
Partial average coordination numbers 
(small-small, small-large, large-small, large-large)
as a function of
volume fraction of the small particles $\h= 1/(1+x r^3)$ for different values of $r$.
Full symbols are numerical data from~\cite{BCPZ09}. Open symbols are experimental
data from \cite{PZYZM98}.
Note that in the lower right panel a different scale is used for $z_{ls}$.
}
\label{fig:contacts}
\end{center}
\end{figure}

In Fig.~\ref{fig:phij}, the jamming density $\f_j$ is reported for different mixtures, putting
together numerical results~\cite{BCPZ09}, experimental data~\cite{YCW65}, 
and the theoretical results. The latter have been obtained by fixing the complexity
to a single value for all mixtures, such that for the monodisperse system $\f_j \sim 0.64$.
Note that a single ``fitting'' parameter allows one
to describe different sets of independent numerical and experimental data.
The prediction of the theory are qualitatively similar to previous ones~\cite{Do80,OT81},
but the quantitative agreement is much better. Interestingly, a similar qualitative behavior
for $\f_{d}$ has been predicted by Mode-Coupling theory~\cite{GV03,FGSTV03}; 
although there is no {\it a priori}
reason why the variation with mixture composition of $\f_{d}$ and $\f_j$
should be related, it is reasonable to
expect that they show similar trends~\cite{FGSTV03}.

The average coordination numbers at $\f_j$ are denoted $z_{\m\n}(\f_j)$, 
and it was checked in~\cite{BCPZ09} that their variations with $\f_j$ are negligible.
They are reported in figure~\ref{fig:contacts} for different mixtures. 
The numerical values have been obtained in~\cite{BCPZ09}.
Experimental data from~\cite{PZYZM98} are also reported
in the right panel of figure~\ref{fig:contacts}.
The total coordination
is close to the isostatic value $z=6$, which is the value predicted by the theory
also for binary mixtures~\cite{BCPZ09}.
As it can be seen from figure~\ref{fig:contacts},
the computed values agree very well with the outcome of the numerical simulation, at
least for $r$ not too large, while some discrepancies are observed in the contacts of the large
particles for large $r$.

\subsection{Discussion}

In summary, replica theory provides a quite satisfactory mean field theory of the jamming transition.
First of all, the very fact that the theory predicts the existence of the transition from first principles is non trivial.
Furthermore, the basic phenomenology of the transition, with the associated diverging peak of $g(r)$ and jump
of the number of contacts to the isostatic value, are reproduced. And finally, quite accurate predictions for the scaling
functions $\ZZ_\pm(x)$ can be obtained, and compare quite well with numerical data.

Despite these successes, the level of accuracy of numerical simulations around the jamming transition is so high
that very detailed measurements can be made, and these reveal several discrepancies with the theory.
In fact, the exponents $\th$ and $\a$ are predicted by the theory to be $\th = \a =0$ which is wrong.
Furthermore, the growth of the number of contacts above the isostatic limit is predicted to be linear,
while numerics indicate a power-law growth with exponent close to $0.5$.
These discrepancies are crucial, because these non-trivial exponents have been related to the mechanical stability
of the packings~\cite{Wyart,Wy12}, and to the proliferation of soft modes that is connected to the isostatic nature of
the force network.

The results of~\cite{CCPZ12} indicate that all the non-trivial exponents are independent of dimension, therefore
the nature of the jamming transition in $d=13$ is exactly the same as in $d=3$. This strongly suggests a mean field
description and indeed it has been suggested~\cite{GLN12} that the ``upper critical dimension" for the jamming transition is $2$
(although an equilibrium glass transition probably does not exist in $d=2$~\cite{SK00,DTS06}).
Moreover, it has been shown~\cite{MKK08} that a model of hard spheres on the Bethe lattice, which is therefore rigorously
mean field in nature, reproduces all the phenomenology of the jamming transition, including the soft modes physics.
It seems therefore that one can construct a refined mean field theory, capable of predicting the non-trivial exponents and the associated
peculiar mechanical features of nearly jammed packings.

How can one do this in practice? Some different paths can be devised.
The first one is to include the effect of non-Gaussianity of the cage shape in the theory. A first step has been performed in~\cite{KPZ12},
but the detailed predictions of the non-Gaussian theory around jamming have not been worked out yet.
Besides this,
there are reasons to suspect that the exact description of the jamming transition within the replica theory requires an infinite number of replica symmetry breakings~\cite{MPV87}. 
First of all, the generic situation with systems which have a glass transition to a one-step replica symmetry breaking (1RSB) solution describing the glass 
is to have a further transition to a phase with more -- eventually infinite --
levels of replica symmetry breaking. 
In mean field $p$-spin glasses,
1RSB equilibrium states have a spectrum with no soft modes -- and this is true of 
all but the very highest metastable states. A full replica symmetry breaking scheme would instead naturally bring in soft modes, as happens for example in the case of
the Sherrington-Kirkpatrick model
(the analogy between the physics of soft modes and those of the SK model has been already noted in~\cite{Wy12}). 
Performing an infinite replica symmetry breaking calculation in the context of particle systems seems a difficult task: however, the calculation
might be doable in the limit of $d\to\io$ along the lines of~\cite{KPZ12}. A detailed analysis of the Bethe lattice model of~\cite{MKK08} should help to shed light on this difficult problem.
Also, setting up a field theory description of the jamming transition, using the replica method, along the lines of~\cite{FJPUZ12}, might be possible.



\chapter*{Acknowledgments}
\pagestyle{empty}
\addcontentsline{toc}{chapter}{Acknowledgments}

Because these thesis covers work done in the last eight years, a complete list of acknowledgements would be extremely long.
In an effort to keep it short, I am forced to refrain from mentioning many people to which I am indebted for different reasons. 
I hope that they will not feel offended.

I would like to warmly thank all my collaborators on these works: of course without them none of what I presented would have been
done. Among them, I would like to particularly thank Giorgio Parisi and Jorge Kurchan, who have been
continuous sources of inspiration during all this time;
Ludovic Berthier and Patrick Charbonneau, who provided most of the numerical data, were infinitely helpful in interpreting them,
and gave new ideas and motivations to push the theory forward;
and my former (often co-supervised) students Indaco Biazzo, Francesco Caltagirone, Laura Foini, Hugo Jacquin and Pierfrancesco Urbani, 
who accepted enthusiastically to work on these topics providing 
new energies to solve difficult problems on which I had been stuck for long.
I would also like to thank all the colleagues that provided feedback on this work
 and patiently pointed out problems and limits of the theory.

In addition, I would like to thank Guilhem Semerjian for being a fantastic
office mate during most of these years, extremely patient despite the amount of noise that my collaborators and I produced,
and always available for interesting discussion;
and Marco Tarzia for continuous support and many extremely useful discussions.
I am particularly indebted to both of them, not only for being precious collaborators in related projects, but also for a stimulating conversation
that gave me the main inspiration on how to write this thesis.

Last but not least, I would like to thank all the members of the committee for accepting to be part of it, for 
the time they devoted to reading this thesis and discussing it with me, and for their very useful comments on
the manuscript.

\newpage
\pagestyle{empty}

\mbox{}


\backmatter

\addcontentsline{toc}{chapter}{Bibliography}

\bibliographystyle{is-unsrt}
\bibliography{HS}

\end{document}